\documentclass[pdflatex,sn-mathphys-num]{sn-jnl}

\usepackage{color}
\usepackage{ulem}
\usepackage{siunitx}
\usepackage{amssymb}
\usepackage{float}

\newcommand{\bExp}{\mathbb{E}}
\newcommand{\difd}{\mathrm{d}}
\newcommand{\expe}{\mathrm{e}}
\newcommand{\sgn}{\mathrm{sgn}}
\newcommand*{\argmin}{\mathrm{arg\,min}}

\newcommand{\Exprand}{\mathrm{Exp}}
\newcommand{\Ncyc}{N_{\mathrm{c}}}
\newcommand{\dTcyc}{\Delta T_{\mathrm{c}}}

\newcommand{\dTattb}{\Delta T_{1,2}}
\newcommand{\dTdet}[1]{\Delta T_{#1}}
\newcommand{\CM}{M}
\newcommand{\dCMcyc}{\Delta \CM_{\mathrm{c}}}

\newcommand{\dCMattb}{\Delta \CM_{1,2}}

\newcommand{\dCMdetjump}[1]{\Delta \CM^{(\mathrm{d})}_{#1}}
\newcommand{\dCMattjump}[1]{\Delta \CM^{(\mathrm{a})}_{#1}}

\newcommand{\dMcyc}{\Delta M_{\mathrm{c}}}

\newcommand{\dMattb}{\Delta M_{1,2}}
\newcommand{\dMdet}[1]{\Delta M_{#1}}
\newcommand{\dMdetjump}[1]{\Delta M^{(\mathrm{d})}_{#1}}
\newcommand{\dMattjump}[1]{\Delta M^{(\mathrm{a})}_{#1}}

\newcommand{\Talldet}{T}
\newcommand{\Malldet}{L}
\newcommand{\dettime}{\tau_{\mathrm{d}}}
\newcommand{\atttime}{\tau_{\mathrm{a}}}

\newcommand{\Atstate}[1]{Q^{(#1)}}

\newcommand{\OAM}{1AM}
\newcommand{\TAM}{2AM}
\newcommand{\ttam}{t_{0}}
\newcommand{\ttaut}{t_{\mathrm{taut}}}
\newcommand{\ttauttrue}{\ttaut^{\mathrm{TC}}}
\newcommand{\ttautfalse}{\ttaut^{\mathrm{FC}}}
\newcommand{\truecoop}{truly cooperative}
\newcommand{\truecoopabb}{TC}
\newcommand{\falsecoop}{falsely cooperative}
\newcommand{\falsecoopabb}{FC}
\newcommand{\neutralcoop}{neutrally cooperative}
\newcommand{\neutralcoopabb}{NC}
\newcommand{\Veltam}[1]{\mathcal{V}^{(#1)}}
\newcommand{\Rdet}{R^{(\mathrm{d})}}
\newcommand{\Ratt}[1]{R^{(\mathrm{a})}_{#1}}
\newcommand{\detmot}{J^{(d)}}

\newcommand{\attmot}{J^{(a)}}

\newcommand{\pdet}[1]{p_{\mathrm{d}}^{(#1)}}

\newcommand{\Vstate}[1]{V_{#1}}
\newcommand{\Dstate}[1]{D_{#1}}

\newcommand{\Iatt}{I_{\mathrm{a}}}

\newcommand{\barvtwo}{\Vstate{1,2}}
\newcommand{\detone}[1]{\bar d^{\ast}_{#1}}
\newcommand{\dettwo}[1]{\bar d^{\ast(#1)}_{1,2}}
\newcommand{\dettwoboth}{\bar d^{\ast}_{1,2}}
\newcommand{\dettwoz}[1]{\bar d^{(#1)}_{1,2}}
\newcommand{\E}{\mathbb E}

\newcommand{\Rtimegen}{T}
\newcommand{\Rlengthgen}{L}
\newcommand{\Rlength}[1]{\Rlengthgen^{(#1)}}

\newcommand{\Rtime}[1]{\Rtimegen^{(#1)}}
\newcommand{\Nexp}{S}
\newcommand{\Vpool}{V_{\mathrm{ens}}}
\newcommand{\Vrun}{V_{\mathrm{run}}}
\newcommand{\Dpool}{D_{\mathrm{ens}}}
\newcommand{\Drun}{D_{\mathrm{run}}}

\newcommand{\ip}{i^{\prime}}
\newcommand{\jp}{j^{\prime}}
\newcommand{\totdetprob}{p^{\emptyset}_{\mathrm{d}}}

\newcommand{\Feq}{F_{\mathrm{eq}}}
\newcommand{\req}{R_{\mathrm{eq}}}
\newcommand{\pdetlax}[1]{p_{\mathrm{d}, \mathrm{lax}}^{(#1)}}
\newcommand{\pdettaut}[1]{p_{\mathrm{d}, \mathrm{taut}}^{(#1)}}
\newcommand{\ellrest}[1]{\ell_0^{(#1)}}
\newcommand{\pdfnorm}{\mathcal{N}}
\newcommand{\Zlag}{\Xi}
\newcommand{\kbt}{k_BT}
\newcommand{\RdetJtaut}[1]{R^{(d)}_{#1, \mathrm{taut}}}
\newcommand{\Rdettaut}{R^{(d)}_{\mathrm{taut}}}
\newcommand{\dTattblax}{\Delta T_{1,2}^{\mathrm{lax}}}
\newcommand{\dTattblaxb}{\Delta T_{1,2}^{\mathrm{laxII}}}
\newcommand{\dCMattblax}{\Delta \CM_{1,2}^{\mathrm{lax}}}
\newcommand{\dCMattblaxb}{\Delta \CM_{1,2}^{\mathrm{laxII}}}
\newcommand{\dMattblax}{\Delta M_{1,2}^{\mathrm{lax}}}
\newcommand{\dTattblaxd}{\Delta T_{1,2}^{\mathrm{lax}\rightarrow\mathrm{d}}}
\newcommand{\dTattblaxt}{\Delta T_{1,2}^{\mathrm{lax}\rightarrow\mathrm{taut}}}
\newcommand{\dRattblax}{\Delta R_{1,2}^{\mathrm{lax}}}
\newcommand{\dtWattblax}{\Delta \tilde{W}_{1,2}}
\newcommand{\dTattbtaut}{\Delta T_{1,2}^{\mathrm{taut} }}
\newcommand{\dCMattbtaut}{\Delta \CM_{1,2}^{\mathrm{taut} }}
\newcommand{\dMattbtaut}{\Delta M_{1,2}^{\mathrm{taut} }}
\newcommand{\Fdrag}{F_{\mathrm{drag}}}
\newcommand{\Tdlax}{T_{\mathrm{d,lax}}}
\newcommand{\Tdlaxb}{T_{\mathrm{d,laxII}}}
\newcommand{\Ttaut}{T_{\mathrm{taut}}}
\newcommand{\Rdetlax}{R^{(\mathrm{d})}_{\mathrm{lax}}}
\newcommand{\Rdetlaxatt}[1]{R^{(\mathrm{d})}_{\mathrm{lax};#1}}
\newcommand{\Rsemidet}{\widehat{R}}

\newcommand{\riskfn}{\psi}
\newcommand{\sigrisk}{\sigma_F}
\newcommand{\kappamax}{\kappa_{\mathrm{max}}}
\newcommand{\kappamin}{\kappa_{\mathrm{min}}}
\newcommand{\plax}{p_{\mathrm{d},\mathrm{lax}}}

\newcommand{\Var}{\mathrm{Var}}
\newcommand{\Cov}{\mathrm{Cov}}

\usepackage{amsmath, amsthm}
\usepackage{caption,subcaption}
\usepackage{graphicx}
\usepackage{makecell}

\theoremstyle{thmstyleone}%

\theoremstyle{thmstyletwo}%

\theoremstyle{thmstylethree}%

\begin{document}

\title[Effective motor statistics with tether slack]{Effective transport statistics for cooperative molecular motor ensembles with tether slack}


\author*[1]{\fnm{Joseph} \sur{Klobusicky}}\email{joseph.klobusicky@scranton.edu}

\author[2]{\fnm{John} \sur{Fricks}}\email{jfricks@asu.edu}

\author[3]{\fnm{Peter R.} \sur{Kramer}}\email{kramep@rpi.edu}

\affil*[1]{\orgdiv{Department of Mathematics}, \orgname{The University of Scranton}, \orgaddress{\street{800 Linden St.}, \city{Scranton}, \postcode{18510}, \state{Pennsylvania}, \country{USA}}}

\affil[2]{\orgdiv{School of Mathematical and Statistical Sciences}, \orgname{Arizona State University}, \orgaddress{\street{1151 S Forest Ave.}, \city{Tempe}, \postcode{85287}, \state{Arizona}, \country{USA}}}

\affil[3]{\orgdiv{Department of Mathematical Sciences}, \orgname{Rensselaer Polytechnic Institute}, \orgaddress{\street{110 8th St.}, \city{Troy}, \postcode{12180}, \state{New York}, \country{USA}}}


\abstract{We improve upon our previous work using asymptotic analysis to characterize analytically the effective transport properties of a stochastic model for an ensemble of a cargo being pulled by two molecular motors walking on a microtubule. The governing equations are a system of stochastic differential equations with stochastic switching due to attachment and detachment kinetics of the molecular motors from the microtubule. Our model adopts a nonlinear tether which exerts no force when end-to-end distances are sufficiently small. This challenges some of the scale separation arguments from our previous work, and we modify the asymptotic analysis accordingly. In particular, spatial dynamics of the molecular motors may not always reach their quasi-stationary distribution within a particular attachment state. Rather, molecular motors tend to have a slack connection to the cargo for a while after they reattach, and we incorporate some first passage time analysis to determine if the tether becomes taut again. Through these and other modifications to the asymptotic analysis argument, the theoretical statistics for the run length, run time, and velocity of the motor-cargo ensemble can be brought into reasonable agreement with direct model simulations of two-motor ensembles consisting of kinesin-1 and kinesin-2. Additionally, we propose a new data fitting approach to obtain a double-exponential parameterization for the detachment rate of a motor as a function of the momentary force on its tether to the cargo, which is continuous at low load. Such a modified parameterization is needed to avoid artifacts in theoretical modeling of intracellular transport.
}




\maketitle

\section{Introduction}
\label{sec:intro}
This work addresses two modeling deficiencies in our previous paper~\cite{klobusicky2020cooperative} in which we developed a mathematical framework to analytically express the effective transport statistics of a cooperative system of molecular motors in terms of the biophysical parameters of the motors and the cargo they carry.  Molecular motors are proteins in cells which convert chemical energy into various forms of useful work; molecular motors such as kinesin and dynein in particular provide the means to transport cargo within cells by a chemically-activated stepping process along microtubule tracks.
We focused on pairs of molecular motors attached to a common cargo, with a common direction of motion but of possibly dissimilar types, such as kinesin-1 and kinesin-2 which are two varieties of the kinesin protein. 
In particular, we examined our theoretical model in the context of homogenous ensembles of a pair of kinesin-1 motors and a pair of kinesin-2 motors, together with a heterogenous combination of a kinesin-1 motor with a kinesin-2 motor.  Such systems have been studied experimentally in \textit{in vitro} studies~\citep{feng2018motor,furuta2013measuring,Jamison:2010,rai2013molecular}  with at least two aims:  1) to assess how well motors of a certain type cooperate with each other when they team up~\citep{bovyn2021diffusion,furuta2013measuring,Gross2007twoone,Jamison:2010,kolomeisky2016collective,uppulury2013varying,rai2013molecular,mallik2013teamwork}, and 2) to understand what benefits may accrue to having different types of motors simultaneously carry a cargo in a given direction as is observed in some cellular systems~\citep{arpag2019kin1kin3,feng2018motor,milic2017intraflagellar}.  Our approach relies on the exploitation of separation of time scales in the dynamics of the motors and cargo, including attachment and detachment to obtain analytical expressions for the mean and variance of key transport characteristics of the motor-cargo ensemble, including run length, run time, effective velocity, and effective diffusivity.  
Other theoretical~\citep{klumpp2005cooperative,mckinley2012asymptotic,materassi2013exact,talukdar2016steady,klumpp2015review,WangLi:2009,miles2018analysis,howard2019direction} and computational modeling advances~\citep{arpag2019kin1kin3,kunwar2011mechanical,lipowsky2013network,gross2011arranged} on cooperative motor ensembles were reviewed in~\cite{klobusicky2020cooperative}.  Since that publication, we remark that
\citet{leighton2022performance} also employed stochastic differential equation
models for the motors coupled to the cargo, using a purely linear model for
the tethers as we did in~\cite{klobusicky2020cooperative}, for the purpose of performance and
efficiency analyses as a function of the number of motors attached.

Our model is formulated in Section~\ref{sec:model} as a system of switched stochastic differential equations, with the switching governed by force-dependent attachment and detachment rates of the molecular motors from the microtubule.  Within each attachment state, the motor and cargo positions evolve according to overdamped stochastic differential equations with a prescribed force-velocity relationship for the attached motors and a force-displacement relationship for the tether between the motor and the cargo.  Kinetic stepping models are more common in the literature~\cite{bouzat2016models,reld:rgovm,reld:rsmp,hughes2012kinesins,WangLi:2009,tce:mdbt,abk:mm,driver2010coupling,lipowsky2013network}, and our stochastic differential equation framework can be viewed as a coarse-grained representation of such stepping models which provides a more convenient starting point for the analysis in~\citet{klobusicky2020cooperative}.  Given the scale separation assumptions we use, and their at least ostensible viability, we don't expect the results would differ substantially if we started from a kinetic stepping model, though this remains to be explored systematically.  Here we focus on rectifying two more blatant deficiencies in the practical applicability of our methodology which were revealed in the theoretical predictions made in~\citet{klobusicky2020cooperative}.  The first pertains to the relationship adopted for the motor detachment rates as a function of force experienced by the motor, and the second to the physical force model for the tether connecting the motor and the cargo.  We elaborate next on these central concerns of the present work.

The rate at which molecular motors detach from a microtubule track is clearly observed experimentally to be sensitive to the force applied to the motor's tether.  Within the one-dimensional longitudinal structure of our model described in Section~\ref{sec:model}, we will restrict ourselves to only consider dependence on longitudinal forces, though some experimental work indicates a separate dependence on forces applied in the other two directions~\citep{block2003optical2d,howard2019direction,ostap2020geometry,yildiz2026lowz}.  In~\citet{klobusicky2020cooperative}, we parameterized the dependence of motor detachment rate on tether force by directly using double-exponential fits to the relationship between inferred motor detachment rate and \emph{trap force} on the cargo in~\citet{milic2017intraflagellar,milic2014kinesin}.  Note that what our model requires is the dependence on the \emph{momentary} force on the motor tether, and the applied steady trap force on the cargo would only be the average of this fluctuating force.  Moreover, the theoretical fits to the experimental data in~\citet{andreasson2015elife,milic2017intraflagellar,milic2014kinesin}  lead to a discontinuous dependence of detachment rate on applied force as it switches from the hindering to the assisting direction.  This is not really supported by the data, and arises simply due to a gap in observational data available at small assisting trap forces (due to challenges with the force feedback in this regime~\citep{hancock:pc}) and the fitting methodology.   The discontinuity of that double-exponential model leads to poor results when employed in a model with small trap force because small fluctuations of the tether force lead to large and likely unphysical variations in the detachment rate of the motor as the force fluctuates between small hindering and assisting values. Previous computational modeling work has employed various low-force smoothings of the double-exponential model~\citep{bovyn2020attach} or inference of the underlying relation between instantaneous tether force and detachment rate via simulations~\citep{arpag2014populations}.  We offer in Section~\ref{sec:opt} a systematic attempt to infer a model relationship between instantaneous tether force and motor detachment rate using the data from~\citep{milic2014kinesin,milic2017intraflagellar}, combining our theoretical model's relationship between the trap force on the cargo and probability distribution of tether forces on the motor with a weighted least-squares optimization that emphasizes the quality of fit of the trap force data over a smaller range of trap forces more relevant to natural dynamics in a cell.  

The second modeling concern requires a more elaborate remedy, and makes up most of our new theoretical treatment here.
The mathematical framework for the transport statistics of cooperative motor-cargo complexes was developed in~\citep{klobusicky2020cooperative}  using a linear (Hookean) tether model $F^{(i)}(\ell) = \kappa^{(i)}\ell$ to express the force of the tether as a function of its signed stretch $\ell$.  The index $i$ allows us to consider different tether models for different motor types.  This simplified some of the mathematical calculations, but as noted in the concluding discussion of~\citep{klobusicky2020cooperative}, cannot be taken as an adequate representation of the actual compliance of the tether between a motor and cargo.   By tether, we mean the flexible portion of the molecular motor structure that connects the part that interacts with and walks on the microtubule (which is represented in our model as the motor particle) with the part that attaches to cargo.  The main feature missed by the linear model is that the tether exerts essentially no force when compressed below some slack length $ \ellrest{i}$, and only exerts tension when stretched beyond it.  In the present work, we therefore re-examine our analyses in the context of a nonlinear cable model for the tether, which is the simplest encoding of this tether behavior and widely adopted in other theoretical and computational models~\citep{arpag2019kin1kin3,arpag2014populations,bovyn2021diffusion,gross2011arranged,kunwar2008stepping,kunwar2016anisotropy,furuta2013measuring,korn:2009,grossmiles2024binding}: 
\begin{equation} \label{eq:cable}
F^{(i)} (\ell) = 
\kappa^{(i)} \mathrm{sgn} (\ell) (|\ell| - \ellrest{i})_+,
\end{equation}
where
$ (y)_+ = \max(y,0) $. 
We say the tether is \emph{slack} when extended below its length scale $\ellrest{i}$ and therefore applying no force on the motor or cargo, and that the tether is \emph{engaged} or \emph{taut} when the extension exceeds this length and creates a tensile force in the tether.  $\kappa^{(i)} $ is again a linear spring constant, but now only applicable when the tether is taut.  In~\citet[App. A]{klobusicky2020cooperative}, we gave a general way to extend the linear tether force analysis to nonlinear tether forces, but that assumed the scale separation assumptions still hold.  The slackness in the cable model~\eqref{eq:cable} to be studied here creates departures from some of those scale separation assumptions, and motivates the revised derivations of the transport characteristics of cooperative motor-cargo ensembles which we will present.  In particular,  substantial refinements of the analysis for motor statistics will be necessary due to tracking when and how the tethers engage and relax as the motors attach, detach, and move. 

The derivation of the effective transport of cooperative motor-cargo complexes via averaging arguments and homogenization in~\citet{klobusicky2020cooperative} was based on an assumed well-separated ordering of time scales, with the dynamics of the detached motors the fastest, then the cargo, then the attached motors, then the attachment/detachment dynamics.  These assumptions are motivated in physical terms, at least for \textit{in vitro} experiments, in~\cite{klobusicky2020cooperative}.  With a tether model~\eqref{eq:cable} incorporating slack, we must re-examine those scale separation assumptions.  We find that the averaging arguments used for states with one motor attached to the microtubule should still be valid, and their adaptation to the nonlinear tether model is presented in Section~\ref{sec:oam}.   

But when both motors are attached to the microtubule, the slack in the tether does cause violations in the assumption that the attached motor dynamics is faster than the time scale of motor detachment.  
In~\citet{klobusicky2020cooperative}, this assumption implied the progress of the motor-cargo complex in this state would proceed with a constant effective velocity and diffusivity based on averages with respect to the stationary distribution of the relative motor-cargo positions. 
Depending on the sign of the trap force on the cargo, and whether the motor ensemble is homogenous (i.e., both kinesin-1 or kinesin-2) or heterogenous (i.e., one motor  kinesin-1 and the other kinesin-2), the tether of one or both of the motors may be slack for a period of time we call the `lax phase'.  In this lax phase, no force is acting on at least one of the motors to move it toward mechanical equilibrium, and thus this phase can last for much longer than the nominal time scale of the tether dynamics based on the compliance $\kappa$.  In particular, the lax phase may last longer than the time scale of detachment kinetics, undermining the argument that the tether quickly drives the separation of attached motors into a stationary distribution before detachment.
Eventually, unless a motor detaches first, the system will enter what we call the `taut phase' in which the motors are near a mechanical equilibrium with both tethers actively engaged.  For some cases, the taut phase statistics mirror those of the stationary distribution developed in~\citet{klobusicky2020cooperative}.  However, due to the exponential dependence of detachment rate on tether force~\eqref{detmod}, in some cases detachment occurs too quickly relative to the mechanical relaxation time scale to assume the stationary distribution is achieved.  In these instances, we 
 develop a `semideterministic approximation' to derive the statistical dynamics within the taut phase.   The theoretical development based on this lax and taut phase decomposition is presented  in Section~\ref{sec:laxtaut}.   Note that the failure of the assumption of a quasi-stationary distribution for the cargo dynamics relative to the attached motor dynamics even more so implies the invalidity of the stronger assumption in the modeling framework of~\citet{klumpp2015review} that the cargo is always in a mechanical equilibrium with the forces applied via the tethers to the attached motors.

We review the modeling framework from~\citet{klobusicky2020cooperative} in Section~\ref{sec:model}, incorporating the more realistic nonlinear tether force model~\eqref{eq:cable}.  The methodology for calculating transport statistics in terms of the time and displacement over a cycle of detachment and attachment of motors from the doubly attached state is summarized in Section~\ref{sec:cycle}.  The contributions to these statistics from the state of a single attached motor and from jumps between single and doubly attached motor states are presented, respectively, in Subsections~\ref{sec:oam} and~\ref{sec:jump}.  Up to this point, the results are rather direct generalizations of the calculations based on scale separation from~\citet{klobusicky2020cooperative}, with the extensions described in Appendix A of that work.

The subsequent sections develop the conceptually new methodology to address the technical gaps identified above.  In Section~\ref{sec:opt} we describe a method for deriving the parameterization for the relationship between detachment rates and \emph{instantaneous} force from the experimental results in~\citet{milic2014kinesin,milic2017intraflagellar} which report run lengths and times as a function of \emph{trap} force.  Then in Section~\ref{sec:laxtaut} we rework the statistics for the duration and cargo displacement within the doubly attached motor state to relax the assumption that the attached motors reach a quasi-stationary distribution before one of the motors detach.  This proceeds essentially by decomposing the sojourn in this state into phases where tethers are taut or lax, and employing first passage time arguments and semi-deterministic approximations.   We thereby extend our theoretical calculations from~\citet{klobusicky2020cooperative} for mean run time, mean run length,  velocity, and diffusivity for homogenous ensembles, and present explicit results for all but diffusivity for heterogenous ensembles because in this case the analytical expressions for second order statistics become very ungainly in the extended calculation framework.  Our revised theoretical characterization  is compared in Section~\ref{sec:numeric} with direct Monte Carlo simulations of two-motor ensembles with parameterizations corresponding to kinesin-1 and kinesin-2.  We close in Section~\ref{sec:conc} with an assessment of the performance of the updated theoretical framework and avenues for future improvement and exploration.

Supplementary  material is collected in appendices:  corrections for the actual cargo displacement from the cargo tracking variable used in the main analysis (Appendix~\ref{app:zshift}), a justification for the neglect of drag force on the cargo even in the cable model for the tether (Appendix~\ref{app:drag}), some complicated analytical expressions for some key statistical components (Appendix~\ref{sec:rhointegral}),
the statistical distribution of forces experienced by the motors through their tethers to the cargo (Appendix~\ref{sec:cargoflucs}),  the probability distribution for the separation between two attached motors when one of them detaches within the lax phase (Appendix~\ref{app:laxdetsep}), an illustrative analytical calculation for how tether force fluctuations due to cargo movement affect the effective detachment rate of a motor (Appendix~\ref{app:detcargoflucs}), an indication of how the calculation framework is capable of computing second order ensemble statistics such as diffusivity and why the results would be cumbersome to present (Appendix~\ref{app:secmompain}), a comparison of numerical results with theoretical predictions neglecting the lax phase (Appendix~\ref{app:nolax}), and a comparison of how different parameterizations for the relationship between force and detachment rate affect run length statistics (Appendix~\ref{app:compdetmodels}).

\section{Stochastic switching model and parameters}
\label{sec:model}
We will be focusing on  ensembles consisting of  a single cargo  and two possibly distinct motors traveling along a given microtubule.  Here we summarize the general model setup from~\cite{klobusicky2020cooperative}.  The longitudinal positions of the motors indexed $i = 1,2$ at times $ t \geq 0 $ will be denoted by $X^{(i)}(t)$, and the cargo position by $Z(t)$.    Following many theoretical models, the dynamics are expressed only in the longitudinal direction, with transverse fluctuations neglected~\citep{bouzat2016models,hendricks2009collective,smith2018assessing,klumpp2015review,miles2018analysis,leighton2022performance}.  The simplest picture are for both motors proceeding on the same microtubule but the model would also approximately work for motors moving on parallel nearby microtubules. The governing equations for the motor and cargo positions are a system of stochastic differential equations (SDEs) with switching.  
This formulation facilitates our mathematical analysis, and should be viewed as a coarse-graining of kinetic stepping models commonly used in simulation models~\citep{arpag2019kin1kin3,bovyn2021diffusion,kunwar2011mechanical,lipowsky2013network,gross2011arranged} by summarizing the dynamics in terms of the procession rate of a motor and its randomness as a function of applied force.  Specifically, our stochastic dynamical model takes the form 
\begin{subequations}
\label{eq:motcar}
\begin{align} 
dX^{(i)}(t) &= \left(-\frac{F^{(i)}(X^{(i)} (t)-Z(t))}{\gamma_{m}^{(i)}}dt+\sqrt{\frac{2k_BT}{\gamma_{m}^{(i)}}}dW^{(i)}(t)\right)\mathbf
(1-\Atstate{i}(t))\label{motors}\\&+ \left(v^{(i)}g\left(\frac{F^{(i)}(X^{(i)}(t)-Z(t))}{F_s^{(i)}}\right)dt+
\sqrt{2
D^{(i)}}dW^{(i)}(t)\right)
\Atstate{i}(t) \quad i \in \{1,2\}, \nonumber
\\
\gamma dZ(t) &= \sum_{j = 1}^2 F^{(j)}(X^{(j)}(t)-Z(t)) dt- F_Tdt +\sqrt{2k_BT\gamma
}dW_z(t). \label{cargo}
\end{align}
\end{subequations}

Independent Brownian motions $W^{(i)}$ for $i = 1,2$ and $ W_z(t)$ govern the noisiness of the spatial dynamics of the motors and cargo, together with stochastic switching between motor attachment states to give an overall Markov process. 
We will now briefly explain  the physical meaning of these equations.  

The variables $\Atstate{i}(t)$ are binary valued indicating whether motor $i$ is attached to the microtubule at time $t$.  
When $\Atstate{i}(t) = 0$, motor $i$ is detached, and is modeled as an overdamped Langevin particle with a friction constant  $\gamma_{m}^{(i)}$. The constant $k_BT$ is the Boltzmann constant multiplied
by temperature. The cargo in equation (\ref{cargo}) is also modeled by a Langevin equation, with a friction constant $\gamma$.  The Brownian motion $W_z(t)$ acting on the cargo dynamics is independent of the Brownian motions $W^{(i)}(t)$ acting on the motor dynamics.

When  $\Atstate{i} (t) = 1$, motor $i$ is attached to the microtubule, and will step along it in a particular direction with a maximum velocity $ v^{(i)} > 0 $ that is slowed when its tether to the cargo exerts an opposing force.  Note that our assumption that both  motor velocities are of the same sign indicates we are considering \emph{cooperative} motors that attempt to pull the cargo in the same direction.  In more quantitative detail, the drift on a motor $i$ is determined by multiplying its unencumbered velocity $v^{(i)} >0$ by the value of a nondimensional piecewise linear  force-velocity relation:
\begin{equation} \label{fvelrel}
g(\tilde{F}) = \begin{cases}
1 & \tilde{F}<0, \\
(1-\tilde{F}) & 0\leq\tilde{F}\leq 1, \\
0 & \tilde{F}>1. \\
\end{cases}
\end{equation}  
The ratio of forces evaluated by $g$ in (\ref{motors}) is given by the ratio of the tether force $F^{(i)}(x-z)$ on motor $i$ and the stall force $F_s^{(i)}$ which is the minimum force needed to stop the motor. Together, this expresses the fractional reduction in the motor velocity as a function of the force applied to the motor via the tether.  The tether force model $ F^{(i)} (\ell) $ for motor $i$ as a function of its signed stretch $\ell$ from the cargo position is taken to be positive when the tether force opposes the motor's natural direction of motion, and negative when it assists it. This sign convention has been conventional in the modeling literature~\citep{lipowsky2013network,klumpp2015review,kunwar2011mechanical,bouzat2016models}  but is opposite to that which is typically adopted in recent experimental studies~\cite{milic2014kinesin,milic2017intraflagellar,howard2019direction}.   The specific tether force model analyzed in this work was presented in Eq.~\eqref{eq:cable}.
The diffusivities $D_i$ correspond to  fluctuations associated with the mechano-chemical process of the stepping mechanism of kinesin on the microtubule.

The dynamics of the motor attachment states $\Atstate{i}(t) $ are determined through detachment and attachment rates of motors to the microtubule.  Unattached  motors are assumed to attach at rates independent of motor and cargo positions, and so are given by  constant rates $a^{(i)}$.  These should actually be thought of as \emph{reattachment rates} when the other motor in the ensemble is already attached to a microtubule, which can of course differ from bulk attachment rates from solution~\citep{bovyn2020attach,bovyn2021diffusion,choudhary2026attach}.  Detachment rates of motors are known to be sensitive to the force on the motor, and as in~\citet{klobusicky2020cooperative}, we adopt a  double exponential model~\citep{milic2017intraflagellar,milic2014kinesin}.
When motor $i$ experiences force $ F$ (with positive sign indicating a force hindering the motor's progress), the detachment rate is given by
\begin{equation}
d^{(i)}(F) = \begin{cases}d_0^{(i)} \exp(-F/F_{d-}^{(i)}) & F \le 0, \\
d_0 ^{(i)}\exp(F/F_{d+}^{(i)}) & F>0.  \label{detmod}
\end{cases}
\end{equation}
The constants $F_{d-}^{(i)}$ and $F_{d+}^{(i)}$ are force scales of detachment
for assisting and opposing forces, respectively. Since the force $F$ in (\ref{detmod}) is determined by the random positions of motors and cargo, the detachment rate is subsequently random, so that detachment is an example of a Cox process.

Of note is that we are considering
a single zero force detachment rate $d_0^{(i)}$, which differs from previous
models~\citep{milic2014kinesin,milic2017intraflagellar} with different rates for both small assisting ($d_{0-} ^{(i)}$) and opposing
($d_{0+} ^{(i)}$)  forces. This means that the force detachment curve will
be continuous at zero force.
Section \ref{sec:opt} is devoted to estimating the values of zero force detachment rates $d_{0}^{(i)}$ and force scales for detachment $F_{d\pm}^{(i)}$.  This is carried out through a weighted maximum likelihood estimator of run lengths from the experiments in \cite{milic2014kinesin,milic2017intraflagellar}.

For each value of $(\Atstate{1}, \Atstate{2})$,  Eq.~\eqref{eq:motcar} takes the form of a distinct system of SDEs.  When   $\Atstate{i}(t) = 1$ for $i = 1,2$, the system is in its `fully attached' state, which we denote as state $(1,2)$.  In what follows, we set Eq.~\eqref{eq:motcar}   to have initial conditions in state $(1,2)$, or $(\Atstate{1}(0), \Atstate{2}(0)) = (1,1)$.  The initial positions of cargo and attached motors  corresponds to random positions of a cargo and single unattached motor with respect to a single attached motor.  This is discussed in greater detail in Appendix~\ref{app:zshift}  in terms of the appropriate stationary distributions for displacement between motors and cargo. After detachment of a motor from state $(1,2)$, the ensemble transitions to one of two possible states.  When $(\Atstate{1}, \Atstate{2}) = (1,0)$, we say that the system is in state $(1)$.  Likewise, $(\Atstate{1}, \Atstate{2}) = (0,1)$ corresponds to state $(2)$. At this point, the detached motor can either reattach, returning the system to state $(1,2)$, or else the other motor detaches, placing the ensemble in the absorption state $\emptyset$ and  terminating the run.  We will sometimes use the acronym \OAM\  (``one attached motor'') to refer to a state with a single motor attached ($(1)$ or $(2)$) and \TAM\ (``two attached motor'') to refer the state $(1,2)$ of full attachment.

Directly simulating Eq.~\eqref{eq:motcar} including the attachment/detachment state changes of $ (\Atstate{1} (t),\Atstate{2} (t)) $ involves common schemes for stochastic processes.  Namely, we fix a time step $ \Delta t $ and use the Euler-Maruyama method for simulating the SDEs, and approximate the Cox processes given in (\ref{detmod}) with Bernoulli processes.  

The biophysical parameter values for kinesin-1 and kinesin-2 listed in Table~\ref{paramtable} will guide the assumptions made about relative time scales in the theoretical analysis to follow.  Most of these values are taken from the referenced literature, though we adopt a novel parameterization of the detachment rate based on a reanalysis of experimental data~\citep{milic2014kinesin,milic2017intraflagellar} as explained in Section~\ref{sec:opt}.  We remark that the recent high resolution work of~\citet{schaffer2021germanium} reports somewhat smaller values of the unloaded motor velocity and stall force, and an order of magnitude larger tether stiffness.  The higher stiffness constant would only reinforce our assumption that the time scale of cargo fluctuations (when the tether is taut) is fast compared to that of attached motors.

\begin{table}
\resizebox{\columnwidth}{!}{
\begin{tabular}{|l||l|l|}\hline 
Parameter & Description&Typical values  \\\hline
$F_s^i$ & Motor stall force& 7 pN \cite{visscher1999single} \\\hline
$k_B T$ & Boltzmann constant by temperature& 4.1 pN nm\\\hline
$\kappa^{(i)}$ & Motor-cargo tether stiffness & 0.25 pN/nm
\cite{furuta2013measuring}
\\\hline
$v^{(i)}$  & Unencumbered motor velocity& 790 nm/s (\textbf{500 nm/s}) \cite{andreasson2015mechanochemical}\\\hline
$\gamma$ & Cargo friction& $1\times 10^{-5}$ pN s/nm \cite{mckinley2012asymptotic}\\\hline
$\gamma_{m,i}$ &  Motor friction& $1\times 10^{-6}$ pN s/nm\\\hline
$D^{(i)} $ & Effective motor diffusion& 2500 nm\textsuperscript{2}/s
(\textbf{750 nm\textsuperscript{2}/s})\cite{visscher1999single,andreasson2015mechanochemical}\\\hline
$F_T$ & Optical trapping force& -6 to 6 pN \cite{milic2014kinesin}
\\\hline
$a^{(i)}$ & Motor attachment rate & 4/s (\textbf{16/s}) \cite{feng2018motor}\\\hline
$F_{d+}^i$ & Force scale for detachment (Hindering force) & 4.63 pN (\textbf{1.15
pN}) \\\hline
$F_{d-}^i$ & Force scale for detachment (Assisting force)   & 1.07 pN
(\textbf{1.27
pN}) \\\hline
$d^{(i)}_{0}$ & Zero force detachment rate 
& 0.78/s \textbf{(.84/s)}
\\\hline
$\ellrest{i}$ & Slack length 
& 80 nm  \cite{milic2014kinesin,milic2017intraflagellar} \\\hline
\end{tabular} 
}
\caption{Parameter values for kinesin-1 and kinesin-2.  When two values appear, the plain text refers to the value for kinesin-1 and the bold refers to the value for kinesin-2.  Parameters $F_{d+}^i$, $F_{d+}^i$ , and $d^{(i)}_{0}$ are found through an optimization procedure described in Section~\ref{sec:opt} using run length data in \cite{milic2014kinesin, milic2017intraflagellar}  \label{paramtable}}
\end{table}

\section{Cycle Analysis}
\label{sec:cycle}
The central approach in \cite{klobusicky2020cooperative} is to assume the spatial motor-cargo dynamics within an attachment state reach a statistical equilibrium before a transition to another attachment state.  This relies on an assumption that the spatial dynamics are fast relative to the attachment/detachment dynamics, and implies that the attachment state dynamics $ (\Atstate{1}(t),\Atstate{2}(t))$ can be viewed as a coarse-grained Markov chain.  Then the long time velocity and diffusivity of the cargo can be well estimated by deriving first and second moments for dwell times and cargo displacements within each attachment state $(1), (2)$, and $(1,2)$, and then appealing to theorems of renewal reward theory  for the four state Markov chain.  \citet{ciocanel2020renewal} describes a general framework for such computations, though here the Markov chain topology is simple enough to allow more explicit analytical results and we account for a finite random time after which the Markov chain terminates (upon entry into the state of complete detachment).  In~\cite{klobusicky2020cooperative}, we defined a renewal ``cycle'' as beginning in the state $(1,2) $ of both motors being attached, progress in the \TAM\ state, detachment of one of the motors, progress in the corresponding \OAM\ state,  then reattachment of that motor.  This is natural for heterogenous ensembles, so that the information about the identity of the motor detaching and reattaching is retained in the same cycle, and subsequent cycles can be treated as independent.  In the present work, where we do not assume the motors reach statistical equilibrium within an attachment state, we will find in Section~\ref{combolaxtaut} that correlations arise between the motor-cargo configuration at entry to and exit from the \TAM\ state, which would be split across renewal cycles as just defined.  This would create complications in the renewal reward theory, so we shift the definition of cycles slightly in the present work as beginning not at the  entry into the \TAM\ state but at the end of the \OAM\ state.   That is, the cycle begins with the attachment of the second motor rather than ending with the reattachment of the second motor.  This keeps the  effects of entry into, progress within, and exit out of the \TAM\ state within a cycle, and thus allows the correlations between these to be computable within the renewal-reward framework.  This shift now disconnects between cycles the information concerning which motor was detached in the \OAM\ state and how the cargo jumps upon its reattachment.  This information is irrelevant for homogenous ensembles but clearly important for heterogenous ensembles.  As we explain in Subsection~\ref{sub:rendelayed}, the renewal-reward framework is difficult to apply to heterogenous ensembles for other reasons when we don't assume convergence to the statistical equilibrium, and as we show in Appendix~\ref{app:secmompain}, second moment properties are even difficult to compute within cycles in the present setting.  We thus satisfy ourselves with only mean calculations for heterogenous ensembles in the present work, using a direct recursive approach rather than a renewal-reward framework, and thus the splitting of correlated random variables across cycles is not problematic.  

We proceed next to describe the key random quantities associated to a cycle consisting of attaching from an \OAM\ state to the \TAM\ state, progress within the \TAM\ state, detachment from the \TAM\ state, and then progress within the \OAM\ state until it is terminated by either an attachment event (that initiates another cycle) or detachment (leading to complete detachment and termination of the run).  What follows is a slight adjustment from the framework used in~\citep{klobusicky2020cooperative} due to the  shift in cycle definition.
The time for a single cycle consists of the time spent in state $(1,2)$ and $(\detmot{}')$, where $\detmot{}$ denotes the motor which detaches from the \TAM\ state, and a prime $i' = 3-i$ denotes the index of the ``other motor.''
We write this as
 \begin{equation}\label{tcycleterms}
 \dTcyc = \dTattb + 
\Delta T_{\detmot{}^{\prime}}.
 \end{equation}
The corresponding ``rewards'' in the renewal process are given by displacements of the cargo.  In~\citet{klobusicky2020cooperative} we use the rapid time scale of cargo fluctuations relative to the dynamics of attached motors to represent the cargo in the analysis by its conditional mean position $M(t)$ expressed in terms of the attachment state and positions of the attached motors.  That worked well for the simple linear tether force model used there, where the mean of the quasistationary distribution was simply
a weighted average of the motor positions (based on the tether stiffnesses) with a shift proportional to the cargo trap force, but with the nonlinear cable model~\eqref{eq:cable}, the mean position of the cargo relative to the motors depends nontrivially on the displacement between the motors.   This leads to unhelpful complications in the analysis of the jump statistics, which can be easily avoided by instead defining $ \CM (t)  $ as the arithmetic mean of the positions of the attached motors.  Either way, 
we show explicitly in Appendix~\ref{app:zshift} that the full statistics of the cargo position, conditioned on the motor positions, can be recovered from the cargo tracking variable $ \CM (t) $ together with the  signed displacement of the motors
 $ R (t) = X^{(1)} (t)- X^{(2)} (t) $ in the \TAM\ state.  
The stochastic process $\CM(t)$, which we sometimes refer to as the cargo tracking variable, gives a smoothed way to track the cargo dynamics.  In the \TAM\ state, we have $ \CM_{1,2} (t) = \frac{X^{(1)} (t) + X^{(2)} (t)}{2} $ while in the \OAM\ states we have $ \CM_i (t) \equiv X^{(i)} (t) $.  Then we generally define the cargo tracking variable as $ \CM (t) \equiv \CM_{S(t)} (t) $, where $ S(t) =\{i\in {1,2}: Q_i (t)=1\} $ is the set of indices of attached motors.  The increments of the cargo tracking variable
naturally includes continuous displacements in the \TAM\ and \OAM\ states, denoted by $\dMattb$ and  $\dMdet{\detmot{}^{\prime}}$ respectively, where we recall $ \detmot{}^{\prime} $ gives the index of the motor still attached in the \OAM\ state of a cycle. However, we must account for the rapid recalibrations experienced by the cargo when transitioning between states.  We denote the jump in the cargo tracking variable due to detachment of motor $i$ from the \TAM\ to the \OAM\ state  by  $\dMdetjump{i}$, and the jump due to attachment of motor $i $ from the \OAM\ state to the \TAM\ state  by $\dMattjump{i}$.
  The total displacement of the cargo tracking variable over a cycle is then  the sum 
\begin{align}\label{mcycleterms2}
\dMcyc = \dMattjump{\attmot}+\dMattb +\dMdetjump{\detmot}+\dMdet{\detmot{}^{\prime}},
\end{align}
where $ \attmot $ is the index of the motor that is attaching into the \TAM\ state. 

The mean and variance for the dwell times $\Delta T_{\varpi}$ and tracking variable displacements $\Delta M_{\varpi}$ in each attachment state $ \varpi$ were obtained in~\citet{klobusicky2020cooperative} from averaging arguments and homogenization of the continuous spatial dynamics within each state.   This manuscript departs from \cite{klobusicky2020cooperative} in the calculation of these quantities for  $\varpi=(1,2)$, as the slackness in the tether leads to a substantial violations of the scale separation assumption  in this state.  In Section~\ref{sec:laxtaut}, we replace the invalid averaging and homogenization arguments with another computable approximation scheme taking into account the engagement of the tethers on the two attached motors as their separation varies.  This affects not only the effective model for the continuous spatial dynamics in this state, but also the effective detachment rate and jumps in the cargo tracking variable at detachment.   In place of the constant effective  drift, diffusivity, and detachment rates in the \TAM\ state from~\citet{klobusicky2020cooperative}, we develop more complex effective models which account for the time since the attachment of the second motor.  We similarly develop a more detailed time-dependent model for the statistics of the intermotor displacement $ R=X^{(1)}-X^{(2)} $ at the time a motor detaches and the probability $ \pdet{i}$ that motor $i$ is the one to first detach.  This leaves the question of the probability distribution for the index $ \attmot $ of the motor  attaching to begin the cycle.  For homogenous ensembles, it is irrelevant.  For heterogenous cycles, we develop a simple recursion scheme in Subsection~\ref{laxhetnonz} to determine it.

 

Over a whole run, we  consider the motor-cargo
complex undergoing a random number  $\Ncyc $ of cycles (starting with attachment  into the \TAM\ state, going through detachment of one motor into the \OAM\ state and ending with transition out of the \OAM\ state)
with the last cycle ending in complete detachment. 

We now write the total run time $ \Talldet $ and run length $ \Malldet $ taken by an ensemble of a cargo with two cooperative motors before complete
detachment as random sums
\begin{equation}
T =\Delta T_{\mathbf{a}} + \sum_{j = 1}^{\Ncyc} \dTcyc^j, \quad \Malldet =\Delta M_{\mathbf{a}} + \Delta Z_{\mathbf t}-\Delta Z_{\mathbf a}+  \sum_{j = 1}^{\Ncyc}
 \dMcyc^j, \label{eq:cycsums}
\end{equation}
where $\{\dTcyc^j\}_{j=1}^{\infty}$
and  $\{\dMcyc^j\}_{j=1}^{\infty}$ are streams of  random variables for cycle times and displacements of the cargo tracking variable for the $j$th cycle, if realized.  We naturally exclude from our statistics those situations where one motor attaches and detaches before the second motor has a chance to attach; that just means we condition on the number of cycles $ \Ncyc \geq 1 $.  We have two adjustments from simply summing over the contributions from $ \Ncyc $ full cycles. 
The run begins with a sojourn in the \OAM\ state.  The rates of detaching or reattaching from this state can be treated as constant (Section~\ref{sec:oam}), so conditioning upon the state being terminated by detaching does not affect the statistics of the time $ \Delta T_{\mathbf{a}} $  and displacement $\Delta M_{\mathbf{a}} $ within this state.    The corrections $\Delta Z_{\mathbf
a}$ and $\Delta Z_{\mathbf t}$ in the run length account for  deviations between the cargo and tracking variable at the start and termination of a run, and are independent of each other and the other random variables associated to cycles.  A derivation of the first and second order statistics of these quantities can be found in Appendix~\ref{app:zshift}.

In the remainder of this section, we indicate how to relate fundamental transport statistics of interest to the statistics of progress of time and displacement within each cycle.   
The statistics of individual cycles, which are the components of these formulas, will be derived in Section~\ref{sec:laxtaut}.  The fact that we do not generally assume the \TAM\ state automatically reaches a stationary distribution brings into question whether memory effects will disrupt the renewal process framework from~\citet{klobusicky2020cooperative}.   In particular we will show in Subsection~\ref{laxhetnonz} that the identity of the motor detaching from the \TAM\ state can substantially affect the dynamics in the next visit to the \TAM\ state.  For homogenous ensembles, this is irrelevant and we can apply the  standard renewal-reward relations from~\cite{klobusicky2020cooperative}, recapitulated in  Subsection~\ref{sub:renewal}, to compute the first and second order statistics of run time and run length. In Subsection~\ref{sub:veldiff}, we review from~\cite{klobusicky2020cooperative} how two measures of the velocity and diffusivity of an ensemble can be expressed in terms of cycle statistics.  For heterogenous ensembles, we cannot generally treat the cycle statistics as independent.   Nonetheless, we  indicate in Subsection~\ref{sub:rendelayed} an alternative recursive framework  to calculate the mean run length and run time in additional to effective velocities.

\subsection{Transport Statistics from Cycle Statistics for Homogenous Ensembles}
\label{sub:renewal}
For homogenous ensembles, the statistics of the times  and displacements of the cargo tracking variable,
$ \{\dTcyc^j,\dMcyc^j\}_{j=1}^{\infty}$ can be taken as independent across cycles (as the identity of which motor detached does not affect the next cycle).  
 The number of cycles $ \Ncyc $ is first of all a Markov time with respect to the filtration generated by these random variables viewed as  discrete-time stochastic process.
Moreover,  $\Ncyc$ is a geometric
random variable with mean $ 1/\totdetprob$ 
  where the probability of complete
detachment during an initiated cycle
 is $ \totdetprob = \frac{\detone{}}{\detone{}+a^{}} $

Here $ \detone{} $ is the effective rate at which a motor detaches from the \OAM\ state; its formula is presented in Eq.~\eqref{detavg}.
As such, it immediately follows that
for the total number of cycles 
\begin{equation}
\E[\Ncyc] = \frac{1}{\totdetprob}, \quad 
\Var[\Ncyc] = \frac{1-\totdetprob}{(\totdetprob{})^2}. \label{fullcycstat}
\end{equation}
The random time $ \Delta T_{\mathbf{a}} $ and displacement of the cargo tracking variable $ \Delta M_{\mathbf{a}} $ from the visit to the initial \OAM\ state are random variables independent of the other contributions in Eq.~\eqref{eq:cycsums}, with statistics given in Eq.~\eqref{eq:oamstats}

Using Wald's identity (Th. 14.6 of \cite{dasgupta2011probability}) and the second Wald identity (Th. 2.4.5 of \cite{ghosh2011sequential}) as well as the independence of the cargo position corrections $ Z_{\mathbf a} $ and $ Z_{\mathbf t} $,
we  obtain that the mean run time and length are given  in terms of single-cycle statistics by:
\begin{align} \label{meanrun}
\E[T]  &= \E[\Delta T_{\mathbf{a}}] + \E[\Ncyc]\mu_{\dTcyc},\\ \quad 
\E[\Malldet] &= \E[\Delta M_{\mathbf{a}}]+\E[Z_{\mathbf t}]-\E[Z_{\mathbf
a}]+ \E[\Ncyc]\mu_{\dMcyc},   \nonumber
\end{align}
and the variances and covariances by:
\begin{align} \label{varrun}
 \Var(T) &= \Var(\Delta T_{\mathbf{a}})+  \E[\Ncyc]\sigma^2_{\dTcyc}
+ \Var(\Ncyc)\mu^2_{\dTcyc},\\
\Var (\Malldet) &=
\Var(\Delta M_{\mathbf{a}})+\Var(Z_{\mathbf t})+\Var(Z_{\mathbf
a})+\E[\Ncyc]\sigma^2_{\dMcyc}+
\Var(\Ncyc) \mu^2_{\dMcyc},\nonumber\\
\Cov(T,\Malldet) &=
\Cov(\Delta T_{\mathbf{a}},\Delta M_{\mathbf{a}})+\E[\Ncyc]\sigma_{\dTcyc,\dMcyc}+\Var(\Ncyc)\mu_{\dTcyc}\mu_{\dMcyc}. \nonumber
\end{align}
Statistics of the cycles ($\dTcyc$ and $ \dMcyc$) are expressed by subscripts on  $ \mu $ for means and $ \sigma $ for standard deviations or covariances. The components of these formulas are presented in Subsection~\ref{combolaxtaut} and derived in the preceding subsections in Section~\ref{sec:laxtaut}.  

\subsection{Effective Velocity and Diffusivity of Homogenous Ensembles}
\label{sub:veldiff}
For  a collection of run times $ \Rtime{j} $ and run lengths $ \Rlength{j}
$ over independent experiments $ j=1,\ldots,\Nexp $, we  consider two distinct definitions
for  effective velocity and diffusivity which were formulated in ~\cite{klobusicky2020cooperative}.  First, we may consider the large-sample limits in which
we concatenate run lengths and times to produce \emph{ensemble} velocities and diffusivities
given by

\begin{equation} 
\Vpool \equiv \lim_{\Nexp\rightarrow \infty} \frac{\sum_{j=1}^{\Nexp} \Rlength{j}}{\sum_{j=1}^{\Nexp}
\Rtime{j}},\quad
\Dpool \equiv \lim_{\Nexp \rightarrow \infty} \frac{\sum_{j=1}^{\Nexp}\left(
\Rlength{j}
- \Vpool \Rtime{j}\right)^2}{2 \sum_{j=1}^{\Nexp} \Rtime{j}}.\label{pooleddiff}
\end{equation}
By censoring shorter runs with fewer full cycles, 
we may also define  \emph{long-run} velocity
and diffusivity by\begin{align} 
\Vrun &\equiv \lim_{k\rightarrow \infty}\lim_{\Nexp\rightarrow
\infty} \frac{\sum_{j=1}^{\Nexp} \Rlength{j}\mathbf 1_{\Ncyc^{(j)}>k}}{\sum_{j=1}^{\Nexp}
\Rtime{j}\mathbf 1_{\Ncyc^{(j)}>k}},\label{runvel} \\
\Drun &\equiv \lim_{k\rightarrow \infty}\lim_{\Nexp \rightarrow \infty} \frac{\sum_{j=1}^{\Nexp}\left(
(\Rlength{j}
- \Vrun \Rtime{j})\mathbf 1_{\Ncyc^{(j)}>k}\right)^2}{2 \sum_{j=1}^{\Nexp}
\Rtime{j}\mathbf 1_{\Ncyc^{(j)}>k}}.\label{rundiff}
\end{align}  
where $ \Ncyc^{(j)} $ denotes the number of cycles in the $j$th experimental run.

In~\citet{klobusicky2020cooperative}, we have shown that both ensemble and long-run velocity and diffusivity can be found from computing cycle statistics.  For the case of ensemble statistics,
 
\begin{align}
\Vpool &= \frac{\E[\Malldet]}{\E[T]} =\frac{\E[\Delta M_{\mathbf{a}}]+\E[Z_{\mathbf
t}]-\E[Z_{\mathbf
a}]+\E[N_c] \mu_{\dMcyc}}{\E[\Delta T_{\mathbf{a}}]+\E[N_c] \mu_{\dTcyc}},
\label{vpool}\\
\Dpool &= \frac{\Vpool^2 \Var[T]+ \Var(\Malldet) - 2 \Vpool \Cov(T,\Malldet)}{2\E[T]}
\label{dpool} \\
&=\frac{\Var[\Delta M_{\mathbf{a}}]+ \Vpool^2 \Var(\Delta T_{\mathbf{a}})-2 \Vpool \Cov(\Delta T_{\mathbf{a}}, \Delta M_{\mathbf{a}})+\Var[Z_{\mathbf t}]+\Var[Z_{\mathbf
a}]}{2\left(\E[\Delta T_{\mathbf{a}}]+\E[N_c]
\mu_{\dTcyc}\right)} \nonumber \\
&  +\frac{\E[N_c] \left(\Vpool^2\sigma^2_{\dTcyc}+\sigma^2_{\dMcyc} 
-2 \Vpool \sigma_{\dTcyc,\dMcyc}\right)+\Var[N_c] \left(\mu_{\dMcyc} - \Vpool \mu_{\dTcyc}\right)^2}{2\left(\E[\Delta T_{\mathbf{a}}]+\E[N_c]
\mu_{\dTcyc}\right)}. \nonumber
\end{align}
The long-run versions of velocity~\eqref{runvel} and diffusivity~\eqref{rundiff} are obtained by only retaining the terms multiplying $ \E[\Ncyc] $ in Eqs.~\eqref{vpool} and~\eqref{dpool}.

\subsection{Transport Statistics from Cycle Statistics for Heterogenous Ensembles}
\label{sub:rendelayed}

\begin{figure}
\centering
 \includegraphics[width=.7\textwidth]{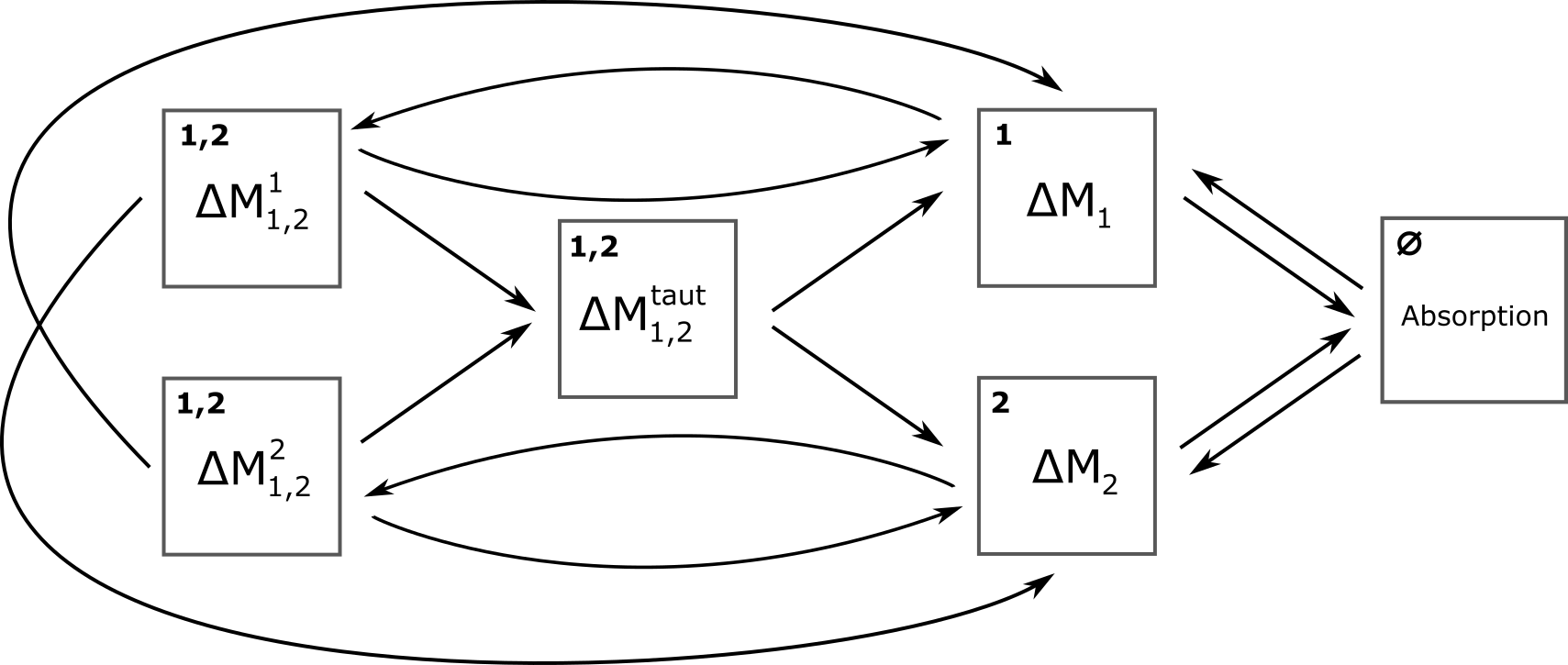}
 \caption{State space and transitions of procession increments for heterogeneous
ensembles with nonzero trap force.} \label{biggerstate}
 \end{figure}

For heterogenous ensembles,  we generally need to augment
the state space by letting $(1,2;j)$ denote a state in which both motors
are attached and motor $j = 1,2$ is the one which most recently attached.   The reason we need this distinction is that the initial distribution of the motor displacements depends on which motor has just attached, and so do the ensuing dynamics while the tether of the recently attached motor remains lax and the other motor is carrying the trap force load on its own.    These distinctions are technically transient, and thus disappear if one can assume the statistics of the intermotor displacement are governed by their stationary distribution, as we did in~\citet{klobusicky2020cooperative}.  As we explain in Subsection~\ref{laxhetnonz}, the distinctions are also irrelevant when trap force is zero and the tether properties are the same for both motors (as in our models for kinesin-1 and kinesin-2) so that the initial distribution of the displacement between the motors (given in Eq.~\eqref{convr}) does not depend on which motor just attached.  For weak trap forces, the differences in the statistics between states $ (1,2;j) $ are small, and thus, provided the tether properties of the two motors are similar, the results of the augmented state space model  should be close to those of the simpler state space model presented in Subsection~\ref{sub:renewal} for homogenous ensembles.

This network structure of the state transitions in the augmented state space model, shown in Fig. \ref{biggerstate}, 
 disrupts the renewal process framework because the state with two motors attached ``remembers'' the previous state of one motor attached.  We thus give up  on the renewal framework for heterogenous ensembles in the present work.  
Even if a renewal-reward approach were applicable for heterogenous ensembles, the analytical computation of the second order statistics for the nonlinear cable tether force model~\eqref{eq:cable}  would produce excessively complicated expressions, as we explain in Appendix~\ref{app:secmompain}.   Thus we content ourselves here with developing only mean statistics of runs for heterogenous ensembles.  
We develop next  a recursive calculation for the probability for which motor is the one attaching into the \TAM\ state in the $k$th cycle, and thus whether substate $ (1,2;1) $ or $(1,2;2) $ is realized in that cycle.  From knowledge of that substate, we compute the means of the time increment and cargo tracking variable increment using formulas to be developed in Section~\ref{sec:laxtaut}.

First we note from  Fig. \ref{biggerstate} that if motor $j $ detaches from the \TAM\ state, 
then substate $(1,2; j)$ will be visited in the subsequent cycle (assuming there is no
complete detachment).   This allows us to develop a recurrence equation to express the probability $p^{(1)}_{\mathrm{a}}[k]$ that the $k$th cycle is initiated with 
 motor 1 being the attaching motor into the \TAM\ state, or equivalently, that $ (1,2;1) $ is the substate of the \TAM\ state visited in cycle $k$.  For initial conditions, we consider the motor ensemble beginning in the completely
detached state $\emptyset$, and compute the probability of motor 2 attaching
followed by motor 1 attaching, conditioned on the ensemble
achieving full attachment. This is given  by 
\begin{equation}
p^{(1)}_{\mathrm{a}}[1] = \frac{\frac{a_2}{a_1+a_2}\frac{a_1}{a_1+\detone{2}}}{\frac{a_2}{a_1+a_2}\frac{a_1}{a_1+\detone{2}}+
\frac{a_1}{a_1+a_2}\frac{a_2}{a_{2}+\detone{1}}} = \left(1+\frac{a_1+\detone{2}}{a_2+\detone{1}}\right)^{-1}. \label{eq:pinitattach}
\end{equation} 
Let $ p^{(1), \mathcal A}_{\mathrm{d};j}$ denote the probability of transitioning from state $(1,2;j)$ to $(1)$ conditioned on  reattachment, which implies that the \TAM\ state will be revisited in substate $ (1,2;1)$.  This probability is computed as part of the cycle statistics in Eq.~\eqref{eq:pdetia}, where the subscript $j$ for the identity of the most recently attached motor was kept implicit.
 
 For $k> 1$ the probability that when the $k$th cycle is realized, it visits substate $(1,2;1)$ (rather than $ (1,2;2) $) is then  
 \begin{equation}
p^{(1)}_{\mathrm{a}}[k] =p^{(1)}_{\mathrm{a}}[k-1] p^{(1), \mathcal A}_{\mathrm{d;1}}+(1-p^{(1)}_{\mathrm{a}}[k-1]
)p^{(1), \mathcal A}_{\mathrm{d;2}}.
 \end{equation}

  This first order linear difference equation can be solved to give for $ k>1$:
\begin{equation}
 p^{(1)}_{\mathrm{a}}[k] = p^{(1)}_{\mathrm{a}}[1] (p^{(1), \mathcal A}_{\mathrm{d;1}}-p^{(1), \mathcal A}_{\mathrm{d;2}})^{k-1}+p^{(1), \mathcal A}_{\mathrm{d;2}}
 \frac{1-(p^{(1), \mathcal A}_{\mathrm{d;1}}-p^{(1), \mathcal A}_{\mathrm{d;2}})^{k-1}}{1-p^{(1), \mathcal A}_{\mathrm{d;1}}+p^{(1), \mathcal A}_{\mathrm{d;2}}}; \label{eq:pattk}
\end{equation}
the edge case $ p^{(1), \mathcal A}_{\mathrm{d;1}} = 1, p^{(1), \mathcal A}_{\mathrm{d;2}} =0 $ is unrealizable for any reasonable model parameter choices.  

The probability $ p_{d}^{\emptyset}[k] $ of the ensemble detaching at the end of the $k$th cycle for $k\ge 1$ is no longer described by a geometric distribution, but rather by the following relation to the previous attachment probability distribution:
\begin{equation}
 p_{d}^{\emptyset}[k] = p^{(1)}_{\mathrm{a}}[k]  p_{d; 1}^{\emptyset}
+(1-p^{(1)}_{\mathrm{a}}[k])p_{d;
2}^{\emptyset},
\end{equation}
with the probabilities $ p_{d;j}^{\emptyset} $ for a cycle which visits substate $ (1,2;j) $ ending in complete detachment computed in Eq.~\eqref{probcompdet}, with the index $j $ of the most recently attached motor suppressed.  From here, we compute the probability  $ \mathbb P(N_c\geq k)= \prod_{k^{\prime} = 1}^{k-1} (1- p_{d}^{\emptyset}[k^{\prime}] ) $, the probability the motor ensemble does enter cycle $k$ (i.e., has not completely detached before cycle $k$).   We recall that we implicitly condition on at least one  cycle occurring.

The expected number of cycles is then given by the infinite sum
\begin{equation}
        \E[\Ncyc ]  
        = \sum_{k=1}^{\infty} \mathbb P(\Ncyc \geq k).
\end{equation}
The mean time and run length are computed as follows:
\begin{align}
        \mu_T &\equiv \E[T] = \E[\Delta T_{\mathbf{a}}] + \sum_{k = 1}^\infty   \mathbb P(N_c \geq k)
        (p^{(1)}_{\mathrm{a}}[k]\E[ \Delta T_{\mathrm{c}; 1}]+ (1-p^{(1)}_{\mathrm{a}}[k] )\E[\Delta T_{\mathrm{c}; 2}]), \\
        \mu_L &= \E[\Delta M_{\mathbf{a}}] + \E[\Delta Z_{\mathbf t}] - \E[\Delta Z_{\mathbf a}]  \\
        & \qquad \qquad +\sum_{k = 1}^\infty   \mathbb P(N_c \geq k) 
        (p^{(1)}_{\mathrm{a}}[k] \E[\Delta M_{\mathrm{c}; 1}] + (1-p^{(1)}_{\mathrm{a}}[k] )\E[\Delta M_{\mathrm{c}; 2}])  \nonumber
\end{align}
with $j = 1,2$ in the subscript for variables denoting  times and displacements  of cycles which visit substate $(1,2;j)$. Their means are computed as described in Subsection~\ref{laxhetnonz}.  The mean contributions $ \E[\Delta T_{\mathbf{a}}] $ and $ \E[\Delta M_{\mathbf{a}}]$  from the visit to the initial \OAM\ state are computed by taking a weighted average of the formulas from Eq.~\eqref{eq:oamstats} with probability $ p^{(1)}_{\mathrm{a}} [1] $ from Eq.~\eqref{eq:pinitattach} giving the probability that motor $2 $ is the motor which attaches first (given the \TAM\ state will be reached).  The mean displacements of the cargo $ \E[\Delta Z_{\mathbf a}] $ and $ \E[\Delta Z_{\mathbf t}] $ from the cargo tracking variable at the beginning and end of the run are computed as in Appendix~\ref{app:zshift}.  The infinite sums are truncated numerically where the approximately geometrically decreasing summands become negligible.

The ensemble velocity~\eqref{pooleddiff} can still be computed via the law of large numbers as $ \Vpool = \mu_L/\mu_ T$, with no simplification in terms of cycle statistics as in~\eqref{vpool}.  The long-run velocity~\eqref{runvel}  can be computed as the ratio of the mean cycle displacement to mean cycle time $ \Vrun = \mu_{\dMcyc}/\mu_{\dTcyc} $ as for homogenous ensembles, but with two changes to the cycle statistics:
\begin{itemize}
\item we use the limiting probability distribution for the \TAM\ substate visited, which is obtained by taking $ k\rightarrow \infty $  in Eq.~\eqref{eq:pattk}:
\begin{align}
p^{(1)}_{\mathrm{a;eq}}  
 =\frac{p^{(1), \mathcal A}_{\mathrm{d;2}}}{1-p^{(1), \mathcal A}_{\mathrm{d;1}}+p^{(1),
\mathcal A}_{\mathrm{d;2}}}. \label{req}
\end{align}
\item some probabilities involving transitions between states within the cycle must be modified by conditioning on the event $ \mathcal{A} $ that the cycle ends in reattachment to start another cycle.  The details are described in Eq.~\eqref{eq:condreat}.
\end{itemize}

\section{Effective Dynamics in States with One Motor Attached}
\label{sec:oam}

\begin{figure} 
\centering
\includegraphics[width=.5\textwidth]{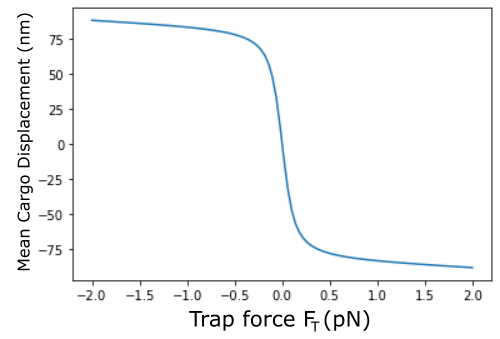}
\caption{
Relation~\eqref{eq:carglag} between the trap force $F_T$ and average displacement of the cargo relative to the attached single motor under cable spring model~\eqref{eq:cable} for kinesin-1 with parameter values given in Table \ref{paramtable}.   Negative displacement values correspond to the cargo lagging behind the motor.  
\label{xifig}}
\end{figure}
For the \OAM\ (one motor attached) states, we adopt the practice~\cite{klobusicky2020cooperative,mckinley2012asymptotic,miles2018analysis,leighton2022performance}  of averaging out the cargo dynamics to obtain an effective constant drift and constant diffusivity.   We justify its applicability even with a tether that can go slack after presenting the framework derived in~\cite{klobusicky2020cooperative}.

Within a given \OAM\ state, the stationary distribution of displacement $ Z(t)-X^{(i)}(t) $ from the attached motor $i$ to the cargo $ Z(t) $ is given through~\eqref{cargo} by:
\begin{align}
\rho_i(y;F_T) = \frac 1{\pdfnorm_i}\exp\left(-[\kappa^{(i)}(|y|-
\ellrest{i})_+^2/2+F_Ty]/k_BT \right) \label{eq:att1stat}
\end{align}
with normalizing constant $\mathcal N_i$.   
We remark that this quasi-stationary distribution for the cargo position, as well as the version~\eqref{rho2} for the \TAM\ state,
has no directional bias with respect to the positions of the attached motors
in the absence of a trap force $ F_T $, and in particular neglects the effect
of drag on the cargo. We argue in Appendix~\ref{app:drag} that the drag force has a small effect for \textit{in vitro} experiments, justifying its omission in our analysis.

The cargo tracking variable, defined in Section~\ref{sec:cycle}, while in an \OAM\ state $(i)$ is simply $ \CM_i(t)= X^{(i)} (t) $.
For the linear spring used in \cite{klobusicky2020cooperative}, $\rho_i$ is a normal distribution and the 
cargo has an average  displacement $ -F_T/\kappa^{(i)} $ from the attached motor.  For the nonlinear cable model~\eqref{eq:cable}, a more complex expression for the mean cargo displacement from the attached motor is derived in Appendix~\ref{app:zlagformula} and plotted in Figure~\ref{xifig}.  A main observation is this substantial change near zero trap force, which can be readily explained by the lack of an opposing force as the cargo is pushed
back or forward until the relative displacement of the cargo from the attached motor reaches $\pm \ellrest{i}$. 

\begin{figure}
\centering
\includegraphics[width=.5\textwidth]{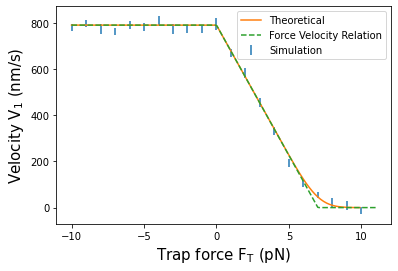}
\caption{Comparison of theoretical effective velocity~\eqref{gbari} (orange solid line), the bare force velocity relation $ v^{(i)} g(F_T/F_s^{(i)}) $ (Eq.~\eqref{fvelrel}) (dashed green line) and direct Monte Carlo simulations (blue  error bars) for each of the trap force values $F_T = -10,-9, \dots, 9,10 \si{\pico\newton}$ for kinesin-1 parameters listed in Table~\ref{paramtable}. 
Error bars describe the standard error about the sample mean over 30 test runs.}
\label{onevelgraph}
\end{figure}

  
By averaging the force-velocity relation against the force distribution on the motor induced by the distribution~\eqref{eq:att1stat} of the tether stretch,  we obtain an  effective velocity of the motor-cargo complex as:
\begin{equation}\label{gbari}
\Vstate{i}=
v^{(i)} \int_\mathbb R g (F^{(i)} (-y)/F_s^{(i)}) \rho_i (y)
\, \difd y. 
\end{equation}
The analytical evaluation of this integral is presented in Appendix~\ref{app:velformula}, and
a graph comparing this function with Monte Carlo simulations under a range of trap forces is given in Fig.  \ref{onevelgraph}. 
In contrast to the linear spring model studied in~\citet{klobusicky2020cooperative,mckinley2012asymptotic}, we find the effective velocity of the motor-cargo complex as a function of trap force is negligibly affected by the fluctuations in cargo force, even at low trap forces:  $ \Vstate{i} (F_T) \approx v^{(i)} g(F_T/F_s^{(i)}) $.   The reason, as described in more detail in Appendix~\ref{sec:cargoflucs}, is that the tether with the cargo is primarily slack at low trap force, while the force-velocity relation is approximately linear over the range of fluctuating forces at larger trap forces.

The asymptotic arguments in~\citet{klobusicky2020cooperative} show the cargo fluctuations to negligibly enhance the motor diffusion.  To make this more directly quantitative, we can use the Kubo formula~\citep[Sec. 1.6]{rk:sp2} to estimate diffusivity enhancement to be no greater than the squared-motor velocity $ (v^{(i)})^2 $ multiplied by the correlation time of the cargo fluctuations, which is $ \gamma/\kappa^{(i)} $ when the tether is engaged.  For kinesin-1 and kinesin-2, this product is no larger than  $ \SI{30}{\nano\metre^2/\sec^{-1}} $ which is one to two orders of magnitude smaller than their bare diffusivities.  When the tether is slack, then the cargo fluctuations are having no effect on the motor, so we carry over the conclusion that the effective diffusivity of the motor-cargo complex in a \OAM\ is approximately the same as the bare value.  Thus, we may approximate the dynamics of the averaged cargo coordinate $ M_i (t) $ when motor $i$ is attached via 
the constant-coefficient effective  SDE:
\begin{equation}
dM_i (t)= V_idt+\sqrt{2D_i}dW(t).
\end{equation}

Continuing with the assumption of scale separation between cargo fluctuations and the motor attachment/detachment time scales, the detachment of the attached motor $i$ can be approximated as being governed by 
the constant effective detachment rate obtained by averaging the force-dependent detachment rate against the distribution of forces endured by the motor as induced by the tether stretch distribution~\eqref{eq:att1stat}
\begin{equation}
\bar d_i^* = \int_\mathbb R d^{(i)} (F^{(i)} (-y))\rho_i(y)dy. \label{detavg}
\end{equation}
Then in the \OAM\ state $ (i)$ we have the classical competition between Markov jump processes of detachment of the attached motor $i$ at constant rate $ d_i^* $
 against attachment of the detached motor $i'$  with constant rate $a_{i'}$, so the first event of either attachment or detachment is  exponentially distributed:
 \begin{equation}\label{timedists}
  \quad \dTdet{i} \sim \Exprand
((\bar{a}^{*(i^{\prime})}+\detone{i})^{-1}).
\end{equation}
We can then compute the mean and variance of the increment of the averaged cargo coordinate $\dMdet{i} = M_i(t+\dTdet{i})-M_i(t)$
 by conditioning on $\dTdet{i}$ and using that increments of  Brownian motion are normally distributed.  Increments in \OAM\ states therefore have the effective statistics:  
\begin{align}
\E[\dTdet{i}
 ]&=  \frac{1}{\bar a^{*(i')}+\detone{i}}, \qquad  \E[\dMdet{i}
 ]=  \frac{\Vstate{i}}{\bar a^{*(i')}+\detone{i}}, \nonumber \\
 \Var(\dTdet{i}
 )&=  \frac{1}{(\bar a^{*(i')}+\detone{i})^2}, \quad  \Var(\dMdet{i} )
= \frac{2D_{i}}{\bar a^{*(i')}+\detone{i}}  +\left(\frac{\Vstate{i}}{\bar
a^{*(i')}+\detone{i}}\right)^2, \nonumber \\
\Cov (\dTdet{i},\dMdet{i}) &= \frac{\Vstate{i}}{(\bar a^{*(i')}+\detone{i})^2}.  \label{eq:oamstats}
\end{align}

We address now the question of whether the scale separation argument used in~\citet{klobusicky2020cooperative} to derive these formulas for the linear spring model should also apply for the cable tether model~\eqref{eq:cable}.  The key assumption, supported by the biophysical parameters in Table~\ref{paramtable} for the linear spring model, is that  the tether dynamics occur on a faster time scale than the attachment/detachment dynamics of the motor.  But with the nonlinear cable model~\eqref{eq:cable} for the tether, we must take into account the equilibration of the tether under phases in which the motor is not engaged with the cargo.  First, we consider the situation with a trap force $F_T $ on the order of a few $ \si{\pico\newton}$, that will itself push the cargo within the slack region toward the mean of its stationary distribution at a velocity $ F_T/\gamma  \sim 10^5 $nm/s.  As the cargo only feels a slack tether within $ \pm 80 \si{\nano\metre}$ of the attached motor, the trap force will push the cargo to engage the tether within $ \SI{1E-3}{\sec} $. 
When the tether is engaged, the time scale of the cargo dynamics is $\gamma/\kappa^{(i)}= 4 \times 10^{-5}$s.  Both times are orders of magnitude smaller than the dwell time  scale of the \OAM\ state $1/(a_{i'}+\bar d_i^*) \ge \SI{.04}{\sec}$ for kinesin-1 and kinesin-2, where we have numerically evaluated the averaged detachment rate $\bar d_i^*$ defined in (\ref{detavg}).  For the case of no trap force $F_T = 0$, the relevant relaxation time scale is the diffusion time scale $ \ellrest{i}{}^2/D =\SI{0.02}{sec}$ of the cargo over the an interval of half-width equal to the tether rest length $ \ellrest{i}=80 \si{\nano\metre}$, with cargo diffusivity $D = k_BT/\gamma$.   
This is comparable to the time scale of attachment or detachment, and thus the scale separation argument nominally fails.  However, this is of little concern, because in the absence of a trap force, the quasi-stationary distribution of the cargo is heavily concentrated in the slack region (see Appendix~\ref{sec:cargoflucs} for further discussion), where its precise position has no effect on the motor.   A similar argument could be made, with less precision, for the case of trap forces $ |F_T| \lesssim \SI{0.1}{\pico\newton} $ weak enough that the drift time scale is comparable or long compared to the diffusion time scale across the slack configuration of the tether.

\section{Effective jump statistics}
\label{sec:jump}
As discussed in Section~\ref{sec:intro}, the scale separation argument in the \TAM\ state $(1,2) $ requires substantial revisitation, which we take up in Section~\ref{sec:laxtaut}.  But the statistics of the jump in the cargo tracking variable between \TAM\ and \OAM\ states carries through with a direct generalization, and we present these now.  
To this end, we still treat the cargo dynamics as being in a quasi-stationary distribution with respect to the attached motors in the \TAM\ state.   The arguments supporting this from~\citet{klobusicky2020cooperative} apply when the tether is engaged with one or both motors, and when both tethers are slack, the argument at the end of Section~\ref{sec:oam} can be applied.  Thus we treat the cargo position as always governed by the quasi-stationary distribution given the positions $x_1, x_2$ of the attached motors, which has probability density
\begin{equation}
\rho_{1,2}(z;x_1,x_2) =  \frac 1{\pdfnorm_{1,2} (x_1-x_2,F_{T})}\exp\left(-\frac{1}{k_BT}\left[\sum_{i
= 1}^2\kappa^{(i)}(|z-x_i|- \ellrest{i})_+^2/2+F_Tz\right]
\right).  \label{rho2}
\end{equation}
The normalization constant $ \pdfnorm_{1,2} (r;F_T) $ can be expressed directly in terms of the motor separation $ r= x_1-x_2 $ via the change of variables relating the cargo position $ z $ to the average attached motor position:  $ y = z-(x_1+x_2)/2 $:
\begin{align}
\pdfnorm_{1,2} (r,F_{T})=\int_{\mathbb{R}}
\exp\left(-\frac{1}{k_BT}\left[\sum_{i
= 1}^2\kappa^{(i)}(|y+(-1)^{i}r/2|- \ellrest{i})_+^2/2+F_Ty\right]
\right)dy. \nonumber
\end{align}


The cargo tracking variable $ \CM (t) $ defined in Section~\ref{sec:cycle}  will jump when motors attach and detach.  To work out these jumps, we use the fact that detached motors are fast relative to the cargo to represent their displacement from the cargo by the quasistationary distribution, with probability density
\begin{align}
\rho_i^{(d)}(y) =\rho_i(y;0).   \label{rhod}
\end{align}
That is, the tether displacement between the cargo and a detached motor has the same quasistationary distribution as between an attached motor (of the same type) in an \OAM\ state and the cargo under zero trap force, essentially because the drag force from the attached motor's motion does not affect the leading order cargo dynamics as argued in~\citet{klobusicky2020cooperative}.  The distribution of the detached motor position is relevant in our model because it is used as the distribution of where the motor attaches.  

The jump in the cargo tracking variable upon detachment of motor $i$ from the state $(1,2)$ to state $(\ip)$ is found by comparing the values immediately before and after a detachment time $\dettime$, yielding
\begin{equation}
\dCMdetjump{i} = \CM_{\ip}(\dettime)-\CM_{1,2} (\dettime^{-}) = X^{(\ip)}(\dettime) - \frac{1}{2} (X^{(1)} (\dettime^{-})+X^{(2)} (\dettime^{-})) = (-1)^{i} \frac{\Rdet_{i}}{2}. \label{eq:detjumpr}
\end{equation}
Here $ \Rdet_{i}  \sim R(\dettime^{-}) $ is a random variable denoting the intermotor displacement just before detachment of motor $i$.  Its statistics will be worked out in Section~\ref{sec:laxtaut}, where we also set the context of possible dependence on the index $j$ of the motor in a heterogenous ensemble that attached into the \TAM\ state.  

For the  jump from a \OAM\ to a \TAM\ state, note that
the position of  unattached motor $i$ immediately before it attaches can
be written by adding the already attached motor position to the displacement of the cargo from that attached motor and the displacement of the detached motor relative to the cargo:
\begin{align} \label{unbitoz}
X^{(i)}(\atttime^-)= X^{(i')}(\atttime^-)+(-1)^{i+1}\Ratt{i} ,\\
 \Ratt{i}  \sim (-1)^{i+1}( \Theta^{(i')}_1(F_T)-\Theta^{(i)}_2(0)), \label{convr}
\end{align}
 where $\Theta^{(i)}_j(F_T),$ $j = 1,2$ are independent and distributed according to $\rho_i(\cdot, F_T)$. Here, $\Ratt{i} $ denotes the intermotor
displacement at the time of the attachment of motor $i$.   
Under our model that motors attach at a position governed by their detached spatial distribution,
the jump in the cargo tracking variable upon attachment of motor $i$ is then
\begin{align}
\dCMattjump{i} &= \frac{X^{(1)}(\atttime)+X^{(2)}(\atttime)}{2} - 
X^{(i')} (\atttime^-) \label{eq:dcMattjump}
\\
&=(-1)^{i+1} \frac{\Ratt{i} }{2}. \nonumber
\end{align}

For homogenous ensembles, the index of which motor is attaching into the \TAM\ state clearly doesn't affect the statistics of the jump upon attachment so we can assume the index of the attaching motor $\Iatt $ is equally likely to take the values $ 1 $ and $ 2 $ (to preserve symmetry of expressions for homogenous ensembles) and we can write more simply the initial separation upon attachment as:
\begin{equation}
\Ratt{}  \sim (-1)^{\Iatt+1} (\Theta_1 (F_T) - \Theta_2 (0)) \label{eq:ratthom}
\end{equation}
where $ \Theta_j (F_T)$, $j=1,2$ are random variables defined as above with the irrelevant motor index label suppressed.   Analytical expressions for the first two moments of $ \Ratt{i} $ are presented in Appendix~\ref{app:rattformula}.

\section{Parameter optimization for force-detachment rate relationship}\label{sec:opt}
In~\citet{klobusicky2020cooperative} we adopted directly the double exponential model fits from~\citet{milic2017intraflagellar,milic2014kinesin}   to describe the rate of detachment of kinesin-1 and kinesin-2 from the microtubule, as a function of trap force strength in those experiments.  This modeling choice has at least two concerns:  1) what we require in Eq.~\eqref{detmod} is a relationship between detachment rate and the momentary longitudinal force on the motor via its tether to the cargo, not the cargo trap force, and 2) the double exponential model fits from~\citet{milic2017intraflagellar,milic2014kinesin,andreasson2015elife} are discontinuous at zero trap force due to the exponential fits being done separately for assisting and hindering trap forces.  The results from zero trap force experiments were used in the fit for the rate of detachment for hindering trap forces but not for assisting trap forces.  To be clear, our concern is not with the statistical analyses adopted in~\citet{milic2017intraflagellar,milic2014kinesin,andreasson2015elife}, but with adopting this particular data product as an input to a theoretical or computational model.  

The fundamental issue is that the motor feels a fluctuating force from its tether which can deviate significantly from what would be assumed based on the trap force on the cargo.  This point has been well appreciated in systems with multiple cooperative motors, where deviations from load sharing of the cargo trap force have been explicated~\citep{driver2011productive,furuta2013measuring,Jamison:2010,kunwar2011mechanical,diehl2009interference}.   But this is also relevant when a single motor is carrying the cargo load, as noted in~\citet{arpag2019kin1kin3,mckinley2012asymptotic}.     We provide a minimalist calculation based on the cable model~\eqref{eq:cable} for the tether in  Appendix~\ref{app:detcargoflucs} to show the importance of taking force fluctuations from the cargo into account when computing the detachment rate as a function of instantaneous force from an experiment with an applied trap force.  We further remark that in~\citet{andreasson2015elife},
the detachment rate is inferred from quotients of mean velocities and run
lengths, not from actual run times, which could lead to some bias when fitting a detachment rate in a theoretical model.  Thus, we argue first that experimental relationships between cargo trap forces and dynamical properties of motors should not be directly employed in dynamical models involving functions of the momentary tether force on a motor.

The force fluctuations also render the discontinuity in detachment rate as a function of force problematic.  If we employed a discontinuous double exponential model for the detachment rate as a function of instantaneous force on the motor, then at zero trap force, we would see force fluctuations to small hindering and assisting values.  The small assisting forces would incur a dramatic, and presumably unrealistic, rise in detachment rate for the motor.  We would thereby find that the run lengths and run times in simulations of single kinesin-1 or kinesin-2 motors would be considerably shorter at zero trap force than those reported in~\citet{milic2017intraflagellar,milic2014kinesin,andreasson2015elife} if we used a detachment rate relation to  trap force on the cargo as our detachment rate relation to the instantaneous tether force.   The reason the experimental data in~\citet{milic2017intraflagellar,milic2014kinesin,andreasson2015elife} is fit to a discontinuous model for detachment rates is likely because assisting force experiments could only be performed for trap forces of $ \SI{2}{\pico\newton}$ and greater, and the run length data is very different at zero trap force and this smallest assisting trap force value.  Presumably a transition occurring at small assisting trap forces could not be resolved by experiments.  \citet{bovyn2020attach} adopted an interpolation to remove the discontinuity in their model.  Mindful of the distinction between the detachment rate model for instantaneous tether force used in their simulation, \citet{arpag2019kin1kin3,arpag2014populations}
describe an inverse inference of an exponential relation between detachment
rate and instantaneous force from the experimental data in~\citet{andreasson2015elife}
for use
in their computational models, 

\subsection{Methodology}

Rather than simulating model output as in~\citet{arpag2014populations}, we use the theoretical predictions presented in Section~\ref{sec:oam} for the \OAM\ state, together with some analytical integral evaluations presented in Appendix~\ref{sec:rhointegral} to expedite computations, to fit the three-parameter \emph{continuous} double exponential model~\eqref{detmod} to the experimental data 
in \citet[Fig. 1]{milic2014kinesin} for  kinesin-1 and \citet[Fig. 4]{milic2017intraflagellar}  for kinesin-2.  These report sample means and standard errors  of run lengths for a large range of assisting and hindering trap forces.  We will only use their data for trap forces between $-6$ pN and $6 $ pN.  Note that \citet{milic2014kinesin} and~\citet{milic2017intraflagellar} use the convention of positive trap forces being assisting, which is counter to our sign convention, and we tacitly make this sign conversion in what follows.


For a given motor type $i$ and choice
of parameters $\theta \in \Theta^{(i)}$ in the parameter space
\begin{equation}
 \Theta^{(i)} = \{(d_0^{(i)},F_{d-}^{(i)},F_{d+}^{(i)}): d_0^{(i)},F_{d-}^{(i)},F_{d+}^{(i)}>0\},
\end{equation}
and each value $ f_j $ of trap force in the experimental data set, we compute our parameterized theoretical model prediction for run length as 
the quotient of the effective velocity $\Vstate{i} (f_j)$~\eqref{gbari} and effective detachment rate $\bar d_i^*(f_j;\theta)$~\eqref{detavg}.  In this section we temporarily make explicit the dependence of these quantities on the trap force and detachment rate model parameters.
 
\begin{figure}
\centering
\includegraphics[width = .5\textwidth]{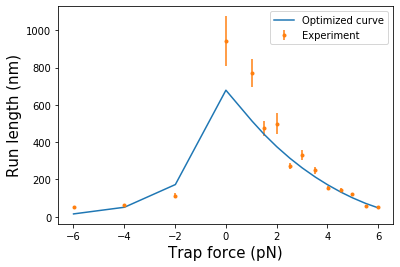}
\caption{Optimized curve for run lengths of kinesin-1 as a function of trap force under
heteroskedastic least squares minimization.  Experimental data from~\citet{milic2017intraflagellar,milic2014kinesin} is plotted with standard error bars.  The continuous curve indicates the theoretical prediction for run length using the detachment rate model with the optimized parameters:
$\widehat d_0^{(1)} = .61$/s, $\widehat F_{d-}^{(1)} = .62$ pN, and
$ \widehat F_{d+}^{(1)} = 2.95$ pN.\label{fig:milicfit}}
\end{figure}

For a given motor type $i$ we take from the previously cited figures a dataset $\{(f_1, l_1,se_1),
\dots, (f_n, l_n,se_n)\}_{j=1}^n$ of reported trap forces $ f_j $ with reported run length mean $ l_j $ and standard error $ se_j $.  We then set up an optimization problem  to compare $l_j$ with the
run length estimate $\bar
V_i(f_{j})/\bar d_i^*(f_j;\theta)$.  A first attempt  was to consider
a weighted (heteroscedastic) least squares minimization, where estimators
are given by\begin{align}
(\widehat d_0^{(i)},\widehat F_{d-}^{(i)}, \widehat F_{d+}^{(i)}) = \argmin_{\theta
 \in \Theta^{(i)}} \sum_{j = 1}^n \frac{1}{\mathrm{se}_j^2}(\bar
V_i(f_{j})/\bar d_i^*(f_j;\theta)-l_j)^2. \label{milell1fit}
\end{align}See Fig.~\ref{fig:milicfit} for the optimized
curve under the minimization (\ref{milell1fit}) for kinesin-1.

The disappointing fit seems to be driven by the data at large trap forces,
which have smaller standard errors presumably because more experimental trials
were performed than for smaller trap forces.  But especially for the purpose
of understanding the dynamics of motor-cargo complexes in actual cells, where
the viscous drag force should correspond to a relatively small trap force,
we are more interested in an accurate representation of the force-detachment
rate relation at small forces.  Therefore we modify the loss function to
be optimized as follows 
\begin{align}
(\widehat d_0^{(i)},\widehat F_{d-}^{(i)}, \widehat F_{d+}^{(i)}) = \argmin_{\theta
 \in \Theta } \sum_{j = 1}^n \riskfn (f_j)(\bar
V_i(f_{j})/\bar d_i^*(f_j;\theta)-l_j)^2. \label{milell1fitweight}
\end{align}
where we have introduced the function $ \riskfn (F) $ which provides
a subjective weight to how important the quality of fit should be at various
trap force values.  We take $ \riskfn (F) = \frac{1}{\sqrt{2\pi \sigrisk^2}}
\expe^{-F^2/(2\sigrisk^2)} $ to be a centered Gaussian with width $ \sigrisk
$, to prioritize the fit at lower values of trap force.   We choose $  \sigrisk
$ to correspond to the typical motor-cargo tether forces we expect to see
in the ``natural'' setting in the absence of a trap force ($F_T = 0 $). 
From Appendix~\ref{sec:cargoflucs}, we expect such tether forces to fluctuate
in the range $ \pm \sqrt{k_B T \kappa^{(i)}} $ in a state when only motor
$i $ is attached.  During the state when both motors are attached, taking
motor $i=1$ to be the one with the faster unloaded speed, this typically
leading motor $i=1$ will have a tether force distribution roughly centered
at $ F_s^{(1)} (1-v^{(2)}/v^{(1)})$ with width
 \begin{equation*}
\sqrt{\frac{k_BT (\kappa^{(1)})^2}{\kappa^{(1)}+\kappa^{(2)}} + \frac{(D^{(1)}+D^{(2)})
\kappa^{(1)}\kappa^{(2)} F_s^{(1)}} {v^{(1)}(\kappa^{(1)}+\kappa^{(2)})}},
\end{equation*}
while the typically trailing motor $ i=2 $ will have a tether force distribution
roughly centered at $ -F_s^{(1)} (1-v^{(2)}/v^{(1)}) $ with width
\begin{equation*}
\sqrt{\frac{k_B T (\kappa^{(2)})^2}{\kappa^{(1)}+\kappa^{(2)}} + \frac{(D^{(1)}+D^{(2)})
\kappa^{(1)}\kappa^{(2)} F_s^{(1)}}{v^{(1)}(\kappa^{(1)}+\kappa^{(2)})}}.
\end{equation*}
As both kinesin-1 and kinesin-2 are modeled with the same tether elasticity
constant $ \kappa $, and kinesin-2 has the larger motor diffusivity $ D $,
we choose: 
\begin{equation*}
\sigrisk = \max \left(\sqrt{k_B T \kappamax},F_s^{(1)} (1-v^{(2)}/v^{(1)})
+ \sqrt{\frac{k_B T \kappa}{2} + \frac{D^{(2)} \kappa F_s^{(1)}}{v^{(1)}}}\right),
\end{equation*}
with, just in this formula, motor 1 taken to be kinesin-1 and motor 2 to
be kinesin-2. This produces  a numerical value of $\sigrisk = 
 \SI{1.8}{\pico\newton}$, 
which
we use in the weight factor for the optimization~\eqref{milell1fitweight}.
This should provide a rough upper estimate
to characterize the magnitude of motor-cargo tether forces we expect to emerge
for at zero trap force among the various pairings of kinesin-1 and kinesin-2.

In using this estimate for the weight function in fitting the experimental
observations of run length as a function of trap force, we are of course
transferring a measure of what instantaneous motor-cargo tether forces are
important to what trap force measurements are important and these are distinct
quantities.  But this approach does at least us give us a rational basis
for setting a scale for the weight function as an alternative to completely
subjective manipulation.

\subsection{Revised Parameter Fits}
\begin{figure}
  \includegraphics[width=\linewidth]{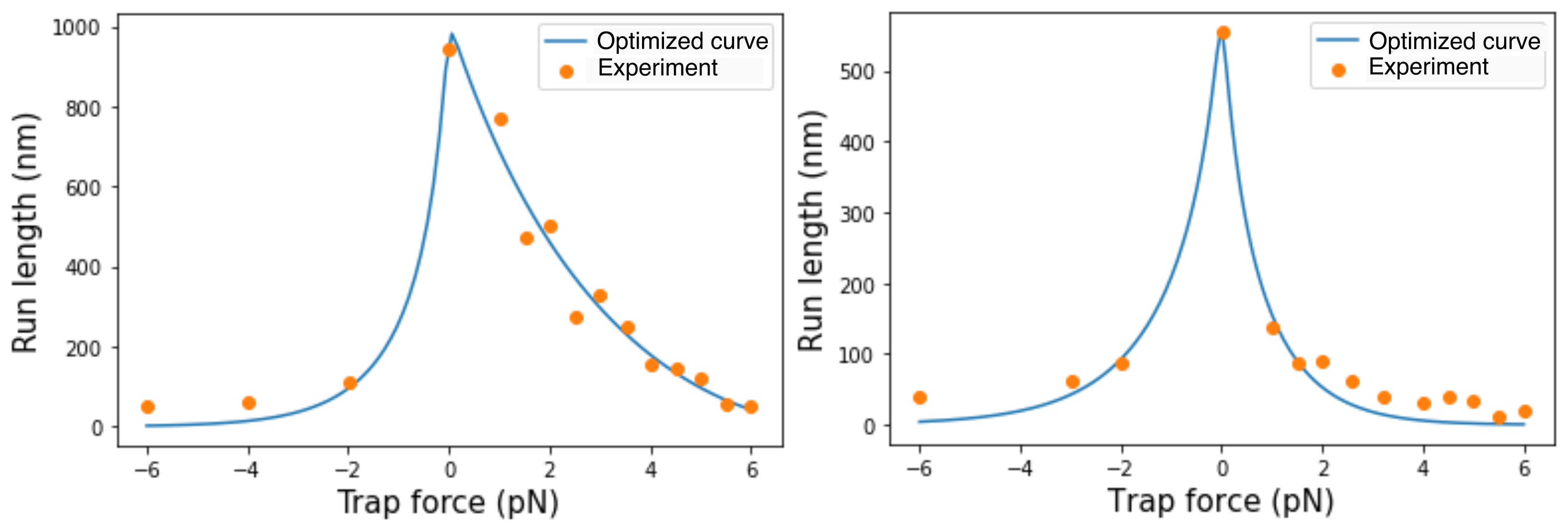}
\caption{Optimized curves for run lengths of kinesin-1 (left) and kinesin-2
(right) as a function of trap force under
loss function minimization with weights.   Experimental data from~\citet{milic2017intraflagellar,milic2014kinesin}
is plotted as discrete points.  The continuous curves indicate the theoretical
prediction for run length using the detachment rate model with the optimized
parameters:  (\textbf{kinesin-1, left}): $ \widehat d_0^{(1)} = .78$/s, $\widehat
 F_{d-}^{(1)} = 1.07$ pN, and
$\widehat  F_{d+}^{(1)} = 4.63$ pN. (\textbf{kinesin-2, right}): $ \widehat
d_0^{(2)} = .84$/s, $\widehat F_{d-}^{(2)} = 1.27 $ pN, and
 $\widehat F_{d+}^{(2)} = 1.15$ pN. } \label{weightfit}
\end{figure}

The fits for optimization with the weight function are given in Fig.~\ref{weightfit}. As expected from using the weight function, the
agreement
between the data and optimal model agrees well for smaller forces. There
are larger differences in run lengths under forces with larger magnitudes,
especially in the case of kinesin-2.  This is likely a result of the drastic
drop in run-lengths as the force ranges from 0 pN to around 2 pN, to which the 
weight
function optimization as formulated gives greater weight and represents with a rapidly decaying exponential curve.

Our fit for a double-exponential model~\eqref{detmod} for the detachment
rate as a function of \emph{instantaneous} force naturally differs, for multiple
reasons, from those obtained from experiments as a function of \emph{applied
trap force} $ F_T $.   First of all, we are imposing continuity at zero force
to avoid artifacts~\citep{klobusicky2020cooperative} from force fluctuations.
  Thus, the assisting force behavior is represented in~\citet{milic2014kinesin,milic2017intraflagellar}
for kinesin-1 and kinesin-2 by a substantial jump increase in the detachment
rate at small assisting loads followed by a slow decrease, on the scale of
$ 10 \si{\pico\newton}$ as a function of increasing assisting load. 
The discontinuity in experimental fits does not seem entirely necessary,
but arising more from underresolution of the variation of run length at small
assisting loads.  Indeed, the smallest assisting load measured in~\citet{milic2014kinesin}
is $\SI{2}{\pico\newton} $, which is from our analysis in Appendix~\ref{sec:cargoflucs},
approximately the scale of fluctuations experienced by 2-motor ensembles
under zero trap force.  Consequently, 
for modeling purposes we would like to resolve the detachment rate at these
force scales, and are less interested in the detachment rate at much larger
trap forces.  Thus we obtain small force scales for the detachment rate as
a function of assisting trap forces.  

Turning next to the zero force detachment rates -- the zero trap force measurements
in~\citet{milic2014kinesin,milic2017intraflagellar,andreasson2015elife} are
all associated to the exponential fit for hindering trap forces, so we find
it most appropriate to compare our zero-force detachment rate $ d_0 $ to
those inferred from experiments at the limit of small hindering forces. 
For kinesin-1,  \citet{andreasson2015elife} inferred detachment rates from
the quotient of velocities and run lengths and obtained a zero-force detachment
rate of $ 1.11 \pm 0.03 \si{\sec}^{-1} $, with the rate increasing under
hindering loads with a force scale of $ 6.9 \pm 0.1 \si{\pico\newton}$. 
Note this force scale is considerably larger than the $ 2.1 \pm 0.1 \si{\pico\newton}
$ inferred from run length as a function of trap force in~\citet{milic2014kinesin},
as this compounds the effects of detachment and the velocity slow-down. 
Our inferred detachment rate at zero \emph{instantaneous} force and force
scale of exponential increase are both approximately two-thirds of the values
from~\citet{andreasson2015elife} fit as a function of \emph{trap} force. 
This is understandable because the force fluctuations from the cargo will
be on the order of $ \sqrt{k_B T \kappa} \sim 1 \si{\pico\newton} $, which
make the measured detachment rate at zero trap force somewhat higher than
the detachment rate at zero instantaneous force~\citep{arpag2019kin1kin3}.
 These fluctuations should also tend to smooth the detachment rate as a function
of trap force, which makes the force scale of increase somewhat larger than
what it is for the instantaneous force law~\eqref{detmod}.  
\citet{arpag2014populations} inferred via their fitting of modeling simulations a zero-force detachment rate $ 0.93 \si{\sec}^{-1}
$ and hindering force scale of increase $ 6.1 \si{\pico\newton} $.  These
are closer to the experimental values as a function of trap force, but we
cannot well examine the cause of the different inferences because the method
of the inference in~\citet{arpag2014populations}  is not clearly articulated,
beyond
the averaging over cargo fluctuations described in \citep[Sec. S3]{arpag2019kin1kin3}
for zero trap force, which is a somewhat simpler version of our averaging
procedure conducted in Appendix~\ref{app:detcargoflucs} for arbitrary trap force.

A similar analysis of run-length and velocity data for kinesin-2 (K1F3AB)
in~\citet{milic2017intraflagellar} inferred a zero-load detachment rate of
approximately
$\SI{3}{\sec}^{-1} $ and did not appear to quantitatively specify the force
scale of exponential increase of detachment rate.  This is oddly much larger
than our inferred detachment rate at zero instantaneous force, contrary to
what we would expect.  This seems however to be more related to the exponential
fitting to the experimental data in \citet[Fig. 5A]{milic2017intraflagellar},
as the detachment rate inferred purely from unloaded experiments (as opposed
to that inferred from an exponential fit to hindering forces) is $ \SI{0.8}{\sec}^{-1}
$, much closer to our inferred detachment rate at zero instantaneous force.
  The force scale of detachment under hindering trap forces in the exponential
fit for \emph{run length} as a function of hindering trap force is $ 2.0\pm0.1\si{\pico\newton}
$, which is actually larger than our detachment force scale for hindering
instantaneous forces.  This would seem again contrary to what we would expect
based on our previous discussion for kinesin-1, but seems to arise again
due to the poor agreement in~\citet{milic2017intraflagellar}  between the exponential
fits over $ \SI{6}{\pico\newton} $ of hindering trap forces to the detachment
rate (or run length) at zero trap force.  With our optimization prioritizing
the fit at smaller force values, we obtain a smaller detachment force scale
to resolve the sharp drop between no trap force and a hindering trap force
of $ \SI{1}{\pico\newton} $.  Our fit for kinesin-2 in Figure~\ref{weightfit} does somewhat
undershoot the run length at trap forces larger than $\SI{2}{\pico\newton}
$.  This seems to be a necessary tradeoff with a single exponential model
for the detachment rate as a function of each sign of the force.   In Appendix~\ref{app:compdetmodels}, we show that using parameterizations for the double-exponential relation between detachment rate and \emph{instantaneous force} based the parameter fits  from~\cite{milic2014kinesin,milic2017intraflagellar} produces considerably worse agreement between their data and  theoretical predictions for run length as a function of \emph{applied trap force}.

More complex detachment rate models could be contemplated.  
\citet{howard2019direction} advocate for fitting the relationship between detachment rate and
trap force in terms of two exponentials with different
force scales for the force along and transverse to the microtubule.  In fact, \citet{howard2019direction,schaffer2021germanium} argue that the run time at constant force should not even be modeled as exponentially distributed, but rather involve two rates rather than one force-dependent rate.  A meaningful fit to such higher-parameter models, as well as others  would require
access to the raw experimental data and probably higher resolution of observations at smaller trap forces.

\section{Lax and taut statistics in two attached motors state}
\label{sec:laxtaut}
The theoretical analysis in \cite{klobusicky2020cooperative} for the effective
dynamics of a cargo transported by two possibly dissimilar cooperative motors
is premised on an assumption of scale separation between the dynamical components.
 The detached motors have the shortest dynamical time scale, then the cargo,
then the attached motors, then the attachment/detachment dynamics.  Under
this idealization, we successively average over the dynamics from fastest
to slowest scales to obtain effective rates of attachment and detachment
together with the updates of the cargo and motor positions upon these attachment
state changes.  As shown in \cite{klobusicky2020cooperative}, the separation
of time scales is nominally supported by the biophysical parameters in at
least in vitro studies, with some complications emerging from the exponential
rise of the detachment rate as a function of the force on the tether. Though
the theoretical framework was implemented on a simple linear model for the
tether force in \cite{klobusicky2020cooperative}, we sketched in Appendix A of that work  how the theory generalizes to nonlinear tether force models.
 A direct implementation to the cable force model~\eqref{eq:cable} under
present consideration, however, led to poor agreement between theoretical
predictions and model simulations (Appendix~\ref{app:nolax}).  

The culprit is evident upon examination of the model simulations, which show
that the motors are often experiencing no force from the tether because the
tether extension is less than the slack length $ \ellrest{i}$.  The assumption
in \cite{klobusicky2020cooperative} that the cargo dynamics are  fast compared
to the dynamics of the attached motors relies on the dynamical time scale
of the tether bringing the cargo dynamics into a stationary distribution
with respect to the current attached motor positions.  This is clearly not
relevant when the tether is slack and therefore not exerting force on the
cargo.  In particular, a motor with a slack tether should proceed not as
a motor feeling a stationary distribution of cargo forces, but rather as
an unloaded motor -- until such time as the motors reposition themselves
to make the tether taut.  This is a new time scale not considered in \cite{klobusicky2020cooperative}, and as we shall argue in detail below, can be comparable to or even exceed  the time until one of the motors detaches.
   Thus we must revisit the effective dynamics of the attached motors, which
in \cite{klobusicky2020cooperative} were assumed to experience a stationary
distribution of forces from the cargo along their tether.

We have previously discussed in Section~\ref{sec:oam} that these new considerations do not dramatically disrupt the general use of the quasi-stationary approximation for the cargo position relative to the attached motors, which is all that is needed in the analysis of the the \OAM\ states.  But we must reconsider the assumption that the relative displacement of the motors in the \TAM\ state reaches a quasi-stationary distribution.  

In this section, 
we refine the calculation of effective statistics in the \TAM\ state by separate consideration of the lax and taut phases within this state.  In mathematical notation to follow, `lax' will refer to statistics related to ensembles before cargo engagement of both motors, and `taut' will refer to the statistics subsequent to the moment when both motors start carrying some force (including later reversions to one or both motors becoming slack again).   The taut phase of course may not be reached before detachment of a motor.  
We label the
time spent in the lax and taut phases as
$\dTattb^{\mathrm{lax}}$ and $\dTattb^{\mathrm{taut}}$,  respectively.  The
total time spent in state $(1,2)$ is then
\begin{equation}
\dTattb = \dTattb^{\mathrm{lax}}+\mathbf 1_{\mathrm{taut}}\dTattb^{\mathrm{taut}},
\end{equation}
where $\mathbf 1_{\mathrm{taut}}$ is an indicator function of whether the
taut phase occurs.  Because of the strong Markov property of the stochastic
dynamics, we can compute $ \dTattb^{\mathrm{taut}} $ as conditionally independent
of $ \dTattb^{\mathrm{lax}} $ given that the taut phase is reached.  
Thus we can  analyze the total time $ \dTattb $ spent in the state with both
motors attached by separately computing 1)  the probability the taut state
is reached before a motor detaches, 2) the conditional probability distribution
for $ \dTattb^{\mathrm{lax}} $ given the taut state is reached (or not) before
a motor detaches, and 3) the probability distribution for $ \dTattb^{\mathrm{taut}}
$ assuming the taut state is reached.  This is why we split out consideration
of taut and lax phases in the subsections to follow.  Similar considerations apply for the decomposition of the displacement $ \dCMattb $ of the cargo tracking variable in the \TAM\ state into its increment $ \dCMattb^{\mathrm{lax}} $ in the lax phase and further increment $ \dCMattb^{\mathrm{taut}} $ in the taut phase, if it is reached.

After collecting some basic technical preliminary statements in Subsection~\ref{sub:tamsetup}, we must separately consider 
the nature of the motor behaviors in the lax and taut phases depending on whether the ensembles are heterogenous or
homogeneous  and the value of the trap force (Subsection~\ref{sub:tamdecomp}).  The qualitative characterization of the onset and dynamics within the taut phase depends on the capacity of the two motors to effectively work cooperatively.   We thus define classes of ``truly cooperative,'' ``falsely cooperative,'' and ``neutrally cooperative'' ensembles in terms of the relation between the individual biophysical properties of the motors.  For heterogenous ensembles, we will without loss of generality assume the motors are indexed so that $ v^{(1)} > v^{(2)} $ (the first motor has a faster unloaded velocity than the second motor).

The taut phase of homogenous ensembles can be reasonably approximated for truly and neutrally cooperative ensembles using the stationary distribution of the \TAM\ state, and the relevant formulas from~\citet{klobusicky2020cooperative} are reviewed in Subsection~\ref{tauthom}.  
In Section \ref{tauthet}, we argue that the stationary distribution is not generally appropriate for falsely cooperative heterogenous ensembles in the taut phase due to the difference in motor properties tending to drive the motor separation to a point where both tethers experience substantial force, so the detachment time scale is not slow compared to the time scale of the spatial motor separation dynamics.  Instead, we
introduce  a `semideterministic' model for
calculating taut statistics under heterogeneous ensembles by dropping the
noise term for the dynamics of the intermotor distance $R$.  This simplification allows for a closed analytical expression for the detachment rate as a function of time as well as the other statistics associated to the taut phase for heterogenous ensembles as a more defensible alternative to the expressions in Subsection~\ref{tauthom} based on the stationary distribution.  

Next we analyze the transient lax phase dynamics, which involve a first passage time calculation for the intermotor separation variable $ R(t) $.  We analyze separately the particular first passage time setup  for truly cooperative ensembles in Subsection~\ref{laxhomnonz},  for falsely cooperative ensembles in Subsection~\ref{laxhetzero}, and for neutrally cooperative ensembles in Subsection~\ref{sec:neutcoop}.  
Section \ref{combolaxtaut}  addresses how to combine the lax and taut phases
to produce statistics over the entirety of state (1,2).  The methods used
here are straightforward, relying on standard conditioning laws over a partition
of two events.   
While our techniques in principle allow the calcluation of the motor-cargo ensemble diffusivity for both homogenous and heterogenous ensembles, the correlation calculations that would be involved for heterogenous ensembles produce such cumbersome expressions as to undercut their utility (Appendix~\ref{app:secmompain}).   We thus satisfy ourselves here with the calculations needed to compute the primary processivity statistics of mean run length, mean run time, and long-run and ensemble velocities for both homogenous and heterogenous ensembles, and present the calculation formulas for diffusivity only for homogenous ensembles.

Finally, we handle   heterogeneous ensembles with nonzero
trap force in Section \ref{laxhetnonz}.  For such ensembles, we augment the
state space to specify whether state $(1,2)$ begins with the attachment of
motor 1 or 2.  Unfortunately, cycles under this model no longer behave independently, and we cannot apply Wald's theorem directly as presented in Subsection~\ref{sub:renewal} to achieve a compact expression for the run length and time statistics.  Still, we can represent at least the first order statistics  in terms of geometrically convergent infinite sums of explicitly computable terms.

\subsection{Preliminary considerations}
\label{sub:tamsetup}
We begin by writing some expressions for the dynamics of the motors in the \TAM\ with the fast cargo fluctuations averaged out, in analogy to the expressions in Section~\ref{sec:oam} for the \OAM\ states, but the expressions now depend on the momentary intermotor displacement $r =x^{(1)}-x^{(2)}$.   The effective velocity for motor $ i $ is obtained by integrating against the quasi-stationary distribution~\eqref{rho2} $\rho_{1,2}$ of the cargo in the \TAM\ state: 
\begin{align}
&\Veltam{i}( r)= \frac{v^{(i)}}{\pdfnorm_{1,2}^{(i)} (r,F_{T})}\int_{\mathbb{R}} g(F^{(i)}
(-y) /F_s^{(i)}) \nonumber\\
&\times \exp\left(-\frac{1}{k_BT}\left[\frac{\kappa^{(i)}(|y|- \ellrest{i})_+^2+\kappa^{(i')}(|y+(-1)^{i+1}r|-
\ellrest{\ip})_+^2}{2}+F_Ty\right]\right) dy,\nonumber
\end{align}
with
normalizing constant\begin{align}
&\pdfnorm_{1,2}^{(i)} (r,F_{T}) \\
& \qquad =\int_{\mathbb{R}}
\exp\left(-\frac{1}{k_BT}\left[\frac{\kappa^{(i)}(|y|- \ellrest{i})_+^2+\kappa^{(i')}(|y+(-1)^{i+1}r|-
\ellrest{\ip})_+^2}{2}+F_Ty\right]\right) dy \nonumber
\end{align}We can characterize the behavior of these velocity functions by combining the
following observations (with $ \kappamin \equiv \min(\kappa^{(1)},\kappa^{(2)}) $):
\begin{itemize}
\item $ F^{(i)} (y) = 0 $ for $ |y| \leq \ellrest{i} $, 
\item the portion of the exponential factor not involving the trap force
$ F_T $ takes its maximal constant value of $0$ over the interval $ [\max(-\ellrest{i},(-1)^{i}
r-\ellrest{\ip}),\min(\ellrest{i},(-1)^i r+\ellrest{\ip})]$ and then falls
off rapidly on either side over a length scale no larger than $ \sqrt{k_B
T/\kappamin} $. 
\item If the interval just described is empty, then the exponential factor is maximized at the unique mechanical equilibrium for the cargo displacement from motor $ i $: $ y = (-1)^{i}\sgn(r)(\ellrest{i}+\kappa^{(\ip)}(|r|-\ellrest{\ip}-\ellrest{i}))/(\kappa^{(1)}+\kappa^{(2)})-F_T/(\kappa^{(1)}+\kappa^{(2)})$, and  falls
off rapidly on either side over a length scale no larger than $ \sqrt{k_B
T/\kappamin} $.
\end{itemize}

We can thereby conclude:
\begin{itemize}
\item For zero trap force $ F_T
=0$, we have that $ \Veltam{i} (r) \approx v^{(i)} $ whenever the motor separation satisfies $
\ellrest{1}+\ellrest{2}-|r| \gg \sqrt{k_B T/\kappamin}
$ because then the stationary distribution of the cargo is primarily associated to both tethers being slack.
\item For nonzero trap forces $ |F_T|>0$, we find that the exponential factor will
be maximized at values of $ y $ corresponding either to mechanical equilibrium of the cargo with just motor $i $, just motor $ \ip $, or both motors.  Which of these cases are realized depends on self-consistency based on the relative displacement of the motors $ r $ and their tether properties.  Thus,
\begin{itemize}
\item the cargo will have its tether to motor $ \ip $ mostly slack and be essentially in a one-motor stationary distribution~\eqref{eq:att1stat} with respect to motor $i$, so $ \Veltam{i} \approx V_i $, the effective velocity~\eqref{gbari} for this motor carrying the cargo as if it were the only motor attached and $ \Veltam{\ip} \approx v^{(\ip)} $, the unencumbered velocity, provided 
$ (-1)^{i+1} \sgn(F_T) r + \ellrest{\ip}-\ellrest{i} -|F_T|/\kappa^{(i)} \gg \sqrt{k_B
T/\kappa^{(i)}} $ and $ \ellrest{1}+\ellrest{2} + |F_T|/\kappa^{(i)}  -|r| \gg \sqrt{k_B T/\kappa^{(i)}}$,  
\item the cargo will in other cases have substantial engagement with both motors in its stationary distribution.  Away from the
maximum, the exponential factor falls off rapidly on a length scale no larger
than $ \sqrt{k_B T/\min(\kappa^{(i)},\kappa^{(\ip)})} $
along the trap force direction and exponentially on a length scale $ k_BT/F_T
$ against the trap force direction.  
\end{itemize}
\end{itemize}


\subsection{Decomposition into Lax Phase and Taut Phase}
\label{sub:tamdecomp}

Here we give precision to the notion of lax and taut phases of the \TAM\ state, based on the key length scales of motor separation described in Subsection~\ref{sub:tamsetup}.  We define the time $ t_{\mathrm{taut}} $ at which the lax phase terminates and the taut phase begins in terms of the motor separation $ R(t) $ reaching a certain range of values.  The separation length depends on whether the motor dynamics induces them to eventually reach configurations where 1) the cargo is on the same side of both, so the motors are essentially sharing the load cooperatively, and 2) the cargo is between the motors, so the motors are eventually experiencing a perverse tug of war despite their ostensibly cooperative nature, with one motor feeling an assisting force and the other motor a hindering force.  We label the motor ensembles that generally approach the first scenario as ``\truecoop'' (\truecoopabb) and those that generally approach the latter scenario as ``\falsecoop'' (\falsecoopabb).  Motor ensembles that have a substantial likelihood to achieve either of these configurations with both motors engaged will be labeled as ``\neutralcoop'' (\neutralcoopabb).  
  \begin{figure}
  \centering
\includegraphics[width=.7\textwidth]{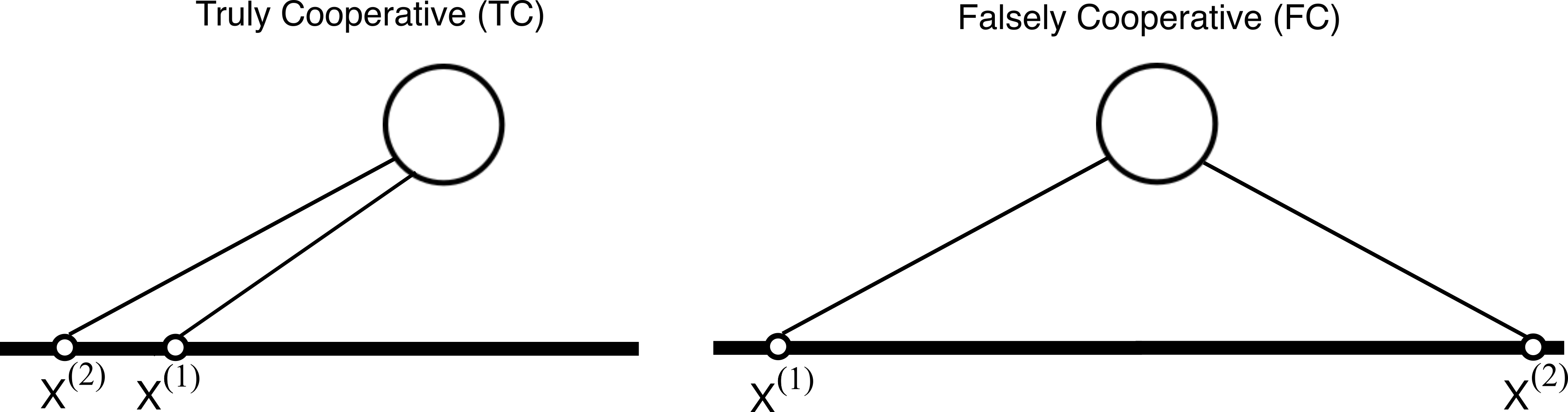} 
\caption{Motor configurations.  Left: Truly cooperative (TC), where both motors (with positions represented by  $X^{(1)}$ and $X^{(1)}$) reach the taut state on the same side of the cargo.  Right: Falsely cooperative (FC), where motors achieve tautness on opposite sides of the cargo.  A neutrally cooperative (NC) ensemble has substantial probability to reach the taut state in either way.}
\end{figure}

With our standing assumption that the motor of index $ 1 $ has an unencumbered velocity at least as large as that of the motor with index $ 2 $ ($v^{(1)} \geq v^{(2)}$), we see that the \truecoop\ scenario applies whenever $ \Vstate{1} < v^{(2)} $.  Note that according to the discussion in Subsection~\ref{sub:tamsetup}, we treat $ \Vstate{1} \approx v^{(1)} $ at zero and assisting trap forces because the hindering force fluctuations are uncommon in the cable model for tether force.  
 For heterogenous motors with $ v^{(1)} >v^{(2)} $, the \truecoop\ scenario should apply at sufficiently large hindering trap forces and the \falsecoop\ scenario apply at sufficiently small hindering, zero, and all assisting trap forces.   At the transitional trap forces we would expect a \neutralcoop\ behavior. 
 The criterion is readily understood by self-consistency of the two scenarios.  We consider first the case of a hindering trap force $ F_T > 0 $.  If motor 1 is slack, it will move faster than motor 2 and eventually have its tether engage.  If the ensuing hindering force can slow motor 1 below the unencumbered velocity of motor 2, then the latter must either share part of the hindering force of the cargo or progress faster than motor 1 until the tether on motor 2 engages.  If on the other hand the hindering force from the cargo does not slow motor 1 below the unencumbered velocity of motor 2, then motor 1 will continue to increase its lead on motor 2, pulling the cargo ahead of motor 2, and eventually pulling the tether between the cargo and motor 2 taut with an assisting force and initiating the tug of war.  The force on both motors 1 and 2 increase beyond what they would experience as single motors carrying a cargo, until they both stall or one detaches.  
For zero or assisting trap forces $ F_T \leq 0 $, the tether on each motor will typically be either slack or pulling in an assisting direction.  In either case, the motors move at their unencumbered velocity and thus inexorably separate until they stretch the tethers into a tug of war. 

Similar considerations applied to the case $ v^{(1)} = v^{(2)} $ suggest homogenous motor ensembles would be \neutralcoop\ at zero and assisting trap forces, and  \truecoop\ at hindering trap forces.  Actually the demarcation would really be determined by a trap force value below which the dynamics of motor separation on the scale of the tether length are diffusion-dominated and above which they are dominated by the difference in the velocity of an unloaded and loaded motor.  We will discuss this criterion in more quantitative detail at the end of this subsection.

Together with the key separation length scales identified in Subsection~\ref{sub:tamsetup}, we define the onset of the taut phase as occurring at the time
\begin{subequations}
\begin{equation}
\ttaut = 
\begin{cases}
\ttauttrue \equiv \argmin_{t \geq \ttam} \{r^{\ast(2)}_{\rightarrow\rightarrow} \leq R(t)\leq r^{\ast(1)}_{\rightarrow\rightarrow}\} & \text{ for \truecoopabb\ case}, \\
\ttautfalse \equiv \argmin_{t \geq \ttam} \{|R(t)|\geq r^{\ast}_{\leftarrow\rightarrow}\} & \text{ for \falsecoopabb\ case; \neutralcoopabb\ case with }  F_T = 0, \\
\min(\ttauttrue,\ttautfalse) & \text{ for \neutralcoopabb\ case with }  F_T \neq 0 ,
\end{cases} \label{eq:tauttimes}
\end{equation}
after the time $ \ttam $ of the most recent entry into the \TAM\ state, and the length scales are defined: 
\begin{align}
r^{\ast(i)}_{\rightarrow\rightarrow} &\equiv \ellrest{1}-\ellrest{2}+ (-1)^{i+1} \left(|F_T|/\kappa^{(i)}
+\sqrt{k_B T/\kappa^{(i)}}\right), \label{rstaronesided} \\
 r_{\leftarrow\rightarrow} &\equiv \ellrest{1}+\ellrest{2} +|F_T|/\kappa^{(3-\sgn(F_T))/2}- \sqrt{k_B T/\kappamin}.
  \label{rstartwosided}
 \end{align}
\end{subequations}
Note that $ \ttaut = \ttam $ if the the \TAM\ state is entered with both motors already in a position to engage with the cargo, meaning no lax phase occurs.

In what follows, we will be examining the time scale over which motors remain in the lax state, whether they proceed to the taut state, and how they behave when they reach the taut state.  For the last purpose,  we consider $\widetilde R(t) =  R(t+t_{\mathrm{taut}}) $, time-shifted relative to the moment
$ t_{\mathrm{taut}} $ when the taut state is first reached, and define  $\widetilde \CM(t)$ similarly. Writing $U_+(r) = (\Veltam{1}(r)+\Veltam{2}(r))/2$ 
and $U_-(r) = \Veltam{1}(r)- \Veltam{2}(r)$, and replacing the drift terms in Eq.~\eqref{motors} by their cargo-averaged versions as in~\citet{klobusicky2020cooperative}, we then
obtain the system 
\begin{subequations}
\begin{align}
d\widetilde \CM &=U_+(\widetilde R(t)) dt +\frac{ \sqrt{2
D^{(1)}}}2 dW_{1}(t)+\frac{ \sqrt{2
D^{(2)}}}2 dW_{2}(t), \label{eq:M}\\
d\widetilde R &= U_-(\widetilde R(t))dt+ \sqrt{2
D^{(1)}} dW_{1}(t)- \sqrt{2
D^{(2)}} dW_{2}(t). \label{eq:R}
\end{align}
\end{subequations}

Before proceeding, we provide some orientation to relate the ensuing theoretical discussion to the particular ensembles of kinesin-1 and kinesin-2, with parameter values reported in Table~\ref{paramtable}, to which they will be applied in Section~\ref{sec:numeric}.
We recall that homogenous ensembles are generally neutrally cooperative at zero and assisting trap forces, and heterogenous ensembles are generally falsely cooperative at zero and assisting trap forces.  Thus, we are just left to characterize the cooperativity of the ensembles at the hindering trap force of $ F_T = 2 \si{\pico\newton} $, specifically, whether this is enough of a trap force to induce truly cooperative behavior. 
For the kinesin-1/kinesin-2 heterogenous ensemble, the faster kinesin-1 motor when loaded by this trap force has velocity $V_1 =564 \si{\nano\metre/\sec}$ relative to the unloaded velocity $ v^{(2)}= 500 \si{\nano\metre/\sec}$ of the other motor, and so this trap force is still small enough for falsely cooperative behavior to persist.  For homogenous ensembles, the question is whether the dynamics in the lax phase is dominated by the relative drift  or diffusion of the loaded and unloaded motors. For all ensembles considered, the relative separation would have to increase by on average  $r^{\ast(2)}_{\rightarrow
\rightarrow} - \mu_{R^a}= 76 \si{\nano\metre} $ to reach the taut phase.  For the kinesin-1/kinesin-1 ensemble,   the displacement between motors changes with velocity $V_1- v^{(2)} = 225 \si{\nano\metre/\sec}$, while for kinesin-2/kinesin-2 ensembles, the relative velocity between motors is  $V_1- v^{(2)} = 143 \si{\nano\metre/\sec}$,  The diffusivity of the relative separation is $ \sqrt{D_1+D_2} = 57 \si{\nano\metre^2/\sec} $.  We thus find the time scale to drift to the taut phase $(r^{\ast(2)}_{\rightarrow
\rightarrow} - \mu_{R^a})/(V_1-v^{(2)}) $ is much faster (less than a second) relative to the time scale  $(r^{\ast(2)}_{\rightarrow
\rightarrow} - \mu_{R^a})^2/(2(D_1+D_2))= 56 \si{\sec}  $ for diffusion to cause the relative separation to reach the taut phase.  The kinesin-1/kinesin-1 and kinesin-2/kinesin-2 ensembles are thus both truly cooperative under the hindering force of  $ F_T = 2 \si{\pico\newton} $. 

\begin{table}
\begin{tabular}{|c|c||c|c|c|}\hline
 $F_T  (\si{\pico\newton})$ & Ensemble & $\plax$ & $\E[\dTattblax]$ (\si{\second})& $\E[\dTattbtaut]$ (\si{\second}) \\\hline
2 & K1/K1 & .493 & .22& .50 \\\hline
0 & K1/K1 & .873& .56 & .23\\\hline
 -2 & K1/K1 & .993 & .12& - \\\hline
 2 & K2/K2 & .976 & .12& .16\\\hline
 0 & K2/K2 &.989 &.59 & .44\\\hline
 -2 & K2/K2 & .997 & .16 & - \\\hline
 0 & K1/K2 & .561 & .35& .20 \\\hline
\end{tabular}
\caption{Probabilities $ \plax $ for various ensembles (K1=kinesin-1, K2=kinesin-2) to detach from the lax phase (and thus not enter the taut phase), together with expected times spent in the lax and taut phase (conditional upon being reached), respectively in the last two columns.  The reported values are based on theoretical expressions developed for various cases in Subsections~\ref{tauthom}-\ref{sec:neutcoop}, starting with no initial separation $r_0 = 0 $ between the motors. Durations for homogeneous ensembles with $F_T = \SI{-2}{\pico\newton}$ are omitted because the taut phase is rarely reached and is thus functionally irrelevant.  The \TAM\ state requires further decomposition for heterogenous ensembles with trap force, so those cases are deferred to Table~\ref{phasetable}.   \label{tab:laxtauttimes}}
\end{table}

To see the role which lax and taut phases play in practice,  Table~\ref{tab:laxtauttimes} reports the expected times spent in the lax and taut phases, based on the theoretical expressions to be developed, for certain ensembles of kinesin-1 and kinesin-2 .   Heterogenous ensembles under trap force have some subtleties and a similar analysis is deferred until Subsection~\ref{laxhetzero}.  For the ensembles presented, we see first of all that all of them have a significant probability to never reach the taut phase, and even if they do, a substantial portion of the time spent in the \TAM\ state is within the lax phase.  This underscores the importance of going beyond the assumption from~\citet{klobusicky2020cooperative} that the spatial dynamics within a state can be approximated by the stationary distribution when we use the cable model~\eqref{eq:cable} with slack for the tether.  We see moreover that many ensembles, particularly homogenous ensembles of kinesin-2, are very unlikely to reach the taut phase at all.  Homogenous ensembles under zero trap force are less likely to reach the taut phase than under load because they fall in the neutrally cooperative category which relies on a slower diffusion process to reach a taut phase.  One caveat is that these calculations are done with an initial motor separation equal to the average (0); random fluctuations would generally start the motor separation ``closer'' to the taut phase so the probability to reach it would be somewhat higher.

Under a hindering trap force of $ \SI{2}{\pico\newton} $, kinesin-1/kinesin-1 ensembles have roughly even odds to reach the taut phase while kinesin-2/kinesin-2 ensembles are very likely to detach from the lax phase.  A key reason for this is the relatively large effective detachment rate for the kinesin-2/kinesin-2 ensembles $d^{(2)}_{0}+ \detone{1} =7.02\mathrm{s}^{-1}$, whose inverse is not much larger than the average time spent in the lax phase.  On the other hand, the effective detachment rate for kinesin-1/kinesin-1 ensembles under the same hindering load is  $d^{(2)}_{0}+ \detone{1} =2.01\mathrm{s}^{-1}$, whose inverse is sufficiently long relative to the time scale $(r^{\ast(2)}_{\rightarrow
\rightarrow} - \mu_{R^a})/(V_1-v^{(2)}) =0.34 \mathrm{s}$ to permit a substantial probability for the taut phase to be reached before detachment from the \TAM\ state.


\subsection{Taut statistics for truly and neutrally cooperative ensembles}  
\label{tauthom}
The simplest scenario is when a homogenous ensemble  reaches the taut state.  The averaging  theory outlined in \cite{klobusicky2020cooperative} appears reasonable to apply here, as the separation dynamics will tend to relax the build-up of tension in the tethers as in passive physical systems.   Thus the motors are not likely to be loaded beyond what they carry in an \OAM\ state, so the detachment rates will not be unduly large.  The same considerations apply to truly or neutrally cooperative heterogenous ensembles.

We therefore summarize the results from the averaging theory for the \TAM\ developed in~\citet{klobusicky2020cooperative}.
The effective drift is taken as: 
\begin{equation} \label{eq:barvtwo}
\Vstate{1,2} = \int_\mathbb R  U_+(r) p_{\widetilde R}(r)dr.
\end{equation}
where \begin{equation}\label{statr}
p_{\widetilde R}(r) = C_{\widetilde R} \exp \left [\frac{1}{ D^{(1)} +D^{(2)}}\int_0^r
U_-(r')dr'\right
], \quad -\infty<r<\infty,
\end{equation}
with normalizing constant $ C_{\widetilde R} $, is the stationary probability density
for $\widetilde R$.  
The effective diffusivity $\Dstate{1,2}$
 is found by using homogenization methods outlined in \cite{pavliotis2008multiscale} applied to the system (\ref{eq:M})-(\ref{eq:R}).  The computation is somewhat involved, with the main complicating factor being the correlated diffusion terms for  $\widetilde M$ and  $\widetilde R$. It is shown in~\cite{klobusicky2020cooperative} (in nondimensional coordinates) that 
is given by:
\begin{align}
& \Dstate{1,2} = \int_{-\infty}^\infty \frac{1}{ D^{(1)} +D^{(2)}}\left(\int_{-\infty}^r
( U_+(r^{\prime}) -\barvtwo)p_{\widetilde R}(r^{\prime})d r^{\prime}\right)^2\frac{1}{p_{\widetilde
R}(r)}\,
dr \nonumber\\&+ 
\left(\frac{D^{(1)}-D^{(2)}}{D^{(1)}+D^{(2)}}\right)\int_\mathbb
R ( U_+(r) -\barvtwo)p_{\widetilde R}(r)\cdot rdr+\frac{D^{(1)}+D^{(2)}}4.
\nonumber
\end{align}

The averaged attachment rate of the $i$th motor over cargo
fluctuations is 
\begin{align}
\dettwoz{i}(r) &= \frac 1{\pdfnorm_{1,2}^{(i)} (r,F_{T})}\int_{\mathbb{R}}
d^{(i)}
(F^{(i)} (-y)) \\
&\times \exp\left(-\frac{1}{k_BT}\left[\frac{\kappa^{(i)}(|y|- \ellrest{i})_+^2+\kappa^{(i')}(|y+(-1)^{i+1}r|-
\ellrest{\ip})_+^2}{2}+F_Ty\right]\right) dy,\nonumber
\end{align}
 and then further averaging over the distribution of intermotor distances
gives the effective detachment rate
\begin{equation}\label{deffmotori}
\dettwo{i}=\int_\mathbb{R} \dettwoz{i}(r) 
 p_{\widetilde R}(r)dr.
\end{equation}
The total detachment rate is given as the sum 
\begin{equation}\label{deff}
\dettwoboth = \dettwo{1}+\dettwo{2}.
\end{equation}

The distribution of the intermotor separation just before
first detachment
$ \Rdet_ \mathrm{taut} = \widetilde R(\dettime^-)  $ will be reweighted via Bayes' rule by the detachment rate, yielding
the probability
density\begin{align}\label{rdetach}
p_{\Rdet_ \mathrm{taut}} (r) 
= \frac{(\dettwoz{1}(r)+\dettwoz{2}(r))p_{\widetilde R}(r)}{\dettwoboth}.
\end{align} 
The distribution of $\Rdet$
conditioned on $ \detmot  = i$, where $ \detmot $ is the
index of the motor which first detaches
from the two-motor-attached state (at the random time $ \dettime $),
is given by 
\begin{align} \label{rdetj}
p_{\Rdet_{i, \mathrm{taut}}} (r) 
= \frac{\dettwoz{i}(r)p_R(r)}{\dettwo{i}}.
\end{align}

Moments related to $\Rdet_{i, \mathrm{taut}}$ and $\Xi _{1,2}(\Rdet_{i,
\mathrm{taut}})$ can  then be directly numerically found through (\ref{rdetj}).
From
 standard theory of continuous-time Markov chains,
we find
that the probability that motor $i$ is the motor which detaches in the taut phase is\begin{equation}
\pdettaut{i} = \frac{\dettwo{i}
}{\dettwoboth}. \label{eq:pdettauti}
\end{equation}
In Appendix~\ref{app:nolax}, we compare the simulated distribution of motor separation upon detachment with that in Eq.~\eqref{rdetach}implied by the stationary distribution, and show this theoretical expression is reasonably well suited to the distribution conditioned upon detaching from the taut state, but not to the unconditional distribution. 

  The distribution
of time $ \dTattb^{\mathrm{taut}} $  spent in the taut phase is exponentially
distributed as
\begin{equation}
 \dTattb^{\mathrm{taut}} \sim \Exprand ((\dettwoboth )^{-1}),
\end{equation}
with mean 
\begin{equation}
\E[\dTattb^{\mathrm{taut}}] = \frac 1 {\dettwoboth}.
\end{equation}

The  progress of the cargo tracking variable can be calculated in terms of the effective velocity in the \TAM\ state by conditioning on the time spent in the taut phase:
\begin{align}
 \dCMattb^{\mathrm{taut}}|\dTattb^{\mathrm{taut}} &\sim N(\barvtwo \dTattb^{\mathrm{taut}},2D_{1,2}\dTattb^{\mathrm{taut}}). \label{m12cond}
 \end{align}  


Finally, the first and second order statistics  of the times and cargo tracking variable displacements in the taut phase of the \TAM\ for truly and neutrally cooperative ensembles are
\begin{align}
\E[\dTattb^{\mathrm{taut}}] = \frac 1 {\dettwoboth},     \quad
\E[\dCMattb^{\mathrm{taut}}] = \frac{  \barvtwo }{\dettwoboth},\quad \Var(\dTattb^{\mathrm{taut}}) =\frac1 {(\dettwoboth)^2},  \\
\Var(\dMattb^{\mathrm{taut}})=  \frac{2D_{1,2}}{\dettwoboth} +\left(\frac{\barvtwo}{\dettwoboth}\right)^2, \quad
\Cov(\dTattb^{\mathrm{taut}}, \dMattb^{\mathrm{taut}})= \frac{\barvtwo}{(\dettwoboth)^2}.
\end{align}

\subsection{Taut statistics for falsely cooperative heterogenous ensembles} \label{tauthet}
Representing the taut phase for falsely cooperative heterogenous ensembles with their stationary distribution in the \TAM\ state is questionable because the taut phase involves a tug of war that progresses in intensity until the leading motor feels sufficient force to slow it to the velocity of the lagging motor.  This implies a sustained buildup in the hindering force on the leading motor and an assisting force on the lagging motor which can accelerate the detachment process to the point its rate is comparable to the time scale on which the separation dynamics reach statistical equilibrium.

To give some numerical characterization for this concern, we consider a heterogeneous
ensemble of kinesin-1 and kinesin-2 with zero trap force.  In this ensemble, 
$U_-(\tilde{R}(t)):= u_- =    290$ nm/s 
remain approximately constant as  $\tilde{R}(t)$ increases until it approaches the value
 $ r^{\ast}_{\leftarrow\rightarrow}= 155.95 \hbox{ nm}$,
the displacement value~\eqref{rstartwosided} at which the cargo becomes significantly engaged with both motors, with the tethers pulling in opposite directions. 
Making a linear approximation of $ U_-(r) $ between $ r^{\ast}_{\leftarrow\rightarrow}$ and the stalling value $ U_-^{-1}(0)$, we have the approximate Langevin dynamics for the motor separation:
\begin{align}
d\tilde{R} (t) &= \frac{u_{-}(U_-^{-1}(0)^{}- \tilde{R})}{U_-^{-1}(0)^{}- r^*_{\leftarrow\rightarrow}}dt+\sqrt{2D^{(1)}+2D^{(2)}} dW (t):=
(-\alpha \tilde{R}+\beta )dt+\theta dW (t), \label{oder}\\  \tilde{R}(0) &= r^*_{\leftarrow\rightarrow}, \quad \alpha = 14 \hbox{
s$^{-1}$},
\quad \beta = 2500 \hbox{
nm/s}, \quad \theta^2 = 2D^{(1)}+2D^{(2)} = 6500 \hbox{ nm$^2$/s}.  \quad \nonumber
\end{align} 
The  difference between the equilibrium point $U_-^{-1}(0) = 181$ nm of (\ref{oder})
and $r^*_{\leftarrow\rightarrow}$  is given by $u_-/\alpha \approx 25$ nm. The time scale for
reaching equilibrium is given by $1/\alpha = .07$ s. The detachment rate is  $\SI{2.0}{\sec^{-1}}$ at motor displacement $ r = r^*_{\leftarrow\rightarrow} $ and $ \SI{8.8}{\sec^{-1}}$ at motor displacement $r= U_-^{-1}(0)$.  Thus, we see that the detachment occurs on a time scale comparable to the relaxation of the separation dynamics to their equilibrium, even once the taut phase is reached.  The presence of a trap force would further increase the detachment rates.  These observations would seem to render unviable the application of the stationary distribution for the taut phase of heterogenous ensembles in general, as we did for truly and neutrally cooperative ensembles in Subsection~\ref{tauthom}.

This would suggest tracking the time-dependent probability distribution of $ \widetilde R (t) $ within the taut phase, but then the detachment process would become a Cox process with random rate $ \dettwoz{i}(\widetilde
R(t)) $, for which computing the statistics of the detachment time would become analytically unwieldy.  We therefore propose instead a 
a semideterministic approximation $\widehat
R(t)$ for $ \widetilde R (t)$, obtained by  removing
the noise term in (\ref{eq:R}) to produce   the  autonomous differential
equation \begin{equation} \label{semidet}
\frac{d\Rsemidet}{dt} = U_-(\Rsemidet), \quad \Rsemidet (0) = r^*_{\leftarrow\rightarrow}.
\end{equation} 
This makes the detachment process an inhomogeneous Poisson process, for which we can readily compute
the distribution time for detachment  as
the derivative of the survival function:\begin{equation}
p_{\dTattb^{\mathrm{taut}} }(t) =\dettwoz{i}( \Rsemidet (t))\exp\left(-\int_0^t\dettwoz{i}(\Rsemidet (s))ds\right). \label{taut12semidet}
\end{equation}
We note that the expected time with this approximation, $\E[\dTattb^{\mathrm{taut}}] = .20$ s,  is consistent with the 
approximate time scale for detachment $0.11$ s in the previous discussion.   
To get a sense of the semi-deterministic approximation, we note the diffusive component in Eq.~\eqref{oder} would induce fluctuations of size $ \sim \theta \E[\dTattb^{\mathrm{taut}}]  \sim 4 $ nm  for the kinesin-1/kinesin-2 ensemble at zero trap force, which is about a fifth of the  range of separations between mechanical equilibrium and the onset of taut dynamics. Thus we cannot claim the noise fluctuations in $ \widetilde R $ are truly negligible in the taut phase, but neither are they dominant over the deterministic dynamics.   While the semideterministic approximation of the taut phase statistics for falsely cooperative motors is thus rough, we would generally expect falsely cooperative ensembles to spend a  shorter time in the taut phase than the lax phase, so this approximation should not have a substantial impact on the overall transport statistics.   

Proceeding now to develop the statistics associated to spatial dynamics within the taut phase under the semideterministic approximation~\eqref{semidet}, a solution for the SDE of  $\widetilde \CM(t)$ can  be written explicitly as 

\begin{equation}
\widetilde \CM(t) -\widetilde \CM(0)= \int_0^t U_+(\Rsemidet (s)) ds +\frac{ \sqrt{2
D^{(1)}}}2 W_{1}(t)+\frac{ \sqrt{2
D^{(2)}}}2 W_{2}(t).\label{mifrnoise}
\end{equation}
 The expected displacement
is then\begin{equation}
\E[\dCMattb^{\mathrm{taut}}] =\int_0^\infty \int_0^t U_+(\Rsemidet (s))
ds \cdot p_{\dTattb^{\mathrm{taut}}}(t)dt.
\end{equation}



Since we are not assuming that intermotor distance equilibrates quickly,
even in the taut phase,   the distribution of $\tilde R$ at detachment is
dependent on detachment time, or\begin{equation}
\Rdet_{\mathrm{taut}} = \Rsemidet (\dTattb^{\mathrm{taut}}). \label{taut12inrhat}
\end{equation}
It does not however depend on the identity of which motor attached into the \TAM\ state as it does not (with high probability) affect how the taut phase for falsely cooperative ensembles is entered.  
The probability that a  motor which detaches in the taut phase is motor $i$
 is 
\begin{equation}
\pdettaut{i}= \int_{0}^\infty \frac{\dettwoz{i}(\Rsemidet (t))}{\dettwoz{1}(\Rsemidet (t))+\dettwoz{2}(\Rsemidet (t))} p_{\dTattb^{\mathrm{taut}}}(t) dt.
\end{equation}

To find  the probability density of $\dTattb^{i,\mathrm{taut}}$, the  detachment time $\dTattb^{\mathrm{taut}}$
conditioned on the identity of the detaching motor $\detmot = i$, we use
Bayes' rule as in the derivation of Eq.~\eqref{rdetj} in~\citet{klobusicky2020cooperative}, obtaining
\begin{equation}
p_{\dTattb^{i,\mathrm{taut}}}(t) = \frac{p_{\dTattb^{\mathrm{taut}}}(t)}{\pdettaut{i}}\cdot
\frac{\dettwoz{i}(\Rsemidet (t))}{\dettwoz{1}(\Rsemidet (t))+\dettwoz{2}(\Rsemidet (t))}.
\end{equation}
Moments of the separation of motors  at detachment from the taut state, conditioned on motor $i $ being the detaching motor, denoted $ \Rdet_{i, \mathrm{taut}}$,   are computed as
\begin{align}
 \E[(\Rdet_{i, \mathrm{taut}})^m] &= \int_0^\infty p_{\dTattb^{i,\mathrm{taut}}}(t)(\Rsemidet (t))^m dt.
\end{align}


\newpage

\subsection{Lax statistics for \truecoop\ ensembles}
 \label{laxhomnonz}
We recall from the discussion in Subsection~\ref{sub:tamdecomp} that truly cooperative ensembles only occur with sufficiently strong hindering trap forces $ F_T > 0 $.
In calculations that follow, we set $ j $ to be the index of the most recently attached motor, which will typically be in a slack state while the other motor with index $ \jp = 3-j $ is already engaged with the cargo.  
Thus, as discussed in Subsection~\ref{sub:tamsetup}, motor $\jp$ acts as if it were the only attached motor, with an effective
velocity of $\Vstate{\jp}$, while motor $i$ is essentially unencumbered, moving
at its maximum velocity of $v^{(j)}$:  
\begin{subequations}
\begin{align}
\difd X^{(j)} (t) = v^{(j)} \, \difd t + \sqrt{2 D_j} \, \difd W^{(j)} (t), \\
\difd X^{(\jp)} (t) = \Vstate{\jp} \, \difd t + \sqrt{2 D_{\jp}} \, \difd W^{(\jp)} (t). \label{eq:leadmot}
\end{align}
\end{subequations}
 For \truecoop\ ensembles, we have  $ v^{(j)} > \Vstate{\jp} $ so the motor separation $ |R(t)| $ will tend to decrease until both motors are engaged.
The associated
SDE
is  given by 
\begin{equation*}
\difd R(t) = u_- \, \difd t + (-1)^{j+1} ( \sqrt{2 D_j} \, \difd W^{(j)} (t) -\sqrt{2 D_{\jp}} \, \difd W^{(\jp)} (t)),
\end{equation*}
or equivalently via a different standard Brownian motion $ W(t)\equiv \left(\frac{\sqrt{2D_1}}{\theta} W^{(1)} (t) - \frac{\sqrt{2D_{2}}}{\theta} W^{(2)} (t)\right) $:
\begin{equation}
d R (t)= u_- \, \difd t + \theta \, \difd W(t), \quad  R(\ttam) = r_0\sim
\Ratt{j}, \label{eq:Rcc}
\end{equation} 
with constant coefficients:
\begin{equation}
u_- \equiv (-1)^{j+1}(v^{(j)}- \Vstate{\jp}), \qquad \theta \equiv \sqrt{2D_1+2D_2}. \label{eq:relsepvel}
\end{equation}
The lax phase terminates at the time $ \ttauttrue $ defined in Eq.~\eqref{eq:tauttimes}.   In practice we approximate this by a one-sided first passage time:
\begin{equation*}
\ttauttrue \approx \argmin_{t \geq \ttam} \{(-1)^{j+1}(R(t)-r^{\ast(\jp)}_{\rightarrow\rightarrow})\geq 0\}, 
\end{equation*}
which is consistent with our above assumptions based on neglecting the rather unlikely possibility that the most recently attached motor lands in a position to make it immediately relieve the previously attached motor from the cargo load.  (That is we assume with high probability $ (-1)^{j+1} (R(\ttam) - r^{\ast(j)}_{\rightarrow\rightarrow}) \leq 0 $; for our kinesin-1 and kinesin-2 models this probability is $0.95$). 
In calculations that follow, we consider all statistics to be conditioned
under a fixed value of $r_0$ and index $j $ of the most recently attached motor.  For ease of exposition, we mostly suppress
this conditioning  in our notation.  

We denote $\Ttaut = \ttauttrue  -\ttam$ as the time until
reaching the taut phase under the absence of detachment.  
We use the well-known fact that the probability density of the one-sided hitting time for a Brownian motion with drift 
is distributed 
as an inverse Gaussian \cite{chhikara1988inverse} to obtain the distribution
for the time $\Ttaut $:
\begin{align}
p_{\Ttaut} \sim \mathrm{IG}(|r^{\ast(\jp)}_{\rightarrow\rightarrow}-r_0|/u_-), ((r^{\ast(\jp)}_{\rightarrow\rightarrow}-r_0)/\theta)^2), \text{ for } (-1)^{j+1} (r_0 - r^{\ast(\jp)}_{\rightarrow\rightarrow}) < 0.
 \label{fpostime1}
\end{align}
where the two parameter family of inverse Gaussian densities $p(t)\sim \mathrm{IG}(\mu,
\lambda)$ is defined by\begin{align} 
p(t;\mu,\lambda) \sim \sqrt{\frac{\lambda}{2\pi
t^3}}\exp\left[- \frac{\lambda(t-\mu)^2}{2\mu^2 t}\right], \quad \mu, \lambda>0. \label{eq:ig}
\end{align}



Now let us consider the time $\Tdlax$ until detachment occurs if both motors
always remained in lax phase.  Detachment rates for the engaged motor can then  be found by averaging the
force detachment function over a single attached motor, which gives a rate
of $\detone{\jp}$.  For motor $j$, we simply use the unencumbered force
detachment rate of $d_{0}^{(j)}$.   Detachment times will then follow  
\begin{equation}
\Tdlax \sim \mathrm{Exp}(d_0), \quad d_0 = d_{0}^{(j)}+\detone{\jp}. \label{fpostime2}
\end{equation}
When detachment occurs in the lax phase, the probability that motor $ j$ (the most recently attached motor)  detached is
 then
\begin{equation}
\pdetlax{i} = \frac{d_{0}^{(j)}}{d_{0}^{(j)}+\detone{\jp}}. \label{eq:detprobs}
\end{equation}

The probability that detachment occurs first can  
be found  from the law of total probability, yielding 
\begin{equation}
\plax = \mathbb P(\Tdlax< \Ttaut) = \int_0^\infty(1-e^{-d_0t}) p_{\Ttaut}(t)
dt
=1-\int_0^{\infty} e^{-d_0 t} p_{\Ttaut}(t)\, \difd t. \label{plax}
\end{equation} 
The actual duration spent in the lax phase will be governed by either detachment or cargo engagement by both motors, so the dwell time in the lax phase is given by 
\begin{equation}
\dTattblax=\min(\Ttaut, \Tdlax).
\end{equation}
Its probability density is a mixture of the probability densities based on whether or not the motors reach the taut phase before detaching
or not:
\begin{align}
p_{ \dTattblaxd}(t)&:=p_{\dTattb^{\mathrm{lax}}}(t|\Ttaut>\Tdlax)
= \frac{d_0e^{-d_0t}\mathbb
P(\Ttaut>t)}\plax. \label{detfirst} \\
p_{ \dTattblaxt}(t)&:= p_{ \dTattb^{\mathrm{lax}}}(t|\Ttaut<\Tdlax)
= \frac{p_{\Ttaut}(t)e^{-d_0t}}{1-\plax}, \label{tautfirst}
\end{align}
We note that the survival
function
of the inverse Gaussian distribution in Eq.~\eqref{eq:ig} can be computed explicitly from
the CDF $\Phi(x)$ of a standard Gaussian, so:
\begin{align*}
\mathbb
P( \Ttaut
>  t) &= 1-\Phi(\sqrt{\lambda/t}(t/\mu-1)) - e^{2\lambda/\mu}\Phi(-\sqrt{\lambda/t}(t/\mu+1)),
\\ \mu &= |r^*_{\leftarrow\rightarrow}-r_0|/u_-, \quad \lambda = ((r^*_{\leftarrow\rightarrow}-r_0)/\theta)^2.
\end{align*}
The probability distribution for the intermotor displacement upon detachment when it occurs in the lax phase is derived in Appendix~\ref{app:laxdetsep} to be: 
\begin{align}
 p_{\Rdetlaxatt{j}}(r) &= \int_{0}^{\infty}\frac{d_0 \expe^{-d_0 t}}{\sqrt{2\pi\theta^2 t} \mathbb P(\Ttaut>t)} 
 \left[\expe^{-\frac{(r-r_0-u_- t)^2}{2\theta^2 t}}-\expe^{\frac{2(r^{\ast(\jp)}_{\rightarrow \rightarrow}-r_0)u_-}{\theta^2}}
\expe^{-\frac{(2r^{\ast(\jp)}_{\rightarrow \rightarrow}-r-r_0+u_- t)^2}{2\theta^2 t}}\right]dt \nonumber \\
& \text{ for }  (-1)^{j+1} (r- r^{\ast(\jp)}_{\rightarrow\rightarrow}),  (-1)^{j+1} (r_0- r^{\ast(\jp)}_{\rightarrow\rightarrow})< 0.\label{eq:rdetlax}
\end{align}
As the detachment rate is independent of the motor separation $ R(t) $ in the lax phase,  we generally have that the motor separation at detachment from the lax phase of the \TAM\ state  is independent of the identity of the motor which detached.  However, we have indicated explicitly that it does depend on the identity $j$ of the motor which most recently attached, in particular through the definition of $ u_- $ in Eq.~\eqref{eq:relsepvel}, since this motor will tend to be unloaded  while the other motor is carrying the trap force load.

 The mean and variance of $ \dTattblax $ is obtained by using the probability~\eqref{plax} for how the lax phase terminates to weight the means and variances of $ \dTattblaxd $ and $ \dTattblaxt $, which are in turn obtained by numerical integration against their probability densities in Eq.~\eqref{detfirst} and~\eqref{tautfirst}, respectively. 
During the lax period, we compute the change in the cargo tracking variable from Eq.~\eqref{eq:M} using the constant speeds of each motor 
because the tether to motor $ j $ is slack.  
 From Eq.~\eqref{eq:leadmot}, the cargo tracking variable velocity takes the constant value  $u_+\equiv \frac{1}{2} \left(\Vstate{\jp} + v^{(j)}\right) $  while motor $ \jp $ already attached is solely carrying the cargo load.  While the noisy component of the cargo tracking variable increment is correlated with the increment in the motor separation and thus with the random time $ \dTattblax$, for the purpose of computing the mean increment we can appeal to the optional stopping theorem applied to the martingale $\CM (t) - u_+ t $ with respect to the filtration generated by both attached motor positions to obtain:
 \begin{align}
 \label{laxstat}
 \E[\dCMattblax] &= 
 u_+ \E[\dTattblax].
 \end{align}
 For heterogenous ensembles, it is difficult to say much more about the cargo tracking variable increment due to the correlations in the noisy components of $ \CM (t) $ and $ R (t) $, as discussed in Appendix~\ref{app:secmompain}.  But as we are only computing means for heterogenous ensembles, this is adequate.   For homogenous ensembles (with a common motor diffusivity $D$), the noise in $ \CM(t) $ and $ R(t) $ is independent, and these complications don't arise.  We can thus write out second moment properties:
 \begin{align}
 \Var( \dMattblax) &=  
D  \E[\dTattblax] +u_+^2 \Var(\dTattblax),  \\
  \E[\dTattblax\dMattblax] &=
 u_+ \E[(\dTattblax)^2].
 \end{align}
 These equations, as well as the relationship~\eqref{laxstat}, also hold when we condition on how the lax phase terminates.
  
 As noted at the beginning of this subsection, these expressions are all implicitly conditional on how the \TAM\ state is entered:  the index $ j $ of the most recently attached motor and $ r_0 $, the initial separation of the motors.  The averaging over the randomness of the initial separation is straightforward and described in Subsection~\ref{combolaxtaut}.  For homogenous ensembles, the index of the most recently attached motor is actually irrelevant. For heterogenous ensembles, we need to expand our Markov chain structure to condition on which motor just attached, and we explain in Subsection~\ref{laxhetnonz} how this is handled.

\subsection{Lax statistics for \falsecoop\  ensembles}
\label{laxhetzero}

  \begin{figure}
  \centering
\includegraphics[width=.7\textwidth]{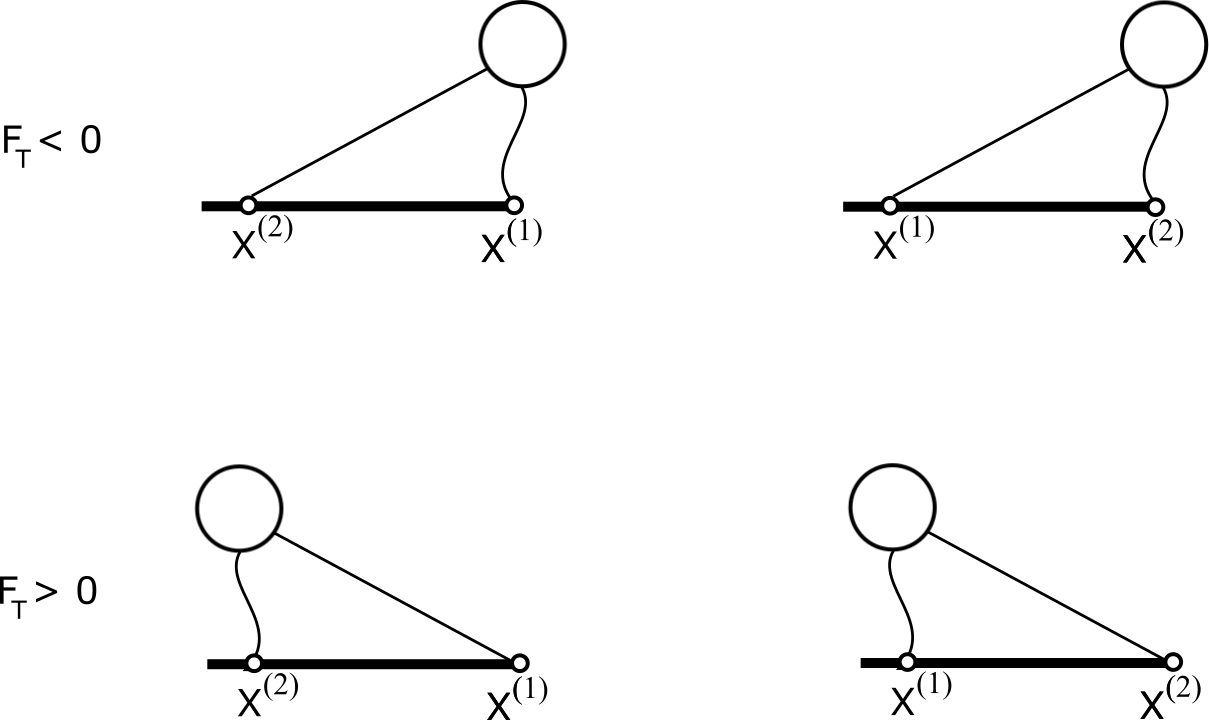} 
\caption{Four types of initial positions in state $(1,2)$ \label{fourmotors}.
 The motor connected to the loose tether is the one most recently attached.}
\end{figure}
Recalling that only heterogenous ensembles can be \falsecoop\, we can carry over the technical analysis for \truecoop\ ensembles in Subsection~\ref{laxhomnonz} with slight modifications when either:
\begin{itemize}
\item no trap force (or a negligible trap force) is applied $ F_T \approx 0 $, or
\item the \TAM\ is entered with the faster motor ahead of the slower motor, and one of the motors slack.   This would be rather likely for (substantial) hindering trap forces $ F_T > 0 $ if the slower motor most recently attached ($j=2 $ in the notation of Subsection~\ref{laxhomnonz}), and for (substantial) assisting trap forces $ F_T< 0 $ if the faster motor most recently attached ($j=1$); see Figure~\ref{fourmotors}.
\end{itemize}
The reasoning is as follows.  Based on the discussion in Subsection~\ref{sub:tamsetup},
the relative separation velocity formula~\eqref{eq:relsepvel} and the cargo tracking variable velocity 
resolves approximately to the following  constant values under the stated conditions:
\begin{itemize}
\item $ u_- = v^{(1)}-v^{(2)}$ and $ u_+ = \frac{1}{2} \left(v^{(1)}+v^{(2)}\right) $ for $ F_T=0$ and $ -r^{\ast}_{\leftarrow\rightarrow} < R(t) < r^{\ast}_{\leftarrow\rightarrow} $, 
\item $ u_-=\Vstate{1}-v^{(2)} $ and $ u_+ = \frac{1}{2} \left(\Vstate{1}+v^{(2)}\right) $ for $ F_T > 0$ and $  r^{\ast(1)}_{\rightarrow\rightarrow} < R(t) < r^{\ast}_{\leftarrow \rightarrow} $,
\item $ u_-=v^{(1)}-\Vstate{2}$ and $ u_+ = \frac{1}{2} \left(v^{(1)} + \Vstate{2}\right) $ for $ F_T<0 $ and $ -r^{\ast(2)}_{\rightarrow\rightarrow} < R(t) < r^{\ast}_{\leftarrow \rightarrow} $.
\end{itemize}   
 The onset of the taut phase could be approximated as
$ 
\ttautfalse \approx \argmin_{t \geq \ttam} \{R(t)\geq r^{\ast}_{\leftarrow\rightarrow}\} $; the likelihood of a fluctuation driving $ R(t) < - r^{\ast}_{\leftarrow\rightarrow} $ with the slower motor leading a tug of war with the faster motor being taken as remote. The probability that $ \ttautfalse=\ttam $ is assumed small, and is $ 0.008 $ for the kinesin-1 and kinesin-2 models.  If we further neglect the possibility of a negative fluctuation in $ R(t) $ causing the faster motor to lose ground to the point where both motors share the cargo load (in presence of trap force), then we would expect $ R(t) $ to continue to fall in the indicated ranges throughout the lax phase. The arguments from Subsection~\ref{laxhomnonz} would carry over by substitution of the appropriate values of relative separation velocity $ u_- $ and cargo tracking variable velocities $ u_+ $ listed above,
together with the use of the constant detachment rates
\begin{itemize}
\item $ d_0^{(1)} $ and $ d_0^{(2)} $, with total detachment rate $ d_0 = d_0^{(1)}+d_0^{(2)} $ for $ F_T =0 $, 
\item $ \detone{1} $ and $ d_0^{(2)} $, with total detachment rate $ d_0 = \detone{1}+d_0^{(2)} $ for $ F_T > 0 $, 
\item $ d_0^{(1)} $ and $ \detone{2} $, with total detachment rate $ d_0 = d_0^{(1)}+\detone{2} $ for $ F_T < 0, $
\end{itemize}
and the associated direct modification to $ \pdetlax{i} $ in Eq.~\eqref{eq:detprobs}.

Calculations for dwell times follow from similar calculations in the previous
subsection:
\begin{align}
p_{\Ttaut} (t)\sim \mathrm{IG}((r^*_{\leftarrow\rightarrow}-r_0)/u_-, ((r^*_{\leftarrow\rightarrow}-r_0)/\theta)^2). 
\end{align}
The probability distribution for the intermotor displacement when detachment occurs in the lax phase is, from Appendix~\ref{app:laxdetsep}:
\begin{align}
 p_{\Rdetlaxatt{j}}(r) &= \int_{0}^{\infty}\frac{d_0 \expe^{-d_0 t}}{\sqrt{2\pi\theta^2 t} \mathbb P(\Ttaut>t)} 
 \left[\expe^{-\frac{(r-r_0-u_- t)^2}{2\theta^2 t}}-\expe^{\frac{2(r^{\ast}_{\leftarrow \rightarrow}-r_0)u_-}{\theta^2}}
\expe^{-\frac{(2r^{\ast}_{\leftarrow \rightarrow}-r-r_0+u_- t)^2}{2\theta^2 t}}\right]dt\\
& \text{ for }   r,r_0<r^{\ast}_{\leftarrow\rightarrow}. \end{align}
For the same reason as for truly cooperative ensembles; this motor separation depends on the identity $j $ of which motor attached into the \TAM\ state but not on the identity of the detaching motor.
We note from Table~\ref{tab:laxtauttimes} that entering the taut regime is common for heterogenous ensembles at zero trap force $ F_T =0$, and that the expected
time
spent in  the taut phase is nontrivial. 

When, on the other hand, the \TAM\ state is entered for a \falsecoop\ heterogenous ensemble with a substantial trap force $ F_T $ with the slower motor initially ahead of the faster motor, then in absence of detachment, the faster motor will catch up with the slower motor so that they are sharing load if they weren't already and then the faster motor will move ahead until one of the motor tethers (depending on the sign of the trap force) goes slack and proceed in this way until the separation is wide enough to re-engage both motors into a cooperative tug of war.  That is, the motors pass through a regime of small separation where both motors are engaged, but this is not what we have been calling the taut state because it does not persist in a way we can treat in as a terminal phase according to Subsection~\ref{tauthet}.  
This means the separation velocity would not be approximated as constant in the lax phase, and we have no real hope to obtain an analytically tractable distribution for the amount of time until the taut regime is reached.  We therefore seek an empirically supported approximation.  We label the phases described above, in order, as Lax I, Taut I, Lax II, and Taut II. For the first three phases, we approximate the separation velocity and detachment rate as suitable constants, and compute in Table~\ref{phasetable}, via direct modifications of the formula~\eqref{plax}, the probability that both motors in a kinesin-1/kinesin-2 ensemble remain attached until their separation $ R(t) $ enters the next phase.  The Taut 1 phase is not really well approximated with a constant separation velocity or detachment rate, as the load is shifting between motors, but for the purpose of this rough analysis, we just use the constant value corresponding to the midpoint separation of the Taut 1 phase (which is equal to $0$  for our kinesin-1/kinesin-2 model where the tether properties are taken to be the same so $ r^{\ast(2)}_{\rightarrow\rightarrow}=-r^{\ast(1)}_{\rightarrow\rightarrow}$.   We recall from the above discussion that the slower motor typically attaches ahead of the faster motor when the faster motor has most recently attached into the \TAM\ state under hindering force, or when the slower motor has most recently attached into the \TAM\ state under assisting force.  In the notation we will develop in Subsection~\ref{laxhetnonz}, these substates of the \TAM\ state are respectively labelled (1,2;1) and (1,2;2).   Table~\ref{phasetable} presents a summary of the probabilities to complete the various phases within the \TAM\ state.  Clearly, the Taut II phase, where the motor separation $ R(t) $ in the  \TAM\ state can be said to have reached its stationary distribution, is rarely reached because the change in $ R(t) $ required is large enough to give ample time for one or the other motor to detach.  We will therefore neglect it.   To achieve a two-phase level of complexity comparable to our analysis of the \TAM\ state in other scenarios, we will also neglect the Taut I phase.  The justification is that it is relatively short and the effective motor detachment rate in the Taut I phase is not dramatically different from those of the Lax I and Lax II phases.  Note that the Lax I and Lax II phases differ primarily in which motor is engaged and which is slack.  We define their separation to be midway through the Taut I phase, which is essentially where the dominant load switches from one motor to the other.

\begin{table}
\resizebox{\columnwidth}{!}{
\begin{tabular}{|c|c|c|c|c|}\hline
$F_T = \SI{2}{\pico\newton}$, (1,2;1) & Sep. Change (nm)  & Detachment rate (1/s) & Sep. Velocity
(nm/s)
& Completion Prob. \\\hline
Lax I &  $r^{\ast(2)}_{\rightarrow
\rightarrow} - \mu_{R^a}= 75.90 $ & $d^{(1)}_{0}+ \detone{2} = 7.80 $ & $v^{(1)}- V_2  =  432.86$ & .25 \\\hline
Taut I & $r^{\ast(1)}_{\rightarrow
\rightarrow} -r^{\ast(2)}_{\rightarrow
\rightarrow}= 24.1$  & $\dettwoz{1}(0)+\dettwoz{2}(0)= 3.39$ & $U_-(0) = 248.57$ & .72 \\\hline
Lax II 
  & $r^{\ast}_{\leftarrow \rightarrow}-r^{\ast(1)}_{\rightarrow
\rightarrow}  = 143.90 $ & $d^{(2)}_{0}+ \detone{1} = 2.07$ & $ V_1-v^{(2)}=64.29$ & .01\\\hline
Taut II   &- & $\dettwoboth = 4.78$ &- & - \\ \Xhline{2\arrayrulewidth}
$F_T = -\SI{2}{\pico\newton}, $ (1,2;2)  & Sep. Change (nm)  & Detachment rate (1/s) & Sep. Velocity
(nm/s)
& Completion Prob. \\\hline
Lax I  & $r^{\ast(2)}_{\rightarrow
\rightarrow} - \mu_{R^a} = 75.90  $ & $d^{(2)}_{0}+ \detone{1} = 8.72 $& $ V_1-v^{(2)}  =290.00 $ & .10 \\\hline
Taut I & $r^{\ast(1)}_{\rightarrow
\rightarrow} -r^{\ast(2)}_{\rightarrow
\rightarrow}= 24.1 $ & $\dettwoz{1}(0)+\dettwoz{2}(0)=4.60  $ & $U_-(0) =290.00 $ & .68 \\\hline
Lax II 
  & $r^{\ast}_{\leftarrow \rightarrow}-r^{\ast(1)}_{\rightarrow
\rightarrow}  =143.90  $ & $d^{(2)}_{0}+ \detone{1} =6.34$  & $ v^{(1)}- V_2 =290.00 $ & .04 \\\hline
Taut II  & -& $\dettwoboth =123.75 $ & - & - \\\hline
\end{tabular}
}
\caption{Analysis of kinesin-1/kinesin-2 ensembles under trap force for scenarios where the slower motor (kinesin-2) is typically ahead of the faster motor (kinesin-1) at the onset of the \TAM\ state.  The upper component of the table reports results for the \TAM\ state when the faster motor (kinesin-1) has most recently attached, under hindering force $ F_T = \SI{2}{\pico\newton}$.  The lower component of the table reports results for the \TAM\ state when the slower motor (kinesin-2) has most recently attached, under assisting force $ F_T = \SI{-2}{\pico\newton}$.  Other notation in the left column is explained in the main text.  The ``Sep. Change'' column indicates the deterministic or average change in motor separation during the phase.  The ``Detachment Rate'' and ``Sep. Velocity'' columns indicate the constant values assumed for these quantities in each phase.  ``Completion Prob.'' reports the probability, computed from formulas similar to Eq.~\eqref{plax}, that the phase will complete before either motor detaches. The Taut II rows are presented only to give a comparison for the effective detachment rates in each phase.  The other entries
in the Taut II phase are not well defined and omitted.}
\label{phasetable}
\end{table}



Thus, for falsely cooperative motor ensembles, we represent the dynamics in the \TAM\ state as consisting of a Lax I phase when $ R(t) < \frac{1}{2}(r^{\ast(1)}_{\rightarrow\rightarrow}+r^{\ast(2)}_{\rightarrow\rightarrow}) $ and a Lax II phase when $ R(t) > \frac{1}{2}(r^{\ast(1)}_{\rightarrow\rightarrow}+r^{\ast(2)}_{\rightarrow\rightarrow}) $.   Considering first the substate (1,2;1) under hindering trap force $ F_T> 0 $:  In the Lax I phase, the slower motor is carrying the load while the faster motor is slack, so the relative separation velocity is $ u_{-,I} = v^{(1)}-\Vstate{2} $, the cargo tracking variable velocity is $ u_{+,I}= \frac{1}{2} \left(v^{(1)}+\Vstate{2}\right) $, and the total detachment rate is $ d_{0,I} = d_0^{(1)} + \detone{2} $.   In the Lax II phase, the faster motor is carrying the load while the slower motor is slack, so the relative separation velocity is $ u_{-,II} = \Vstate{1}-v^{(2)}$, the cargo tracking variable  velocity is $ u_{+,II}= \frac{1}{2} \left(\Vstate{1}+v^{(2)}\right)$, and the total detachment rate is $ d_{0,II} = \detone{1} + d_0^{(2)}$.   For substate (1,2;2) under assisting trap force $ F_T < 0 $, the expressions for the separation velocities and detachment rates are swapped between the Lax I and Lax II phase because which motor is slack in each phase is the opposite of the hindering trap force case.

The formulas from Subsection~\ref{laxhomnonz} are readily adapted to determine the statistics for whether, when, and how a motor detaches in the Lax I phase.  If no detachment happens in the Lax I phase, then rather than proceeding to the taut phase, the dynamics are now governed by the Lax II phase, which is treated as being terminated only by detachment.  The time spent in the Lax II phase is exponentially distributed with mean 
\begin{equation*}
\E[\Tdlaxb] = d_{0,II}^{-1}.
\end{equation*}
As we are neglecting fluctuations out of the Lax II phase back into the Lax I phase, $ R(t) $ is simply a Gaussian process with drift $ u_{-,II} $ and diffusivity $ \frac{1}{2}\theta^2 = D_1+D_2 $, so the distribution of motor separation from the Lax II phase follows by simple conditioning on $ \Tdlaxb $:
\begin{equation*}
 p_{\Rdetlax}(r) = \int_{0}^{\infty}d_{0,II} \expe^{-d_{0,II} t} 
 \exp\left[-\frac{\left(r-\frac{1}{2}(r^{\ast(1)}_{\rightarrow\rightarrow}+r^{\ast(2)}_{\rightarrow\rightarrow})-u_{-,II}t\right)^2}{2\theta^2 t}\right] \, \difd t.
\end{equation*}
The mean cargo tracking variable increment is simply
\begin{equation*}
\E[\dCMattblaxb] = u_{+,II} \E[\dTattblaxb] = \frac{u_{+,II}}{d_{0,II}^{-1}}.
\end{equation*}
Second order statistics are not presented as we find in Appendix~\ref{app:secmompain} they are too unwieldy to calculate for heterogenous ensembles, which covers all falsely cooperative cases.

\subsection{Lax statistics for \neutralcoop\ ensembles}
\label{sec:neutcoop}
\subsubsection{Homogenous ensembles at zero trap force $ F_T = 0 $}
\label{laxhomzero}
As both motors move at their common unencumbered velocities, the separation velocity in the lax phase is $u_- = 0$ so the SDE~\eqref{eq:Rcc} for the separation has no drift term.
Furthermore,  the termination of the lax phase is determined by exit from the two-sided interval $ [-r^{\ast}_{\leftarrow\rightarrow}, r^{\ast}_{\leftarrow\rightarrow}]$, for which we can only obtain analytical solutions in the form of series. 
  For an initial displacement $r$ and time
$t\ge 0$, the survival function $S(t) = \int_t^\infty
p_{\Ttaut}(t^{\prime};r)dt^{\prime} $ takes the form of the backward Kolmogorov 
equation with pure diffusion:
\begin{align*} 
\partial_t S = (D_1+D_2) \partial_{r}^2S, \quad  S(r,0) = \mathbf 1_{[-r^*_{\leftarrow\rightarrow},r^*_{\leftarrow\rightarrow}]},
\quad S(-r^*_{\leftarrow\rightarrow},t) = S(r^*_{\leftarrow\rightarrow},t) = 0.
\end{align*}
Through standard methods, the Fourier series solution is
\begin{equation*}
S(r_{0}, t) = \frac 4\pi\sum_{n = 0}^\infty \frac {(-1)^n}{2n+1} \exp\left(-\frac{(2n+1)^2
\pi^2 (D_1+D_2) t}{4(r^*_{\leftarrow\rightarrow})^2}\right) \cos\left(\frac{(2n+1)\pi r_0}{2r^*_{\leftarrow\rightarrow}}\right). \end{equation*}
We can then recover $p_{\Ttaut} = -\partial_t S$, which has the series form
\begin{equation*}
p_{_{\Ttaut} }(t; r_0) = \frac{\pi \theta}{(r^*_{\leftarrow\rightarrow})^2} \sum_{n = 0}^\infty
(-1)^n
(2n+1)\expe^{-\frac{(2n+1)^2
\pi^2 (D_1+D_2)t}{4(r^*_{\leftarrow\rightarrow})^2}}\cos\left(\frac{(2n+1)\pi
r_0}{2r^*_{\leftarrow\rightarrow}}\right).
\end{equation*}

For intermotor distance at detachment occuring during the lax phase, we  must consider the two-sided
conditional probability
\begin{equation}
\mathbb P(\Rdetlax\leq r)  = \mathbb
P\left(R(T_{d}) \leq r \Big|\sup_{0\le
s\le T_d} |R(s) |\le r^*_{\leftarrow\rightarrow}\right).
\end{equation}
The dependence of $ \Rdetlax$ on the identity of the most recently attached motor is trivial for homogenous ensembles, which we indicate by dropping this index.
 The density of $R$ at time $t$ if there were no detachment can be expressed
by a forward equation
with initial conditions of a point mass:
\begin{align} 
&\partial_t p(r,t;r_0) = (D_1+D_2) \partial_{rr}p, \\  &p(r,0;r_0) =\delta_{r_{0}}(r),
\quad
p(-r^{*}_{\leftarrow\rightarrow},t;r_0) =p(r^{*}_{\leftarrow\rightarrow},t;r_0) = 0  .
\end{align}
It is  straightforward to derive the series solution
\begin{align*}
p(r,t;r_0) &= \frac{1}{r^*_{\leftarrow\rightarrow}} \sum_{n = 1}^\infty \exp \left(-\frac{n^2
\pi^2 (D_1+D_2) t}{(4(r_{\leftarrow \rightarrow}^*)^2}\right)
\sin \left(\frac{n \pi( r_{0}+r^*_{\leftarrow\rightarrow})}{2r^*_{\leftarrow\rightarrow}}\right)\sin\left(\frac{n \pi(r+ r^*_{\leftarrow \rightarrow}}{2r_{\leftarrow\rightarrow}^*}\right). 
\end{align*}
From the law of total probability, we then obtain (with implicit conditioning as usual on the initial condition $ R(0)=r_0 $):
\begin{equation}
p_{\Rdetlax}(r) =   \int_{0}^{\infty}\frac{ 
 p(r,t;r_0)d_{0}e^{-d_0 t}}{\int_{-r^*_{\leftarrow\rightarrow}}^{r^*_{\leftarrow\rightarrow}} p(r',t;r_0) dr'}dt.
\end{equation}
where $ d_0 = d_0^{(1)} + d_0^{(2)} $ is the total rate of detachment of the two unencumbered motors.  Naturally the motor separation has no dependence on the index of detaching motor in a homogenous ensemble.  The formulas for statistics of the lax phase are otherwise straightforwardly adapted from those presented in Subsection~\ref{laxhomnonz} with effective cargo tracking variable velocity $ u_+ = v^{(1)} = v^{(2)}$.

\subsubsection{Homogenous ensembles at assisting trap forces $ F_T < 0 $}
Both motors still travel at their common unencumbered velocities in the lax phase, but now both motors become taut at small separations $ |R(t)| \leq r^{\ast(1)}_{\rightarrow \rightarrow} = -r^{\ast(2)}_{\rightarrow\rightarrow} $ where they share load, in addition to large separations $ |R(t)| \geq r^{\ast}_{\leftarrow \rightarrow} $  where they temporarily pull against each other.  One could thereby modify the first passage calculations in Subsection~\ref{laxhomzero} in a straightforward way, but in fact detachment
rates due to an assisting force greatly increase the probability   that a motor
will detach before reaching the taut phase. In fact, this occurs often enough
to assume that  $\plax = 1$. For an informal justification of this approximation,
we compare the time  scales related to detachment and diffusion. For
$F_T = -2 \si{\pico\newton}$ with two kinesin-1 motors,  the average initial separation of motors upon entry into the \TAM\ state is $ \bExp |\Ratt{}| = \SI{88}{\nano\metre} $, with a slack motor with detachment rate $ d_0^{(1)} = \SI{0.78}{\sec^{-1}} $ and the engaged motor having an enhanced detachment rate of $ \detone{2} = \SI{7.88}{\sec^{-1}}$.  The     average time until first detachment is then  $1/(d_0^{(1)}+\detone{2}) = .13$ s. To reach the taut phase, the separation must either increase to $ r^{\ast}_{\leftarrow\rightarrow} =\SI{156}{\nano\metre} $ or decrease to $ r^{\ast(1)}_{\rightarrow\rightarrow} = \SI{20}{\nano\metre} $, requiring a time scale $ (\SI{68}{\nano\metre})^2/(\theta^2/2) = \SI{1.4}{\sec} $
which is approximately an order of magnitude larger than the detachment
time scale. 

Within the lax phase, we take detachment rates $ d_{0}^{(1)} $ and $ \detone{2} $ to take into account one motor is typically carrying the assisting load (the index of which doesn't matter for homogenous ensembles).  This gives a total detachment rate of $ d_0 = d_0^{(1)} + \detone{2} $, with the natural modifications to the detachment  probabilities $ \pdetlax{i} $ in Eq.~\eqref{eq:detprobs}.

We approximate the intermotor displacement at detachment by removing the conditioning on $ R(t) $ not achieving the taut regime, since this typically has low probability.  Thus, from the purely diffusive evolution of $ R(t) $ starting from $ r_ 0$:
\begin{equation*}
 p_{\Rdetlax}(r) = \int_{0}^{\infty}d_{0} \expe^{-d_{0} t} 
 \exp\left[-\frac{\left(r-r_0\right)^2}{2\theta^2 t}\right] \, \difd t.
\end{equation*}
Of course, this might lead to values of $\Rdetlax $ which are within the taut regime, but our above arguments indicate this should have small probability, and such realizations would create no catastrophes.  The mean cargo tracking variable increment can be computed simply:
 \begin{equation}
 \E[\dMattb] = v^{(2)} /d_{0}.
 \end{equation}
 The formulas for statistics of the lax phase are otherwise straightforwardly adapted from those presented in Subsection~\ref{laxhomnonz} with effective cargo tracking variable velocity $ u_+ = v^{(1)} = v^{(2)}$.

\subsection{Cycle statistics for homogenous ensembles} \label{combolaxtaut}

In the previous subsections, we have determined statistics for the increments in time and progress of the cargo tracking variable while in the \TAM\ state  under lax and taut phases, together with the jump in the cargo tracking variable upon detachment from this state.  
In this subsection, we provide  statistics for the overall time spent and progress made in the \TAM\ state  for homogenous ensembles through simple conditioning arguments, which are 
 then in turn  used to compute statistics for cycles in conjunction with the results regarding the \OAM\ state contribution from Section~\ref{sec:oam} and the cargo tracking variable jump upon attachment into the the \TAM\ state from Section~\ref{sec:jump}.  Since cycle times and displacements are independent across cycles for homogenous ensembles, effective ensemble and run velocities can then be found from Wald's identity as described in Subsection~\ref{sub:renewal}.




 In~\citet{klobusicky2020cooperative}, cycle statistics were obtained from the law of total expectation, variance, and covariance for the contributions in Eqs.~\eqref{tcycleterms} and~\eqref{mcycleterms2} from the \TAM\ state, the \OAM\ state visited, and the jumps out of and back into the \TAM\ state, which are conditionally independent given the identity $ \detmot $ of the motor which detached out of the \TAM\ state.  Now since we don't assume the stationary distribution is necessarily achieved in the \TAM\ state, we must further condition on whether the \TAM\ state terminates in the lax state or in the taut state, as this affects both the statistics of the change in cargo tracking variable in the \TAM\ state $ \dCMattb $ as well as the jump in the cargo tracking variable upon exiting the \TAM\ state, $ \dMdetjump{\detmot} $ (through its dependence on the motor separation at detachment $ \Rdet_{\detmot} $).  Because the detachment rates of each motor depend on the time-dependent motor separation, in a strict sense, the identity of the detaching motor $ \detmot $ would be correlated with the motor separation at detachment as well as the time spent in the \TAM\ state.  Under the approximation of our lax and taut phase decomposition, however, we can neglect these correlations once we condition on whether the detachment occurred in the lax or taut phase, because the detachment rates are modeled as constant within each phase.  We do need to account for the correlation between time spent in the \TAM\ state and the motor separation at detachment when this occurs within the lax state.
  
Moreover, we note that the calculations in the preceding subsections were conditioned on a particular value $ \Ratt{}= R(\ttam)=r_0 $ of the motor separation upon entry into the \TAM\ state.  Averaging over the probability distribution of $ \Ratt{} $ will in general introduce correlations between the jumps into and out of the \TAM\ state as well as the continuous progress within the \TAM\ state.  Thus we make this dependence on the initial separation explicit in this subsection via subscripting the expectation and variance symbols with $ r_0 $ to indicate conditioning on $\Ratt{}= r_0 $.  We will also write `L' in conditionals to signify that the  \TAM\ state terminated within the lax state, and `LT' to signify the taut phase was realized.  
The conditional statistics for the time spent and continuous progress of the cargo tracking variable in the \TAM\ state can  be written: 
  \begin{align*}
   \E[\dTattb|\mathrm{L}, \Ratt{}=r_0] &=  \E_{r_0}[\dTattblaxd], \\
 \E[\dTattb|\mathrm{LT}, \Ratt{}=r_0] &=  \E_{r_0}[\dTattblaxt]+\E[\dTattbtaut], \\
 \Var(\dTattb|\mathrm{L},\Ratt{}=r_0) &= \Var_{r_0}(\dTattblaxd), \\
 \Var(\dTattb|\mathrm{LT}, \Ratt{}=r_0) &= \Var_{r_0}(\dTattblaxt)+\Var(\dTattbtaut), \\
 \E[\dCMattb|\mathrm{L},\Ratt{}=r_0] &= u_+\E_{r_0}[\dTattblaxd], \\
 \E[\dCMattb|\mathrm{LT},\Ratt{}=r_0] &= u_+\E_{r_0}[\dTattblaxt]+\E[\dCMattbtaut], \\
 \Var(\dCMattb|\mathrm{L},\Ratt{}=r_0) &=  u_+^2\Var_{r_0}(\dTattblaxd), \\
 \Var(\dCMattb|\mathrm{LT}, \Ratt{}=r_0) &= u_+^2 \Var_{r_0}(\dTattblaxt)+\Var(\dMattbtaut), \\
 \Cov(\dTattb,\dCMattb|\mathrm{L}, \Ratt{}=r_0) &= u_+ \Var(\dTattb|\mathrm{L},\Ratt{}=r_0) = u_+ \Var_{r_0}(\dTattblaxd), \\
 \Cov(\dTattb,\dCMattb|\mathrm{LT}, \Ratt{}=r_0) &= 
u_+ \Var_{r_0} (\dTattblaxt)+\Cov_{r_0} (\dTattbtaut,\dMattbtaut).
 \end{align*}
 
 The conditional statistics for the jump of the cargo tracking variable upon entry into the \TAM\ state clearly don't depend on how detachment from the \TAM\ state occurred, since the only connection between attachment and detachment is the initial motor separation $ \Ratt{} $, which is conditioned upon:
   \begin{align*}
 \E[\dCMattjump{}|\mathrm{L}, \Ratt{}=r_0] &= \E[\dCMattjump{}|\mathrm{LT},\Ratt{} =r_0] = \frac{r_0}{2},\nonumber \\
 \Var(\dCMattjump{}|\mathrm{L},\Ratt{}=r_0) &=   \Var(\dCMattjump{}|\mathrm{LT},\Ratt{}=r_0) = 0.  \nonumber \\
  \end{align*}

Regarding detachment probabilities, for homogenous ensembles we have from Eq.~\eqref{eq:pdettauti} that $ \pdettaut{1} =\pdettaut{2} = \frac{1}{2} $, but from Subsection~\ref{laxhomnonz}, $ \pdetlax{j} = \frac{d_0}{d_0+\detone{}} $ and $ \pdetlax{\jp} = \frac{\detone{}}{d_0+\detone{}} $, where $ j $ is the index of the most recently attached motor. This just means the motor attaching into the lax phase has a typically lower probability of detaching than the one already carrying load (as $ \detone{} \leq d_0 $).   For homogenous ensembles the index of the detaching motor only affects the jump in the cargo tracking variable upon detachment~\eqref{eq:detjumpr}.   In the lax phase, by assumption, the most recently attached motor is the same as the unloaded motor which will be oriented (relatively to the other motor) in the same direction as the trap force (if present), and is also the one to detach with lower probability.  Thus the detachment jump in the presence of trap force has the higher probability to be of opposite sign to the trap force.   Thus we can suppress the index dependence for homogenous ensembles by re-expressing the jump at detachment in the lax phase as:
$ \dCMdetjump{} =   \frac{S}{2} \sgn(F_T) |\Rdetlax| $ with
\begin{equation*}
S = \begin{cases}
1 & \textrm{ with probability }   \frac{\detone{}}{d_0+\detone{}}, \\
-1 & \textrm{ with probability } \frac{d_0}{d_0+\detone{}}.
\end{cases}
\end{equation*}
 In the absence of trap force, of course, both motors are equally likely to detach, so $ \dCMdetjump{} =  \frac{S}{2} |\Rdetlax| $ with $ S = \pm 1 $ with equal probability.  Which motor detaches in either lax or taut phase can be taken as independent of $ \Rdetlax $ since the detachment rates are constant within each phase.  In the taut phase, the intermotor displacement is a symmetric random variable $ \RdetJtaut{i} = \Rdettaut$ for homogenous ensembles so the sign and index in Eq.~\eqref{eq:detjumpr} depending on the detaching motor is irrelevant.
 We can then obtain: 
  \begin{align*}
 \E[\dCMdetjump{}|\mathrm{L}, \Ratt{}=r_0] &=  \frac{\sgn(F_T)}{2}\frac{\detone{}-d_0}{\detone{}+d_0} \E_{r_0}[|\Rdetlax|], \\
 \E[\dCMdetjump{}|\mathrm{LT},\Ratt{}=r_0] &= 0, \\
 \Var(\dCMdetjump{}|\mathrm{L},\Ratt{}=r_0) & =  \frac 14 \Var_{r_0} (\Rdetlax). \\
  \Var(\dCMdetjump{}|\mathrm{LT},\Ratt{}=r_0) &
 = \frac 14 \Var(\Rdettaut{}).
 \end{align*}

 We also require the covariance between the progress of the cargo tracking variable in the \TAM\ state and the jump upon detachment:  
 \begin{align*}
 \Cov[\dCMattb,\dCMdetjump{}|\mathrm{L},\Ratt{}=r_0] &=\Cov(\dCMattb,
 \frac{S}{2} \sgn(F_T) |\Rdetlax| |\mathrm{L},\Ratt{}=r_0] \\
&  = \frac{\E[S]}{2}  \sgn(F_T) u_+ |u_-| \Var [\dTattb|\mathrm{L}, \Ratt{}=r_0]\nonumber \\
& =  \frac{\sgn(F_T)}{2}  \frac{\detone{}-d_0}{\detone{}+d_0} u_+ |u_-|\Var_{r_0}[\dTattblaxd], \\
  \Cov[\dCMattb,\dCMdetjump{}|\mathrm{LT},\Ratt{}=r_0] &= 0, \\
   \Cov[\dTattb,\dCMdetjump{}|\mathrm{L},\Ratt{}=r_0] &= \frac{\sgn(F_T)}{2}  \frac{\detone{}-d_0}{\detone{}+d_0} |u_-|\Var_{r_0}[\dTattblaxd], \\
     \Cov[\dTattb,\dCMdetjump{}|\mathrm{LT},\Ratt{}=r_0] &= 0.
 \end{align*}
 where we used that both the motor separation and cargo tracking variable have constant drifts, respectively $ u_-$ and $ u_+ $  in the lax phase, with independent Brownian increments.   The first expression, which assumes $ F_T \neq 0 $, also applies for $ F_T =0 $ where $ d_0 = \detone{} $ and $ u_- = 0 $ so the expression on the right hand side evaluates correctly to 0.  The motor separation when detaching from the taut phase is assumed to be drawn independently from its stationary distribution, so the jump in the cargo tracking variable upon detachment is independent of the progress of the cargo tracking variable while in the \TAM\ state.


Combining now the contributions from the \OAM\ state, the attachment into the \TAM\ state, the progress within the \TAM\ state, detachment from the \TAM\ state, we obtain for the conditional statistics of an entire cycle:
\begin{align}
\E[\dTcyc|\mathrm{L},\Ratt{}=r_0] &= \E_{r_0}[\dTattblaxd] + \frac{1}{\bar{a}^{*}+\detone{}}, \\
\E[\dTcyc|\mathrm{LT},\Ratt{}=r_0] &= \E_{r_0}[\dTattblaxt] + \E_{r_0}[\dTattbtaut]+\frac{1}{\bar{a}^{*}+\detone{}}, \\
\Var[\dTcyc|\mathrm{L},\Ratt{}=r_0] &= \Var_{r_0}[\dTattblaxd] + \frac{1}{(\bar{a}^{*}+\detone{})^2}, \\
\Var[\dTcyc|\mathrm{LT},\Ratt{}=r_0] &= \Var_{r_0}[\dTattblaxt] +\Var[\dTattbtaut] + \frac{1}{(\bar{a}^{*}+\detone{})^2}, \\
\E[\dCMcyc|\mathrm{L},\Ratt{}=r_0] &=  u_+\E_{r_0}[\dTattblaxd] +\frac{\Vstate{}}{\bar
a^{*}+\detone{}}  +\frac{r_0}{2}  \\
& \qquad \qquad + \frac{1}{2} \frac{\detone{}-d_0}{\detone{}+d_0} \sgn (F_T) \E_{r_0}[|\Rdetlax|], \nonumber \\
\E[\dCMcyc|\mathrm{LT},\Ratt{}=r_0] &=  u_+ \E_{r_0}[\dTattblaxt]+ \E[\dCMattbtaut]  +\frac{\Vstate{}}{\bar
a^{*}+\detone{}}  +\frac{r_0}{2}, \nonumber \\
 \Var[\dMcyc|\mathrm{L},\Ratt{}=r_0]  &= u^2_+\Var_{r_0}(\dTattblaxd) +\frac{2D}{\bar
a^{*}+\detone{}} +\left(\frac{\Vstate{}}{\bar
 a^{*}+\detone{}}\right)^2 \\
 &  +\frac 14 \Var_{r_0} (\Rdetlax)  
 +\sgn(F_T) \frac{\detone{}-d_0}{\detone{}+d_0}
u_+ |u_-|\Var_{r_0}[\dTattblaxd], \nonumber \\
  \Var[\dMcyc|\mathrm{LT},\Ratt{}=r_0]  &= u_+^2 \Var_{r_0}(\dTattblaxt) + \Var_{r_0}(\dCMattbtaut)\\
 & \qquad +\frac{2D}{\bar a^{*}+\detone{}}
 +\left(\frac{\Vstate{}}{\bar
 a^{*}+\detone{}}\right)^2 +  \frac{1}{4} \Var(\Rdettaut), \\
 \Cov(\dTcyc,\dMcyc|\mathrm{L},\Ratt{}=r_0]  &= 
 \left(u_++\frac{\sgn(F_T)}{2}
 \frac{\detone{}-d_0}{\detone{}+d_0} |u_-|\right)\Var_{r_0}[\dTattblaxd] \nonumber \\
 & \qquad \qquad +\frac{\Vstate{}}{\left(\bar
 a^{*}+\detone{}\right)^2}, \nonumber \\
  \Cov(\dTcyc,\dMcyc|\mathrm{LT},\Ratt{}=r_0] &=  u_+ \Var_{r_0} (\dTattblaxt) \\
  & \qquad \qquad +\Cov_{r_0} (\dTattbtaut,\dMattbtaut)+ \frac{\Vstate{}}{(\bar a^{*}+\detone{})^2} \nonumber \\
  &= u_{+} \Var_{r_0}(\dTattblaxt) + \frac{\barvtwo}{(\dettwoboth)^2}+\frac{\Vstate{}}{\left(\bar
 a^{*}+\detone{}\right)^2}.\nonumber
  \end{align}

Averaging over whether the \TAM\ state is terminated in the lax or taut state, we obtain for mean cycle statistics conditioned on the motor separation upon entry into the \TAM\ state:
\begin{align}
\E_{r_0}[\dTcyc]  &=  \E_{r_0}[\dTattb]   + \frac{1}{\bar{a}^{*}+\detone{}}, \label{eq:cyclemeans}\\
\E_{r_0}[\dCMcyc] &=   \E_{r_0}[\dCMattb]  +\frac{\Vstate{}}{\bar
a^{*}+\detone{}}  +\frac{r_0}{2}  + \frac{\plax}{2}  \frac{\detone{}-d_0}{\detone{}+d_0} \sgn (F_T) \E_{r_0}[|\Rdetlax|]. \nonumber
\end{align} 
where the means within in the \TAM\ state are :  
  \begin{align}
\E_{r_0}[\dTattb]&= \plax \E_{r_0}[\dTattblaxd]+(1-\plax)(\E_{r_0}[\dTattblaxt]+\E[\dTattbtaut]),\label{eq:tammeans}\\
\E_{r_0}[\dCMattb] &= \plax u_+ \E_{r_0}[\dTattblaxd]+(1-\plax)(u_+ \E_{r_0}[\dTattblaxt]+\E[\dCMattbtaut]). \nonumber
 \end{align}


We similarly remove the conditioning on detachment from the lax or taut phase using the law of total variance.  
For the cycle time,
\begin{align}
\Var_{r_0} (\dTcyc) &= \plax \Var(\dTcyc|\mathrm{L},\Ratt{}=r_0)+(1-\plax)\Var(\dTcyc|\mathrm{LT},\Ratt{}=r_0)\\
& \qquad +\plax(1-\plax)(\E[\dTcyc|\mathrm{LT}|\Ratt{}=r_0]-\E[\dTcyc|\mathrm{L},\Ratt{}=r_0])^2 \nonumber \\
&= \frac{1}{(\bar{a}^{*}+\detone{})^2} \\
& \qquad  + \plax \Var_{r_0} (\dTattblaxd) + (1-\plax) (\Var_{r_0} (\dTattblaxt)+ \Var(\dTattbtaut)) \nonumber \\
& \qquad + \plax (1-\plax) (
\E_{r_0}[\dTattbtaut] +\E_{r_0}[\dTattblaxd]-\E_{r_0}[\dTattblaxt])^2. \nonumber
\end{align}

For the cycle increment,
 \begin{align}
 \Var_{r_0}(\dMcyc) = \plax \Var(\dMcyc|\mathrm{L},\Ratt{}=r_0)+(1-\plax)\Var(\dMcyc|\mathrm{LT},\Ratt{}=r_0)\\
 +\plax(1-\plax)(\E[\dMcyc|\mathrm{T},\Ratt{}=r_0]-\E[\dMcyc|\mathrm{LT},\Ratt{}=r_0])^2. \nonumber
 \end{align}
 
 For the covariance between cycle time and increment, we use the law of total covariance to obtain
 \begin{align*}
 \Cov_{r_0}(\dTcyc, \dMcyc) &= \plax \Cov(\dTcyc, \dMcyc|\mathrm{L},\Ratt{}=r_0) \\
 & \qquad \qquad +(1- \plax )\Cov(\dTcyc,
 \dMcyc|\mathrm{LT},\Ratt{}=r_0) \\
 & \qquad \qquad +\plax(1-\plax)( \E[\dMcyc|\mathrm{L},\Ratt{}=r_0] \\
 & \qquad \qquad - \E[\dMcyc|\mathrm{LT},\Ratt{}=r_0])\cdot(
 \E[\dTcyc|\mathrm{L},\Ratt{}=r_0] \\
 & \qquad \qquad - \E[\dTcyc|\mathrm{LT},\Ratt{}=r_0]).
 \end{align*}
In the above expressions, $ \plax $ implicitly depends on the initial separation $ r_0 $.


We finally obtain the unconditional cycle statistics by simply integrating numerically against the probability density for the motor separation upon attachment, $ \Ratt{} $ (Eq.~\eqref{eq:ratthom}).  Thus, for $Y \in \{ \dTcyc, \dMcyc\}$:
\begin{align} 
\mu_{Y} &= \int_{-\infty}^\infty \E_{r_0}[Y] p_{R_a}(r_{0})dr_{0},
\quad \sigma^2_{Y} =  \int_{-\infty}^\infty (\Var_{r_0}[Y]+\E_{r_0}[Y]^2
) p_{R_a}(r_{0})dr_0-\mu_{Y}^2. \label{roavg}
\end{align}

For the unconditioned covariance, 

\begin{equation*}
\sigma_{\dTcyc, \dMcyc} = \int_{-\infty}^\infty (\Cov_{r_0}(\dTcyc, \dMcyc)
+\E_{r_0}[\dTcyc]\E_{r_0}[\dMcyc]) p_{R_a}(r)dr_0-\mu_{\dTcyc}\mu_{\dMcyc}.
\end{equation*}

Together with the probability for total
detachment $\totdetprob =\frac{ \detone{}}{ a^{*}+\detone{}} $, these provide the components of the formulas in Subsection~\ref{sub:renewal} to compute the first and second order statistics of run time and run length  for homogenous ensembles.   


 \subsection{Cycle statistics for heterogeneous ensembles } \label{laxhetnonz}

As we have discussed in Subsection~\ref{laxhetzero}, for heterogenous ensembles, the behavior of the two attached motor state typically
depends on which motor  attaches to begin state $(1,2)$, as indicated in Fig.~\ref{biggerstate}.   The exception is when trap force is negligible $ F_T \approx 0 $ and the tethers of the motors have identical properties ($\ellrest{1}=\ellrest{2}$ and $ \kappa^{(1)}=\kappa^{(2)}$), for then we are in the falsely cooperative case with $ \Vstate{i} \approx v^{(i)} $ and the probability distribution for the initial motor separation $ R_a $ is symmetric and independent of which motor is attaching.  The calculations in Subsection~\ref{laxhetzero} are then not dependent on the identity of which motor has most recently attached.

Otherwise, here we explicitly consider the decomposition of the \TAM\ state in heterogenous ensembles into the substates $(1,2;1)$ and  $(1,2;2)$, corresponding to, respectively, the faster or slower motor being the one which most recently attached.  
The \emph{mean} cycle statistics appearing in the formulas in Subsection~\ref{sub:rendelayed} for the effective transport of the motor-cargo ensemble are computed similarly as in Subsection~\ref{combolaxtaut}, with the following modifications:
\begin{itemize}
\item the attachment jump in the cargo tracking variable depends on which motor is attaching so we go back to the general formula~\eqref{eq:dcMattjump}, 
\item the mean of the detachment jump, progress in the \OAM\ state, and reattachment jump must be averaged over the identity of the detaching motor,
\item the progress in the lax, but not the taut phase, of the  \TAM\ state as well as detachment from the lax phase depends on how the \TAM\ state was entered, and thus on both the initial motor separation $r_0 $ and the index $j$ of the motor which most recently attached.  We indicate these dependencies in the mean statistics with the notation $ \E_{r_0;j}[\cdot]$.
\end{itemize}
With these changes to the arguments from Section~\ref{combolaxtaut}, we have:  
\begin{align}
\E_{r_0;j}[\dTcyc]  &=  \E_{r_0;j}[\dTattb]   +\frac{\pdet{2}}{\bar
a^{*(2)}+\detone{1}} 
+\frac{\pdet{1}}{\bar
a^{*(1)}+\detone{2}},\\
\E_{r_0;j}[\dCMcyc] &=   \E_{r_0;j}[\dCMattb] 
  +\frac{\pdet{2}\Vstate{1}}{\bar
a^{*(2)}+\detone{1}} 
+\frac{\pdet{1}\Vstate{2}}{\bar
a^{*(1)}+\detone{2}} \\
& \qquad \qquad +(-1)^{j+1} \frac{r_0
}{2}  
\nonumber \\
  &+\frac{\plax (\pdetlax{2}-\pdetlax{1})}{2} \E_{r_0;j}[\Rdetlax] \nonumber \\
  &+\frac{1-\plax}{2} \left(\pdettaut{2}\E[\RdetJtaut{2}]-\pdettaut{1}\E[\RdetJtaut{1}]\right).
\end{align} 
where the means within  the \TAM\ state are:  
  \begin{align}
\E_{r_0;j}[\dTattb]&= \plax \E_{r_0;j}[\dTattblaxd]+(1-\plax)(\E_{r_0;j}[\dTattblaxt]+\E[\dTattbtaut]),\\
\E_{r_0;j}[\dCMattb] &= \plax u_+ \E_{r_0;j}[\dTattblaxd]+(1-\plax)(u_+ \E_{r_0;j}[\dTattblaxt]+\E[\dCMattbtaut]).
 \end{align}
The probability for a cycle to end in complete detachment is, as in~\citet{klobusicky2020cooperative}, more complicated for heterogenous ensembles because we must weight the detachment probabilities from the \OAM\ state by the probabilities
$ \pdet{i} = \plax \pdetlax{i}+(1-\plax) \pdettaut{i} $  that motor $i $ is the one that detaches from the \TAM\ state:
\begin{equation}\label{probcompdet}
\totdetprob= \pdet{2}\frac{\detone{1}}{
a^{ (2)}+\detone{1}}+\pdet{1} 
\frac{\detone{2}}{
a^{(1)}+\detone{2}}.
\end{equation}
The various expectations and detachment probabilities are obtained from the results of Sections~\ref{tauthom}-\ref{sec:neutcoop}, with the relevant formulas for the contribution from the lax phase depending on whether the ensemble falls in the truly, neutrally, or falsely cooperative category.  In the above expressions, we have various  probabilities for detachment out of the \TAM\ state  developed in the same subsections above, and all except $ \pdettaut{i} $ are implicitly dependent upon the initial motor separation $r_0 $ in the \TAM\ state and the index $j $ of the most recently attached motor.   To use these results in the framework of Subsection~\ref{sub:rendelayed} to compute overall run statistics, we remove the conditioning on the initial motor separation $ \Ratt{} = r_0 $ in the \TAM\ state through numerical integration against its probability density~\eqref{convr}.   Thus, in particular:
\begin{align*}
\E[T_{\mathrm{c}; j}] &= \int_{-\infty}^{\infty} \E_{r_0;j}[\dTcyc] \, p_{\Ratt{}} (r_0) \, \difd r_0, \\
\E[M_{\mathrm{c}; j}] &= \int_{-\infty}^{\infty} \E_{r_0;j}[\dMcyc] \, p_{\Ratt{}} (r_0) \, \difd r_0. \\
\end{align*}

We note that, as mentioned in Subsection~\ref{sub:rendelayed}, the detachment probabilities for heterogenous ensembles need to be adjusted  for long run calculations by conditioning on the event $ \mathcal A $ of the cycle ending in reattachment and using Bayes' rule:
\begin{subequations}
\label{eq:condreat}
\begin{align}
\plax^{\mathcal A} &= \left(\pdetlax{1}\left(\frac{\bar a^{*(1)}}{\bar a^{*(1)}+\detone{2}}\right)+\pdetlax{2}\left(\frac{\bar a^{*(2)}}{\bar a^{*(2)}+\detone{1}}\right)\right)\frac{\plax}{1-\totdetprob}, \\
\plax^{(i),\mathcal A}&= \frac{\bar a^{*(i)}}{\bar a^{*(i)}+\detone{i'}}\frac{\pdetlax{i}}{\pdetlax{1} \left(\frac{\bar a^{*(1)}}{\bar a^{*(1)}+\detone{2}}\right)+\pdetlax{2}\left(\frac{\bar a^{*(2)}}{\bar a^{*(2)}+\detone{1}}\right)}, \\
p_{\mathrm{d},\mathrm{taut}}^{(i), \mathcal A} &= \frac{\bar a^{*(i)}}{\bar a^{*(i)}+\detone{i'}}\frac{\pdettaut{i}}{\pdettaut{1}\left(\frac{\bar a^{*(1)}}{\bar a^{*(1)}+\detone{2}}\right)+\pdettaut{2}\left(\frac{\bar a^{*(2)}}{\bar a^{*(2)}+\detone{1}}\right)}, \\
p_{\mathrm{d}}^{(i),\mathcal A} &= \frac{\bar a^{*(i)}}{\bar a^{*(i)}+\detone{i'}}\frac{\pdet{i}}{\pdet{1}\left(\frac{\bar a^{*(1)}}{\bar a^{*(1)}+\detone{2}}\right)+\pdet{2}\left(\frac{\bar a^{*(2)}}{\bar a^{*(2)}+\detone{1}}\right)}. \label{eq:pdetia}
\end{align}
\end{subequations}

Expected values that are unconditioned on $r_0$ can be found by integration against the probability density for the motor separation at attachment $ \Ratt{} $, as in (\ref{roavg}), but the probability density $ p_{\Ratt{}} $ now depends on the identity of the motor attaching into the \TAM\ state as indicated by which substate $ (1,2;j) $ is realized in the current cycle.

 \section{Simulations}  \label{sec:numeric} 
 We now compare the effective statistics derived in this paper
against direct simulations of the model described in Section~\ref{sec:model}. Both homogeneous (two kinesin-1 or two kinesin-2) and heterogeneous (one kinesin-1 and one kinesin-2) ensembles of two motors are simulated, along with optical trap forces of $F_T = -2, 0$ and $2$ pN. The parameters for all of the simulations are given in Table \ref{paramtable}, including the optimized parameters related to detachment found in Section~\ref{sec:opt}. For each ensemble considered, we simulate  $S = 1000$ runs.  We begin with a completely detached  ensemble and select motor $i$ to attach with probability $a_i/(a_1+a_2)$. At this point, the ensemble then begins to process in state $(i)$, with the initial conditions of the cargo and detached motor selected with respect to stationary distributions $\rho_i(y, F_T)$.  If  complete detachment occurs before entering state $(1,2)$, we begin the process afresh. Otherwise we run until complete detachment and record the relevant simulated data.  The time step is set to $\Delta t = 10^{-6}$s, which was selected for stability and accuracy  reasons to be smaller than the  time scale $\gamma_m/\kappa$ of drift for unattached motors.

 Comparisons of effective run times and lengths,  are summarized in  Table \ref{resultdexp}. From the direct simulation data,  estimates of  mean number of cycles, run times, and run lengths are found through sample means. For estimating ensemble statistics $ \Vpool $ and $ \Dpool$,  from the dataset $\Theta = (T^{(j)}, L^{(j)})_{j
= 1, \dots,  S}$ of sample run times and run lengths, we drew $B = 1000$ bootstrapped datasets $\Theta^*_b = (T^{(j)}_b, L^{(j)}_b)_{j= 1,\dots,S}$ for $b= 1, \dots,  B$. For long run statistics $\Vrun$ and $\Drun$, we repeat this procedure with a thinned dataset of $\Theta$ of size $S/10 = 100$ containing the top decile of run lengths.

The theoretical results follow from associating each of the test cases to truly, neutrally, or falsely cooperative ensembles as discussed at the end of Subsection~\ref{sub:tamdecomp}.
We compare the same simulations in Appendix~\ref{app:nolax} with the theoretical framework described in \citet[App. A]{klobusicky2020cooperative} which assumes the motor spatial dynamics always reach the stationary distribution within each state quickly relative to the time scale of attachment or detachment.  That is, the theoretical predictions in Appendix~\ref{app:nolax} neglect the finer considerations developed in Section~\ref{sec:laxtaut} which are represented in the theoretical results in Table~\ref{resultdexp}.   The agreement between the refined theory and simulations is far superior across the board to that of the previous theory.  For all but the diffusivities, the discrepancy is within a few percent, though sometimes several standard errors.  This likely points to a quantitatively small but noticable bias incurred in the theory due to the approximations made.  The effective diffusivity of the ensembles are naturally less precisely computed by simulations, and  we simply note their agreement with the theoretical expressions within a few standard errors for all cases except kinesin-1/kinesin-1 ensembles under the $ F_T = 2 \si{\pico\newton} $ trap force.  There the theory overpredicts the diffusivity seen in simulations by about 50\%.  We do see for most other ensembles that the theoretical predictions for the diffusivity tend to be above those of the simulations to a lesser extent.  
\begin{table}
\resizebox{\columnwidth}{!}{
\begin{tabular}{|c||c|c||c|c||c|c|} \hline 

 &  \multicolumn{2}{|c||}{$F_T = -2$ pN }   &\multicolumn{2}{|c||}{$F_T =
0$ pN
}    &\multicolumn{2}{|c|}{$F_T = 2$ pN}\\\hline
 & \textbf{Simulation}&  \textbf{Theory} &\textbf{Simulation}  & \textbf{Theory}&
 \textbf{Simulation} &\textbf{Theory }\\ \hline
  & \multicolumn{6}{|c|}{\textbf{kinesin-1/kinesin-1}} \\ \hline
 $\Vrun$ &$1130\pm20  $& 1190 &$798\pm2 $&808 
&$573\pm 3$ & 567\\
\hline
$\Vpool$ & $1101\pm 8$& 1103&$794\pm 1$ &807 &$576\pm 2$&
567\\\hline
$\Drun$ & $16000\pm2000$ & 18100&$4300\pm600$&$5600$ &$4000\pm 500$&
5900\\\hline
$\Dpool$ &$12900 \pm 700$& 13400 &$4300\pm200 $&4800
&$3700\pm 200$ & 5400\\
\hline
$\E[\Ncyc]$ & $1.52\pm.03$& 1.51&$5.82\pm .2$ &5.79&$4.20\pm .1$&
4.25\\\hline
$\E [\Rtimegen] $ & $.39\pm.01$ & .38&$4.7\pm.1$&$4.5$
&$3.1\pm .1$&
3.1\\\hline
$\E [\Rlengthgen]$ & $431\pm9$ & 424&$3700\pm100$&$3600$
&$1760 \pm 50$&
1770\\\hline  
 & \multicolumn{6}{|c|}{\textbf{kinesin-2/kinesin-2}} \\\hline
 $\Vrun$ &$860\pm10 $& 877 &$494 \pm2 $&498
&$0\pm 10$ & 0\\
\hline
$\Vpool$ & $860 \pm6$& 856&$494\pm1$ &498 &$16 \pm 8$&
22\\\hline
$\Drun$ & $16000 \pm2000$ & 18000&$3200\pm600$&$2700 $ &$17000\pm 2000$&
21000\\\hline
$\Dpool$ &$13900\pm800$& 15900 &$2800\pm300 $&2500
&$16000 \pm 1000$ & 18000\\
\hline
$\E [\Ncyc]$ & $3.7\pm.1$& 3.9&$19.0 \pm.8$ &18.44 &$3.26 \pm .09$&
3.28\\\hline
$\E [\Rtimegen] $ & $ .78\pm.02$ & .83&$11.4\pm.5$&11.7 &$.61\pm .02$&
.62\\\hline
$\E [\Rlengthgen]$ & $670\pm20$ & 710& $5600\pm300$&$5800$ &$10 \pm 5$&
$14$\\\hline
 & \multicolumn{6}{|c|}{\textbf{kinesin-1/kinesin-2}} \\\hline
 $\Vrun$ &$960\pm 20$& 1030 &$673\pm3 $&672
&$429\pm7$ & $442$ \\
\hline
$\Vpool$ & $1000\pm10 $& 1000&$662\pm 2$ &$660$&$401\pm 4$&400\\
\hline
$\E [\Ncyc]$ & $2.24\pm .05$& 2.18&$9.9\pm .3$ &$10.1$ &$3.20\pm
.08$& 2.88\\\hline
$\E[\Rtimegen] $ & $.54\pm.01$ & .49&$5.1\pm.2$&$5.5 $ &$1.39\pm
.04$& 1.38\\\hline
$\E [\Rlengthgen]$ & $520\pm10$ & 490&$3400\pm100$&$3600$
&$560\pm 20$&
550\\\hline  
\end{tabular}
}
\caption{\textbf{Comparison of Model Simulations with New Theory.}
Units of time and distance are measured
in seconds and nanometers, respectively. 
For the simulated estimates of the mean number of complete cycles $\Ncyc$, 
run time $\Rtimegen$,
and run length $\Rlengthgen$ we report the sample mean and standard
error
from 1,000 samples.  The errors in the ensemble and long-run
velocities ($\Vpool$ and $\Vrun$) and diffusivities ($\Dpool$ and $\Drun$)
are obtained
through bootstrap sampling as described in the main text. } \label{resultdexp}
\end{table}

\section{Discussion}
\label{sec:conc}
We have addressed two shortcomings from our original work~\citep{klobusicky2020cooperative} on the analysis of cooperative transport of two molecular motors which could be of the same type (homogenous ensemble) or different types (heterogenous ensembles).  The first, addressed in Section~\ref{sec:opt}, concerns the model for the relation between detachment rates of the motor and the instantaneous force it experiences.  In our previous work as well as others~\citep{bovyn2020attach,bovyn2021diffusion}, this was modeled by some version of a direct fit to the data from~\citet{milic2017intraflagellar,milic2014kinesin,andreasson2015elife}  which relates detachment rates (inferred from measured run lengths and velocities) to the applied trap force.  This neglects the effects of fluctuations in the instantaneous force, and also leads to a discontinuity or sharp transition in the detachment rate as the force on the motor switches from the hindering to the assisting direction.  Applying a model with this sharp transition in detachment rates leads to grossly inaccurate behavior at low hindering or zero trap forces due to force fluctuations into the assisting force direction which dramatically accelerates detachment.  A similar artefact of a discontinuous detachment rate model is shown for gliding assay models in~\citet{hou26gliding}.  We addressed this inconsistency by simply fitting  the data from~\citet{milic2017intraflagellar,milic2014kinesin,andreasson2015elife}  on run lengths as a function of trap forces to the predictions of a model which uses a double-exponential parameterization of the instantaneous force-detachment rate relation and accounts for force fluctuations from the cargo.  The parameterized fit for the instantaneous force-detachment rate based on single molecular motor runs of kinesin-1 and kinesin-2 are then used self-consistently for the instantaneous force-detachment rate relation in our forward predictions for ensembles of two motors.   

More recent experiments on the relationship between detachment rates and  forces on molecular motors have pointed to the importance of considering both the longitudinal force (parallel to the microtubule) and transverse force (perpendicular to the microtubule) on the motor.   In optical trap assays such as the experimental data we reference~\citep{milic2017intraflagellar,milic2014kinesin,andreasson2015elife}, the cargo tends to be rather large (hundreds of nanometers) relative to the motor so trap forces applied longitudinally also induce substantial transverse forces.  In other experimental configurations where the transverse force is reduced, the dependence of the detachment rate on the applied longitudinal force appears to be qualitatively different.  In particular, the detachment rate is found in~\citet{howard2019direction,hancock2026tensio} but not in~\citet{yildiz2026lowz} to exhibit catch bond rather than slip bond behavior, meaning the detachment rate actually decreases with applied longitudinal force when in the hindering direction.    In cells, we do not expect the large cargos and large transverse forces typical of conventional optical trap assays, and thus these more recent experiments are likely to be more informative about how molecular motors including kinesin work in the cell.  We are not aware, though, of studies with reduced transverse force that map out a simply parameterized relationship between detachment rates and applied forces as cleanly and comprehensively as in~\citet{milic2017intraflagellar,milic2014kinesin,andreasson2015elife}, and thus we could not yet adapt our theoretical approach to their data.  We should thus understand our parameterization in Table~\ref{paramtable} as applying to \textit{in vitro} optical trap experiments with large cargo.  Our framework should however be adaptable in principle to other experimental configurations with reduced transverse forces provided we have available a sufficiently rich measurement of detachment rates (or run lengths and velocities) as a function of  longitudinal force.   We could retain our one-dimensional structure by using the approach in~\citet{hou26gliding} of leaving the transverse force implicitly dependent on the longitudinal force via geometric relations presented in~\citet{ostap2020geometry} which assume the cargo stays close to the microtubule, or by simply assuming the transverse force is negligible for example in tensiometer studies~\citep{hancock2026tensio}.  If data became available to model the detachment rate jointly as a function of longitudinal force and transverse force, it could be incorporated into our modeling framework provided we extended the model for the cargo variable to be at least two-dimensional or fully three-dimensional.  This would not be difficult to do by maintaining the cargo as a point particle in higher dimensions, but a proper accounting of the geometry in two or three dimensions would track the dynamics of the point of attachment of the motor to the cargo, as done theoretically in~\citep{choudhary2026attach}  and computationally in~\citep{gross2011arranged,grossmiles2024binding}.

The other unrealistic simplification in the model from~\citet{klobusicky2020cooperative} is the use of a simple linear force law for the tether between the motor and cargo.  In the present work, this was modified to be slack below a rest length and to exert linear tension when stretched, as is common in computational studies of motor-cargo systems~~\citep{arpag2019kin1kin3,arpag2014populations,bovyn2021diffusion,gross2011arranged,kunwar2008stepping,kunwar2016anisotropy,furuta2013measuring,korn:2009,grossmiles2024binding}.  While the extension of the methodology from~\citet{klobusicky2020cooperative} to nonlinear tether forces is in principle straightforward and addressed in the appendix to that work, the fact the tether exerts no force when its length is short enough creates a fundamental challenge to the asymptotic framework based on an assumed well-separated time scales.  In particular, we had assumed the dynamics of the motors is fast compared to the attachment/detachment time scale, which is true for kinesin-1 and kinesin-2 in \textit{in vitro} experiments when the tether is engaged by being stretched above the rest length.  But when one or both tethers are slack, the motor separation is governed by the velocity and diffusivity of the motors, and we find in Table~\ref{tab:laxtauttimes} the time scale for two attached motors to move via such slack tether dynamics to nominally reach a stationary distribution is comparable to the detachment time scale.  

To address this, we relaxed the assumption that the relative separation of the two motors, in the \TAM\ state where they are both attached, reaches the stationary distribution before detachment is likely to occur.  We decomposed this \TAM\ state into a lax phase where at least one motor has a slack tether and a taut phase where the tethers of both motors to the cargo are engaged.  The nature of these phases depends on whether the properties of the two motors create true cooperativity in the sense of leading to load-sharing or whether they tend toward opposition.  Accordingly, the taut phase can be analyzed either in terms of quickly reaching a stationary distribution or of quickly leading to detachment via a semi-deterministic approximation.  The lax phase can be analyzed in terms of constant drift and diffusivity of the motors within this phase.  \citet{diez2023reversals} found that in in vitro tug-of-war configurations between kinesin and dynein, the kinesin often have slack tethers.  Thus, the lax/taut decomposition of attachment states that was developed here for cooperative motor transport may also be relevant for tug-of-war configurations between antagonistic motors.  While conceptually simple, the calculations in Section~\ref{sec:laxtaut} appear cumbersome because of the diversity of approaches needed for different cooperativity scenarios and the assembling of various possible scenarios based on wheter detachment occurs in the lax phase or the taut phase of the \TAM\ state.  Moreover, for heterogenous ensembles in most cases, the lax phase dynamics depend essentially on which motor attached most recently, which is the same as the one that detached from the previous \TAM\ state.  This extra memory breaks the simple renewal structure for the visits to various states of attachment, and required a different recursive approach.  Both because of this loss of the renewal structure and complications regarding correlations between motor separation and cargo process in the lax phase of the \TAM\ state, the computation of second order transport statistics of the cargo such as diffusivity became impractical (if not technically impossible) to compute for heterogenous ensembles.  

In Section~\ref{sec:numeric} we showed that these  extensions give much better agreement between theoretical predictions of transport statistics with direct simulations of our model than the theory from~\citet{klobusicky2020cooperative}.   Moreover, our results in Appendix~\ref{app:nolax} showed that including the complications of the lax phase are important in describing the dynamics in the \TAM\ state.  We do see a general theoretical overprediction of effective diffusivity of the cargo-motor complexes, which is likely due to approximations made in simplifying the dynamics within an attachment state to lax or taut.  The theoretical analysis is sufficiently complicated with these approximations that attempting to improve them by a more detailed model doesn't appear feasible.   Most \textit{in vitro} experimental information on the interaction of cooperative motor pairs appears to be available for kinesin-1/kinesin-1 teams, though they vary considerably.  Both our theory and model predictions in Table~\ref{resultdexp}, by contrast, predict run lengths for unloaded kinesin-1/kinesin-1 (and other ensembles) of several micrometers, which is likely connected to the prediction of several ($\geq 5$) detachment/reattachment cycles.   This appears consistent with the findings in \citet{furuta2013measuring}, but not~\citet{feng2018motor,diehl2009interference}  where the mean run lengths under no load are found to be approximately $ 1.4-1.6 \si{\micro\metre} $, about twice as long as the run length of a single kinesin-1.    \citet{feng2018motor} find that unloaded kinesin-2/kinesin-2 ensembles have run lengths of $ 2.38 \pm 0.26 \si{\micro\meter}$, about 4 times that of single kinesin-2 run lengths.  Our model consistently predicts a greater enhancement for kinesin-2/kinesin-2 ensembles than kinesin-1/kinesin-1 ensembles, but again by too great a factor (likely associated to the large number $ \sim 19 $ of expected detachment/reattachment cycles).   Naturally our model predictions for the mean run time also substantially exceeds those from the experiments of~\citet{feng2018motor}:  $ 1.86 \pm 0.31 \si{\sec} $ for kinesin-1/kinesin-1 ensembles and $ 2.51 \pm 0.39 \si{\sec} $ for kinesin-2/kinesin-2 ensembles.    Our model predicts that the run length for unloaded kinesin-1/kinesin-2 ensembles should be comparable to those of kinesin-1/kinesin-1 ensembles, while~\citet{feng2018motor} find in their experiments that the former have somewhat longer run lengths ($2.18 \pm 0.39 \si{\micro\metre} $ vs. $ 1.62 \pm 0.23 \si{\micro\metre} $).   

On a more positive note, our model prediction that the kinesin-1/kinesin-1 ensemble velocity is comparable to that of a single kinesin-1 is consistent with some experimental observations~\citep{furuta2013measuring,diehl2009interference}, but not others which see a decrease~\citep{Jamison:2010}.  One reason for the variability in experimental results and discrepancy from our idealized model predictions is that the geometry of the coupling of the motor pair can have a strong influence on the dynamics via the effective reattachment rate~\citep{bovyn2020attach,furuta2013measuring,gross2011arranged,bovyn2021diffusion}.  The effective tether lengths could also be shorter  (perhaps 30 to 50 $\si{\nano\metre}$)  in some of the experimental setups relative to our theoretical model value of ($ \SI{80}{\nano\metre} $), which would reduce the role of the lax phase where the motor detachment rates are low.  \citet{feng2018motor} note the finite microtubule lengths in experiment would create a downward bias relative to theoretical values based on infinitely extended microtubules, but this bias is estimated to be about 30\% which is considerably less than the discrepancy from our theoretical predictions.

Finally, we remark that \citet{holmescerfon2022mass} employed similar homogenization approaches as in our previous work~\citep{klobusicky2020cooperative} to compute the effective diffusivity of a ``nanocaterpillar'' system consisting of a particle with ligands that bind and unbind reversibly with constant rates to receptors on a one-dimensional substrate, showing that the particle mass can affect the effective diffusivity of the particle even using overdamped dynamics for the system.  The particle-ligand system in that work can be viewed as a parallel version of the motor-cargo complex studied here, with the particle, ligands, and substrate playing the role of the cargo, motors, and substrate, respectively.  The ligands in~\citet{holmescerfon2022mass} are passive and thus in particular not directional, which is why the effective drift of the particle is zero.  The rather explicit expressions in~\cite{holmescerfon2022mass} for the effective dynamics of the particle with small and large number $N$ of ligands, including a fully inertial treatment, do not appear similarly achievable for motor-cargo complexes due to the complications engendered by the force-dependence of the motor velocity as well as their detachment rates.  

\section{Acknowledgments}
Conversations with William Hancock were helpful in understanding issues with the experimental data used to develop our revised relationship between instantaneous force and detachment rate in Section~\ref{sec:opt}.    The work of JF and PRK was partially supported by National Institutes of Health grant R01GM122082.
The work of JK is partially supported by the National Science Foundation
under Grant No. 2316289.
\newpage
 
 \begin{appendices}

\section{Cargo Position corrections}
\label{app:zshift}

As discussed in Section~\ref{sec:cycle}, run lengths are collected from recording cargo positions at the beginning and ends of runs.  For the initial conditions, we record the cargo position $Z(0)$ immediately after  entry into the \OAM\ state. To recover the cargo position from the cargo tracking variable, we add a term drawn from the stationary distribution between the originally attached motor (which we label as having index $ J_{\mathbf a}$) and cargo:
\begin{equation}
Z(0) = \CM (0)+\Delta Z_{\mathbf a} , \qquad \Delta Z_{\mathbf a} = Y_{J_{\mathbf a}}^{\mathrm{a}}, 
\end{equation}
with $Y_{i}^{\mathrm{a}} \sim \rho_i(y;F_T)$ given in Eq.~\eqref{eq:att1stat}.

Denoting $J_{\mathbf t}$ for the index of the last motor to \uline{detach} in a run,  a similar adjustment for cargo position at the termination of a cycle gives \begin{equation}
Z(T) = \CM (T)+ \Delta Z_{\mathbf t}, \qquad \Delta Z_{\mathbf t} =   Y_{J_{\mathbf t}}^{\mathbf{t}}.
\end{equation}
Here, $Y_{i}^{\mathbf{t}}$ and $Y_{i}^{\mathbf{a}}$ are iid.

 \section{Effects of drag force}
 \label{app:drag}
 In vitro, we can estimate the drag force $ \Fdrag
$ from Table~\ref{paramtable} as no larger than $ \gamma v^{(i)} \lesssim
\SI{0.01}{\pico\newton} $.  Incorporating the drag force in the quasi-stationary
distribution would be equivalent to adding it to the trap force $ F_T \rightarrow
F_T + \Fdrag $, leading to a (further) exponential tilt in the quasi-stationary
distribution.  Under the cable force model~\eqref{eq:cable} for the tether,
the quasi-stationary distribution is supported primarily on tether extensions
up to $ \ellrest{i} + \sqrt{k_B T/\kappa^{(i)}} $, so the drag force would
change the quantity in the exponent of the stationary distribution by $ 
\frac{\Fdrag}{k_B T} (\ellrest{i} + \sqrt{k_B T/\kappa^{(i)}}) \lesssim \frac{\SI{0.01}{\pico\newton}}{\SI{4}{\pico\newton
\nano\metre}} \SI{100}{\nano\metre} \sim \frac{1}{4} $.  Put another way,
a probability distribution proportional to $ \expe^{-\Fdrag y/k_B T} $ on
$ |y| \leq \ellrest{i} + \sqrt{k_B T/\kappa^{(i)}} $ would have a mean $
-(\ellrest{i} + \sqrt{k_B T/\kappa^{(i)}}) \coth (\Fdrag (\ellrest{i} + \sqrt{k_B
T/\kappa^{(i)}})/k_B T)+k_B T/\Fdrag \approx -\frac{1}{3} k_B T/\Fdrag  (\Fdrag
(\ellrest{i} + \sqrt{k_B T/\kappa^{(i)}})/k_B T)^2  $ which is no larger
than a $ \SI{8}{\nano\metre} $ negative shift.
Thus the drag force effect on biasing the cargo to lag behind the motor positions
is small but not entirely negligible in vitro.  We could have included it
in an ad hoc way by the replacement $ F_T \rightarrow F_T + \gamma V $, where
$ V $ is the actual velocity of the motor-cargo complex.  But as this would
only create a small correction to our theoretical predictions at the cost
of additional complexity, and does not appear to generate better agreement
with simulated transport statistics, we will exclude this drag-biasing effect
of the cargo distribution from further consideration.  In cell, of course,
drag effects are stronger and disrupt the scale separation assumptions underlying
our theoretical approach~\citep{mckinley2012asymptotic}.

\section{Analytic expressions for some statistics} 
\label{sec:rhointegral}
Here we collect some evaluations of key statistics associated to the cargo tracking variable in various states.  These all amount to 
evaluations of integrals of piecewise Gaussians, and are thus expressible in terms of the functions:
\begin{align}
\mathcal G(a,b,c,y_{\ell},y_{\mathrm{u}}) \equiv  \int_{y_{\ell}}^{y_{\mathrm{u}}} \exp(-a(y+b)^2+c)dy\\
=  \sqrt{\frac{\pi}{a}}\frac{e^c}2 \mathrm{(erf}((b+y_{\mathrm{u}})\sqrt
a)-\mathrm{(erf}((b+y_{\ell})\sqrt
a)),\\
\mathcal G_1(a,b,c,y_{\ell},y_{\mathrm{u}}) \equiv  \int_{y_{\ell}}^{y_{\mathrm{u}}} y \exp(-a(y+b)^2+c)dy \\=  \frac{e^c}{2a}\left(\exp(-(b+y_{\ell})^2a)-\exp(-(b+y_{\mathrm{u}})^2a)
\right)-b\mathcal G(a,b,c,y_{\ell},y_{\mathrm{u}}), \\
\mathcal G_2(a,b,c,y_{\ell},y_{\mathrm{u}}) \equiv \int_{y_{\ell}}^{y_{\mathrm{u}}} y^2 \exp(-a(y+b)^2+c)dy \\
= \frac{e^c}{2a} \Big[((b+y_{\ell})\exp(-(b+y_{\ell})^2a))-2(b+y_{\mathrm{u}})\exp(-(b+y_{\mathrm{u}})^2a))\Big]\\
-2b\mathcal G_1(a,b,c,y_{\ell},y_{\mathrm{u}})+ \left(\frac{1}{2a}-b^2\right)\mathcal G(a,b,c,y_{\ell},y_{\mathrm{u}}),
\end{align}
the error function 
\begin{equation}
\mathrm{erf}(x) = \frac 2{\sqrt{\pi}}\int_0^x e^{-t^2}dt,
\end{equation}
and elementary functions.  The primary purpose of the analytical expression of the statistics is to expedite the numerical computation, particularly for the parameter optimization in Section~\ref{sec:opt} which involves repeated evaluation of the statistics for different parameter choices for the detachment rate.  

\subsection{Mean Cargo Lag Behind Single Motor}
\label{app:zlagformula}
By subdividing the integral together with the normalizing constant in the distribution~\eqref{eq:att1stat} for the displacement between the cargo and a single attached motor into the intervals $ [\ellrest{i},\infty) $, $[-\ellrest{i},\ellrest{i}] $, and $ (-\infty, -\ellrest{i}) $, and noting the symmetry between the semi-infinite integrals, we obtain the expression for the average displacement of the cargo from the single attached motor $ i $ as:
\begin{align}
&\Zlag_i(F_T)  \equiv \int_{-\infty}^{\infty} y\rho_i (y;F)\,
\difd y =\frac{J_1 (F_T)+J_2(F_T)-J_1(-F_T)}{I_1(F_T)+I_2(F_T)+I_1(-F_T)}. \label{eq:carglag}
\end{align}
where
\begin{align}
I_1 (F_T) &\equiv  \int_{\ellrest{i}}^\infty \exp\left(-[\kappa^{(i)}(y-
\ellrest{i})^2/2+F_Ty]/k_BT \right)dy\nonumber \\
&= \mathcal G\left(a_1,b_1,c_1,\ellrest{i},\infty\right),  \label{eq:abcdefirst} \\
I_2 (F_T)  &\equiv  \int_{-\ellrest{i}}^{\ellrest{i}} \exp\left(+F_Ty]/k_BT \right)dy =\frac{2k_BT}{F_T}
\sinh \frac{F_T\ellrest{i}}{k_BT}, \nonumber \\
J_1 (F_T) &\equiv \int_{\ellrest{i}}^\infty y \exp\left(-[\kappa^{(i)}(y-
\ellrest{i})^2/2+F_Ty]/k_BT \right)dy \nonumber \\
&= \mathcal G_1\left(a_1,b_1,c_1,\ellrest{i},\infty\right), \nonumber \\
J_2 (F_T) &\equiv  \int_{-\ellrest{i}}^{\ellrest{i}}
 y\exp(-(F_T/k_BT) y)dy \nonumber \\
 &= \frac{2(\kbt)^2}{F_T^2}\left[\sinh\left(\frac{\ellrest{i}F_T}{k_BT}
\right)-\frac{\ellrest{i}F_T}{k_BT}\cosh\left(\frac{\ellrest{i}F_T}{k_BT}\right)\right], \nonumber
\end{align}
and the arguments are:
\begin{align}
a_1 \equiv \frac{\kappa^{(i)}}{2\kbt}, \quad b_1 \equiv F_T/\kappa^{(i)}- \ellrest{i},\quad c_1 \equiv \frac{\kappa^{(i)}}{2\kbt}\left( 
(F_T/\kappa^{(i)})^2-2\ellrest{i}F_T/\kappa^{(i)}
\right). \label{eq:firstargs}
\end{align}

\subsection{Explicit formula for moments of motor separation upon attachment}
\label{app:rattformula}
We immediately obtain the mean of $\Ratt{i}$~\eqref{convr}, the signed separation between motors upon attachment of motor $i$, as:
\begin{align}
\mathbb{E}[\Ratt{i}(F_T)]  = (-1)^{i+1}( \mathbb{E}[ \Theta^{(i')}_1(F_T)]-\mathbb{E}[\Theta^{(i)}_2(0)])
 = (-1)^{i+1}\Zlag_{i'}(F_T).  \label{eq:rattformula}
\end{align}

Then we can compute
the second moment as 
\begin{equation*}
\mathbb{E}[(\Ratt{i}(F_T))^2]  = ( \mathbb{E}[( \Theta^{(i')}_1(F_T))^2]+\mathbb{E}[(\Theta^{(i)}_2(0))^2]),
\end{equation*}
where
\begin{align*}
& \mathbb{E}[ \Theta^{(i)}_j(F_T)^2]= \frac{1}{\mathcal N_i}\left( \int_{-\infty}^{-\ellrest{i}}
y^2 \rho_i(y)dy
+ \int_{-\ellrest{i}}^{\ellrest{i}}
 y^2\rho_i(y)dy +  \int_{\ellrest{i}}^\infty y^2 \rho_i(y)dy
\right) \\
&\qquad = \frac{1}{I_1(F_T)+I_2(F_T)+I_1(-F_T)}
\left\{\mathcal G_2\left(a_1,b_1,c_1,\ellrest{i},\infty\right) \right. \\
& \qquad  + \left(\frac{k_B T}{F_T}\right)^3\left[
 \left(2\left(\frac{\ellrest{i}F_T}{k_BT}\right)^2+4\right)\sinh\left(\frac{\ellrest{i}F_T}{k_BT}\right) - 4\left(\frac{\ellrest{i}F_T}{k_BT}\right) \cosh\left(\frac{\ellrest{i}F_T}{k_BT}\right)\right] \\
& \qquad \qquad  \left.  +\mathcal G_2\left(a_1,b_{-1},c_{-1},\ellrest{i},\infty\right)\right\}
\end{align*}
where we introduced arguments corresponding to Eq.~\eqref{eq:firstargs} with the trap force reversed:
\begin{equation*}
 b_{-1} \equiv -F_T/\kappa^{(i)}- \ellrest{i},\quad c_{-1} \equiv \frac{\kappa^{(i)}}{2\kbt}\left( 
(F_T/\kappa^{(i)})^2+2\ellrest{i}F_T/\kappa^{(i)}
\right).
\end{equation*}

%

\subsection{Explicit formula for velocity in single motor state $\Vstate{i}$}
\label{app:velformula}
Since 
\begin{align}
g^{(i)}(F^{(i)}(-y)/F_s^{(i)}) =\begin{cases}0 & y<-\ellrest{i}-F_s^{(i)}/\kappa^{(i)},
\\
(1+\kappa^{(i)}(y+\ellrest{i})/F_s^{(i)}) & -\ellrest{i}-F_s^{(i)}/\kappa^{(i)}<
y \le -\ellrest{i},\\
1 & y >-\ellrest{i}. \\
\end{cases}
\end{align}
the  velocity of a motor averaged over cargo fluctuations~\eqref{gbari} can be written: 
\begin{align*}
\Vstate{i} &=  \frac{v^{(i)}}{\mathcal N_i} \left(\int_{ -\ellrest{i}-F_s^{(i)}/\kappa^{(i)}}^{\infty}
\tilde \rho_i(y)dy + \frac{\kappa^{(i)}}{F_s^{(i)}}\int_{-\ellrest{i}-F_s^{(i)}/\kappa^{(i)}}^{-\ellrest{i}}
(y+\ellrest{i})\tilde \rho_i(y)dy \right) \\
&=\frac{v^{(i)} \left[I_1(F_T)+I_2 (F_T)+\mathcal{G}(a_1,-b_{1},c_{-1},-\ellrest{i}-F_s^{(i)}/\kappa^{(i)},-\ellrest{i})\right]}{I_1(F_T)+I_2(F_T)+I_1(-F_T)} \\
& \qquad + \frac{v^{(i)} \kappa^{(i)}}{F_s^{(i)}} \frac{\mathcal{G}_1(a_1,-b_{1},c_{-1},-\ellrest{i}-F_s^{(i)}/\kappa^{(i)})+\ellrest{i}\mathcal G(a_1,-b_{1},c_{-1},-\ellrest{i}-F_s^{(i)}/\kappa^{(i)})}{I_1(F_T)+I_2(F_T)+I_1(-F_T)},
\end{align*}
where
\begin{equation*}
c_{-1} \equiv \frac{\kappa^{(i)}}{2\kbt}\left((F_T/\kappa^{(i)})^2+2\ellrest{i}F_T/\kappa^{(i)}\right).
\end{equation*}

\section{Force distributions  on motors}
\label{sec:cargoflucs}

In the single attached motor state, the force on attached motor $i$
is given by 
$F^{(i)}(X^{(i)}-Z) $. The distribution for the force is given by the hybrid distribution 
\begin{equation}
\rho_{F}^{(i)}(f) = \rho_{F,c}^{(i)}(f)+\rho_{F,a}^{(i)}(f),
\end{equation}
consisting of 
the continuous part\begin{align}
\rho_{F,c}^{(i)}(f)
&=\rho^{(i)}(-f/\kappa^{(i)}-\sgn(f)\ellrest{i})/\kappa^{(i)}  \\
 &=  \frac 1{\kappa^{(i)}\pdfnorm_i}\exp\left([-f^2/(2\kappa^{(i)}) +fF_T/\kappa^{(i)}
+\sgn (f)F_T \ellrest{i}]/k_BT\right).
\end{align}
and an atom at 0, with
\begin{equation}
\quad \rho_{F,a}^{(i)}(f) = \frac{2k_BT}{F_T\mathcal N_i}\sinh\left(\frac{F_T
\ellrest{i}}{k_BT}\right)\delta(f). \label{atomforce}
\end{equation}
For $ F_T =0 $, the continuous component $ \rho_{F,c} $ has the form of a
Gaussian centered at $ F_T =0$ with standard deviation $ \sqrt{k_B T \kappa_i}$.  
At most times motors feel no force, with the atom in~\eqref{atomforce} having a weight of .94 for both kinesin-1 and kinesin-2.  
  For $ F_T \neq 0$, the half of the Gaussian at forces opposite to the
trap force is suppressed by an exponential factor.  

 Under the change of coordinates $y = z-\frac{x^{(1)}+x^{(2)}}2$
and $r = x^{(1)}-x^{(2)}$, the force can
be written 
as
\begin{align}
F^{(i)}(x^{(i)}-z) &=F^{(i)}(-y+(-1)^{i+1}r/2) \\
&= \begin{cases}-\kappa^{(i)}(y+(-1)^{i}r/2-
\ellrest{i}) & y>(-1)^{i+1}r/2+
\ellrest{i}, \\
-\kappa^{(i)}(y+(-1)^{i}r/2+
\ellrest{i}) & y<(-1)^{i+1}r/2-
\ellrest{i}, \\
0 & \hbox{otherwise.} \ \\
\end{cases}  \nonumber
\end{align}From (\ref{rho2}) the
force for  two motors is then, for a fixed value of $r$,  
\begin{equation}
\rho_{1,2}(y;r) =  \frac 1{\pdfnorm_{1,2} (r)}\exp\left(-\frac{1}{k_BT}\left[\sum_{i
= 1}^2\kappa^{(i)}(|y+(-1)^{i}r/2|- \ellrest{i})_+^2/2+2F_Ty\right] \label{eq:cmsepjoint}
\right).
\end{equation}
Note that with our choice of coordinate change, we do not need to consider
a Jacobian term.

We now use the CDF method to represent the probability distribution for  $F_{1,2}^{(i)}$,  the force felt on motor $i$ in the two-motor attached state after the stationary distribution is achieved.  Defining 
$g(f,r) \equiv -f/\kappa^{(i)}-\sgn(f)\ellrest{i}+(-1)^{i+1}r/2$, we have for $ f \neq 0$
 \begin{align}
\mathbb P(F_{1,2}^{(i)}\le f) &= \int_\mathbb R \mathbb P(F_{1,2}^{(i)}\le
f|R = r))p_R(r)\\
&=\int_\mathbb R p_R(r)\mathbb P(Y>
g(f,r)|R = r))dydr \\
&=\int_\mathbb R p_R(r)\int_{g(f,r)}^{\infty}  \rho_{1,2}(y;r)dydr.   
\end{align}
where $ Y \equiv Z - \frac{X^{(1)}+X^{(2)}}{2} $.
Taking derivatives through the Leibniz rule, the density is then 
\begin{align}
\rho_{F_{1,2}^{(i)}} (f) =\frac{1}{\kappa^{(i)}}\int_{\mathbb R}p_R(r)\rho_{1,2}(g(f,r);r)dr \text{ for } f \neq 0.
\label{eq:twomotorforcegauss}
\end{align}


Finally, we can compute, for $[I_r^l, I_r^u] = [(-1)^{i+1}r/2-
\ellrest{i}, (-1)^{i+1}r/2+ \ellrest{i}],$
\begin{align}
\mathbb P(F_{1,2}^{(i)}=  0) &=\int_\mathbb R \mathbb P(Y \in [I_r^l, I_r^u]|R
= r)p_R(r)dr\\
 &= \int_\mathbb R  p_R(r)\int_{I_r^l}^{I_r^u}   \rho_{1,2}(y;r)dydr. 
\end{align}
The continuous component of the force distribution is a mixture of the densities $ \rho_{1,2} (g(f,r);r)  $ which have the shape of flat cores with Gaussian tails.  
Roughly speaking, for homogenous ensembles, we expect $ p_R (r) $ to be
primarily distributed over $ |r| \leq 2 \ellrest{i} $ over which the tethers
can be slack, with one (but usually not both) of the motor tethers sometimes
under tension, leading to a force distribution with a continuous component
similar to that of the case of a single motor. The estimate $\sqrt{k_BT\kappa^{(i)}}
= 1.0$ pN is a reasonable approximation to the numerical values of the standard
deviation for the continuous component of the force distribution for homogeneous
ensembles, with  .98 pN for both  kinesin-1  and  kinesin-2.   For
heterogenous motors with different speeds, they will tend to separate until
they reach mechanical equilibrium with forces resisting further separation. 
Taking motor $1 $ to have the faster unloaded velocity ($ v^{(1)} > v^{(2)}
$), under zero trap force $ F_T = 0$ the motors will separate until sufficient
force $\Feq $ on motor $ 1$ slows it down to the same speed as the lagging
motor $ 2 $:  $v^{(1)} (1-\Feq/F_s^{(1)}) = v^{(2)} $, which implies $ \Feq
= F_s^{(1)} (1- v^{(2)}/v^{(1)}) $.  We expect the continuous distribution
of the force on motor $ 1 $ to be centered at $ \Feq $ and on motor $2$ to
be centered at $ - \Feq $.  Both motor tethers would correspondingly be expected
to be under tension most of the time, and the continuous force distribution
in Eq.~\eqref{eq:twomotorforcegauss} would be dominated by contributions
from the integral over $ r $ where both quadratic terms in the exponent in Eq.~\eqref{eq:cmsepjoint} are nonzero.  If $ p_R
(r) $ were tightly distributed, we see from this expression that the
cargo fluctuations would induce a force fluctuation on the first motor with
variance $ (\kappa^{(1)})^2 \frac{k_B T}{\kappa^{(1)}+\kappa^{(2)}} $.  The
fluctuations $ \delta_R $ in $ R $ would contribute a fluctuation of size
$ \kappa^{(1)} \frac{1/\kappa^{(1)}}{1/\kappa^{(1)}+1/\kappa^{(2)}}
\delta_R = \frac{\kappa^{(1)}\kappa^{(2)}}{\kappa^{(1)}+\kappa^{(2)}}\delta_R$
to the force on motor 1.  So we expect the width of the continuous component
of the force distribution on motor 1 to be approximately $ \kappa^{(1)}\sqrt{
\frac{k_B T }{\kappa^{(1)}+\kappa^{(2)}} + \left(\frac{\kappa^{(2)} \delta_R}{\kappa^{(1)}+\kappa^{(2)}}\right)^2}$.
We can approximate the fluctuations of
separation~\eqref{eq:R} about the mechanical equilibrium described above
by approximating $ \Veltam{2} (r) $ as locally flat (since the lagging motor
$ 2 $ will be in the assisting force regime) and $ \Veltam{1}(r) $ as locally
linear with response coefficient given by the product of the sensitivity of motor velocity with respect to force and the sensitivity of the force with respect to separation:   $
  -\frac{v^{(1)}}{F_s^{(1)}} \frac{\kappa^{(1)} \kappa^{(2)}}{\kappa^{(1)}+\kappa^{(2)}}
$.  Thus
\begin{equation*}
U_- (r) \approx -\frac{v^{(1)}}{F_s^{(1)}} \frac{\kappa^{(1)} \kappa^{(2)}}{\kappa^{(1)}+\kappa^{(2)}}
(r-\req), 
\end{equation*}
where $ \req $ is the mechanical equilibrium separation of the motors, neglecting
cargo fluctuations.
Thus from the stationary distribution expression~\eqref{statr}:
\begin{equation*}
 \delta_R^2 \approx \frac{D^{(1)} + D^{(2)}}{\frac{v^{(1)}}{F_s^{(1)}} (1/\kappa^{(1)}+1/\kappa^{(2)})^{-1}}
 \end{equation*}
 so the width of the continuous force distribution on motor $1$ should be
approximately
 \begin{equation*}
 \sqrt{\frac{k_B T (\kappa^{(1)})^2}{\kappa^{(1)}+\kappa^{(2)}} + \frac{(D^{(1)}+D^{(2)})
\kappa^{(1)}\kappa^{(2)} F_s^{(1)}}{v^{(1)}(\kappa^{(1)}+\kappa^{(2)})}}.
 \end{equation*}
 Similarly the width of the continuous force distribution on motor $2 $ should
be approximately
 \begin{equation*}
\sqrt{\frac{k_B T (\kappa^{(2)})^2}{\kappa^{(1)}+\kappa^{(2)}} + \frac{(D^{(1)}+D^{(2)})
\kappa^{(1)} \kappa^{(2)} F_s^{(1)}}{v^{(1)}(\kappa^{(1)}+\kappa^{(2)})}}.
 \end{equation*}
Both expressions evaluate to $\SI{.80}{\pico\newton}$ (as $ \kappa^{(1)} = \kappa^{(2)} $),
which is approximately the numerical value of $\SI{.79}{\pico\newton}$ for  the standard deviation
of the continuous part~\eqref{eq:twomotorforcegauss} of the force distribution.  

\section{Motor separation distribution when detaching from the lax phase of \TAM\ state in truly and falsely cooperative ensembles}
\label{app:laxdetsep}
Since the detachment rate of motors is a fixed constant $ d_0 $ independent of their separation $R(t)$ in the lax regime of the \TAM\ state, the distribution of the motor displacement $ \Rdetlax $ at detachment in the lax phase is simply the conditional distribution of $ R(\Tdlax+\ttam) $ given the detachment occurs before the taut phase was reached..  It is moreover independent of which motor detaches first, but as indicated, depends on the index $j $ of which motor most recently attached. We express the lax phase dynamics of $ R(t) $ as:
\begin{equation*}
\difd R(t) = u_- \, \difd t + \theta \, \difd W (t), R(\ttam) = r_0,
\end{equation*}
for constants $ u_-> 0 $ and $ \theta > 0 $, and take the taut phase to correspond to $ \{r:r\geq r^{\ast}\} $ with initial separation $ r_0 < r^{\ast}$.   The results will apply as well to the case $ u_-< 0 $ with the taut phase $ \{r: r \leq r^{\ast}\} $ and initial displacement $ r_0 >  r^{\ast} $, since we can map this to the  calculation below via the substitutions $ u_- \rightarrow - u_- $, $ r \rightarrow -r $, $ r_0 \rightarrow - r_0 $, $ r^{\ast} \rightarrow - r^{\ast} $, which doesn't change the result~\eqref{eq:rdetlaxapp}.   The result thus applies to both truly and falsely cooperative ensembles (with $ u_- \neq 0$) under the one-sided approximation to the first passage time for truly cooperative ensembles in Subsection~\ref{laxhomnonz}.

We compute
the conditional distribution required for the intermotor displacement at detachment by first noting:
 \begin{align*}
\mathbb P(\Rdetlax\leq r) &:= \mathbb P(R(\Tdlax+\ttam) \leq r |\Tdlax< \Ttaut) \\ &=
\mathbb
P\left(R(\Tdlax+\ttam) \leq r \Big|\sup_{0\le
s\le \Tdlax} R(s+\ttam) \le r^{\ast}\right).
\end{align*}
The dependence on $j $ is implicit through the values of the parameters $ u_- $ and $ r^{\ast} $.

Thus we must find a distribution for a Brownian motion with drift, conditioned
on not yet having reached a certain value.  In the absence of drift, this is handled
through the reflection principle. To account for drift,  we use Girsanov's
theorem
to provide a change of probability measure $\mathbb Q$ such that $\tilde
R(t):=  (R(t)-r_0)/\theta$ is
a $\mathbb Q$-Brownian motion.  The relation between $\mathbb P$ and $\mathbb
Q$ is given through the Radon-Nikodym derivative
\begin{equation}
\frac{d\mathbb P}{{d\mathbb Q}} = \exp(cB(t)+(c^2/2) t), \quad  c = u_-/\theta.
\end{equation} 
 Let the differential interval $I_r= [r, r+dr]$  and   $Y(t)$ equal the running
maximum of $\tilde R(t)$. We
compute the joint probability of $Y$ and $\tilde R$ through 
  \begin{align}
&g(r,y, t)dr  :=\mathbb P(\tilde R(t) \in  I_r, Y(t) < y) \\
  &= \mathbb E_{\mathbb P}[\mathbf 1\{\tilde R(t) \in  I_r, Y(t) <y\}] \\
 &=  \mathbb E_{\mathbb Q}\left[\frac{d\mathbb P}{{d\mathbb Q}}\mathbf 1\{\tilde
R(t) \in  I_r, Y(t) <y\}\right] \\
 &= \mathbb E_{\mathbb Q}[\exp(cB(t)+c^2t/2 )\mathbf 1\{\tilde
R(t) \in  I_r, Y(t) <y\}] \\
 &=   \mathbb E_{\mathbb Q}[\exp(c\tilde R(t)-c^2t/2 )\mathbf 1\{\tilde
R(t) \in  I_r, Y(t) <y\}] \\
 &=   \exp(cr-c^2t/2)\mathbb E_{\mathbb Q}[\mathbf 1\{\tilde
R(t) \in  I_r, Y(t) <y\}] \\
 &=  \frac{\exp(cr-c^2t/2)}{\sqrt{2\pi t}}\left(\exp(-r^2/2t)-\exp(-(2y-r)^2/2t)\right)\mathbf1\{r\le
y, y\ge 0\}dr .
 \end{align}
 Combining this result with the conditional probability distribution for the detachment time, given detachment occurs in the lax phase, we obtain: 
  \begin{align}
p_{\Rdetlax}(r) &= \int_{0}^{\infty}\frac{ 
 g((r-r_{0})/\theta,   (r^*-r_0)/\theta, t)d_{0}e^{-d_0 t}}{\theta\mathbb
P( \Ttaut
> t)}dt \nonumber \\
&= 
\int_{0}^{\infty}\frac{d_0 \expe^{-d_0 t}}{\sqrt{2\pi\theta^2 t} \mathbb P(\Ttaut>t)} 
 \left[\expe^{-\frac{(r-r_0-u_- t)^2}{2\theta^2 t}}-\expe^{\frac{2(r^{\ast}-r_0)u_-}{\theta^2}}
\expe^{-\frac{(2r^{\ast}-r-r_0+u_- t)^2}{2\theta^2 t}}\right] dt
\label{pdetrlaxdist} \\
& \text{ for } r < r^{\ast}. \label{eq:rdetlaxapp}
 \end{align}

\section{Exponential force detachment with normal fluctuations}
\label{app:detcargoflucs}
We saw in Figure~\ref{xifig} that the average velocity of a motor depended on trap force in a manner that closely resembled the dependence on instantaneous force, but this is largely attributable to the piecewise linear form of the force-velocity model~\eqref{fvelrel} which causes the effects of fluctuations to cancel out except near zero force where the linear pieces meet. We argue here that the force fluctuations will play a more substantial role in the effective detachment rate model~\eqref{detmod} due to its piecewise exponential dependence on force.
As a simple illustration of how an exponential model of force detachment is affected by cargo fluctuations, we take a symmetric detachment rate function $ d(F) = d_0 \exp(|F|/F_{d})$ and average it against a  force fluctuation profile which superposes a
slack force $ F=0$ with probability $P_0$ with a normal force distribution
with mean $ F_T
$ and standard deviation $ \sigma = \sqrt{k_B T \kappa} $, 
we obtain 
\begin{align}
\langle d\rangle = P_0+(1-P_0)\frac{d_0}{\sqrt{2 \pi}\sigma }\int_{-\infty}^\infty
 \exp(|F|/F_{d}-(F-F_T)^2/(2\sigma^2))dF. \label{drateint}
\end{align}

The exponent in the integrand can be written using a  difference of squares
as
\begin{align}
-\frac{1}{2\sigma^2}(F-F_+)^2+ a_+   \qquad F>0, \\
-\frac{1}{2\sigma^2}(F-F_{-})^2+ a_- \qquad F<0,
\end{align}
where  $F_\pm = F_T\pm  \sigma \alpha$ and $a_\pm = \alpha^2/2 \pm \beta$, with force ratios defined:
\begin{align}
\alpha = \sigma/F_d, \quad \beta = F_T/F_d.
\end{align}

Integrating (\ref{drateint}) separately over positive and negative forces gives
\begin{align}
\frac{d_0}{\sqrt{2 \pi}\sigma }\int_{0}^\infty  \exp(|F|/F_{d})\exp(-(F-F_T)^2/(2\sigma^2))dF
 \\
= \frac{d_0}2  \exp( \alpha^2/2 + \beta)(1+ \mathrm{erf}(\beta/\alpha+
 \alpha) )
\end{align}
and
\begin{align}
\frac{d_0}{\sqrt{2 \pi}\sigma }\int_{-\infty}^0  \exp(|F|/F_{d})\exp(-(F-F_T)^2/(2\sigma^2))dF
= \\
= \frac{d_0}2\exp( \alpha^2/2 - \beta)(1- \mathrm{erf}(\beta/\alpha-  \alpha)
).
\end{align}
The ratio between the averaged and zero force detachment rate is then
\begin{equation}
\begin{split}
\langle d\rangle/d_{0}  &=   P_0+\frac 12 (1-P_0)(\exp( \alpha^2/2 + \beta)(1+
\mathrm{erf}((\beta/\alpha +\alpha)/\sqrt 2) )\\
& \qquad \qquad +\exp( \alpha^2/2 - \beta)(1-
\mathrm{erf}((\beta/\alpha -\alpha)/\sqrt
2)
 )).
 \end{split}
\end{equation}

This calculation was done for zero trap force in Subsection
S3 of~\citet{arpag2019kin1kin3}.   

\section{Sketch of Second Moment Calculations with Lax-Taut Decomposition}
\label{app:secmompain}
We illustrate here, in the context of truly cooperative ensembles studied in Subsection~\ref{tauthom},  the approach and  technical complexities in computing the second order statistics of cycle times and displacements in order to obtain the variances and covariances of run time and run length in~\eqref{varrun} for heterogenous ensembles.   Since, in contrast to~\citet{klobusicky2020cooperative}, we do not assume the motor separation quickly reaches a stationary distribution in the \TAM\ state, we must take into account the correlations between the motor separation $ R(t) $ and the cargo tracking variable $ M(t) $.  
This can be readily done  by decomposing the Brownian motion in the SDE~\eqref{eq:M} for $ \tilde{M}(t) $  in terms of its correlation with the noise $ W(t) $ in the SDE for $ \tilde{R}(t) $ and the independent standard Brownian motion $ \tilde{W} (t) \equiv  \frac{\sqrt{2D^{(1)}}}{\theta} W^{(2)} (t) + \frac{\sqrt{2D^{(2)}}}{\theta} W^{(1)} (t) $:
\begin{align}
\difd \tilde{M} (t) &=  u_+ \, \difd t + \frac{D^{(1)}-D^{(2)}}{\theta} \, \difd W(t) + \frac{2\sqrt{D^{(1)}D^{(2)}}}{\theta} \, \difd \tilde{W} (t)
\label{eq:MintermsR}\\
&= u_+ \, \difd t +\frac{D^{(1)}-D^{(2)}}{\theta^2} \, (\difd \tilde{R} (t)  - u_- \, \difd t) + \frac{2\sqrt{D^{(1)}D^{(2)}}}{\theta} \, \difd \tilde{W} (t).
\nonumber
\end{align}
Thus the tracking variable increment during the lax phase can be expressed in terms of the time increment $ \dTattblax $ and motor separation increment $ \dRattblax \equiv R(\ttam + \dTattblax) -r_0 $ as: \begin{equation*}
\dMattblax =  \left(u_+  + \frac{(D^{(2)}-D^{(1)})u_-}{2(D^{(1)}+D^{(2)})}\right)\dTattblax +\frac{D^{(1)}-D^{(2)}}{2(D^{(1)}+D^{(2)})} \, \dRattblax + \sqrt{\frac{2 D^{(1)} D^{(2)}}{D^{(1)}+D^{(2)}} } \, \dtWattblax
\end{equation*}
where $ \dtWattblax|\dTattblax \sim N(0,\dTattblax) $ is a conditionally independent random variable.

Now the second moment statistics involving the cargo tracking variable need to include the covariances between the random variable associated to a cycle:  
 \begin{align}
  \Var( \dCMattblax) &=  \left(\frac{D^{(1)}-D^{(2)}}{2(D^{(1)}+D^{(2)})}\right)^2  \Var[\dRattblax] + \frac{2 D^{(1)} D^{(2)}}{D^{(1)}+D^{(2)}}  \, \E[\dTattblax] \\
&  \qquad \qquad + \left(u_+ +\frac{(D^{(2)}-D^{(1)})u_-}{2(D^{(1)}+D^{(2)})}\right)^2 \Var[\dTattblax]  \\
& \qquad \qquad  + 2\left(u_+  + \frac{(D^{(2)}-D^{(1)})u_-}{2(D^{(1)}+D^{(2)})}\right)\frac{D^{(1)}-D^{(2)}}{2(D^{(1)}+D^{(2)})}  \Cov (\dTattblax, \dRattblax).
\\
  \E[\dTattblax\dMattblax] &=
   \left(u_+  + \frac{(D^{(2)}-D^{(1)})u_-}{2(D^{(1)}+D^{(2)})}\right)\E[(\dTattblax)^2] 
+   \frac{D^{(1)}-D^{(2)}}{2(D^{(1)}+D^{(2)})}\E[\dTattblax \dRattblax].
 \end{align}
 For heterogenous ensembles, which generally have different motor diffusivities $ D^{(1)} \neq D^{(2)}$, we would not only have several complex terms in these expressions, but we would also have to calculate the nontrivial covariance between the time in the lax phase $ \dTattblax $ and the change in motor separation $ \dRattblax $ in the lax phase.  This subcalculation itself would be quite complicated, due to the need to decompose into whether the lax phase is exited via detachment or entering the taut phase, and  the theoretical expression that would ensue would be tedious and of questionable value.  We therefore only compute second order statistics for homogenous ensembles ($D^{(1)} = D^{(2)} $), for which the increment in center of mass is independent of the change in motor separation, and the second order statistics have the tractable forms presented in the main text.   

\section{Experiments under purely stationary taut statistics}
\label{app:nolax}

\begin{figure}
\centering
  \includegraphics[width=\linewidth]{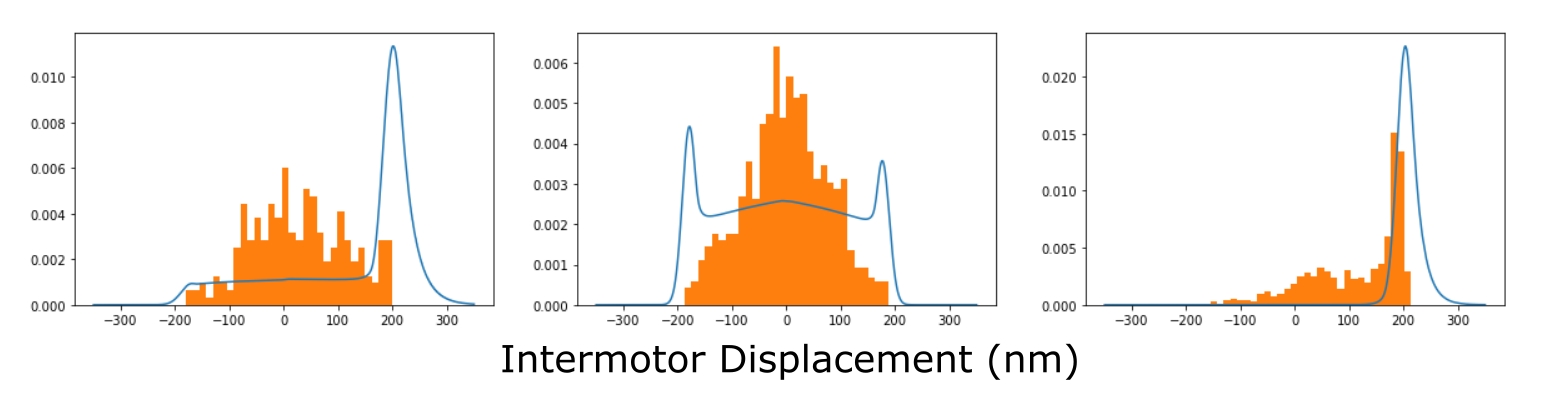}
\caption{\textbf{Distribution of motor separation upon detachment from \TAM\ state with zero trap force $F_T = 0$.}
Data for histograms are generated by collecting, over 100 simulations,  all detachment events from the \TAM\ state  until reaching the terminal state $\emptyset$. The probability densities in blue are calculated using the stationary distribution (\ref{rdetj}). \textbf{Left}: Kinesin-1/Kinesin-1 ensemble. \textbf{Center}: Kinesin-2/Kinesin-2 ensemble. \textbf{Right}:  Kinesin-1/Kinesin-2
ensemble.} \label{statrdisp}
\end{figure}

\begin{figure}
\centering
\includegraphics[width=.5\textwidth]{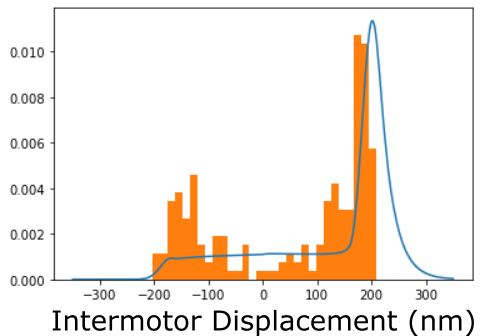}
\caption{\textbf{Distribution of motor separation upon detachment from taut phase of  \TAM\ state with zero trap force $F_T = 0$ for Kinesin-1/Kinesin-1 ensemble.}
Data for histograms are generated by collecting, over 100 simulations,  all detachment events from the \TAM\ state  until reaching the terminal state $\emptyset$, conditioned under the event that motors reach the taut state $|R| \geq  r^{\ast}_{\leftarrow\rightarrow}= 155.95 \hbox{ nm}$.  The probability density in blue is calculated using the stationary distribution (\ref{rdetj}). } \label{semirdisp}
\end{figure}

We present here some further analysis of the assumption that the motor-cargo complex reaches its stationary distribution within each attachment state, particularly including the \TAM\ state.   In Figure~\ref{statrdisp}, we show that the simulated distribution of motor separations upon detachment from the \TAM\ state are very different from the theoretical stationary distributions, particularly for the homogenous ensembles.  Figure~\ref{semirdisp} shows that the simulated distribution of motor separations upon detachment from a taut \TAM\ state agrees much better with the stationary distribution for homogenous kinesin-1 ensembles.  These figures support the arguments of the main text that we cannot assume that the stationary distribution is usually reached in the \TAM\ state before detachment, but a more defensible assumption is that the stationary distribution is achieved after the taut phase is reached.  Table~\ref{resultsde} compares the simulations from Section~\ref{sec:numeric} with theoretical calculations that assume the stationary distribution is always reached within each attachment state.  The agreement between theory and simulation is much less good than that of the main theory presented in Table~\ref{resultdexp}.  

\begin{table}
\resizebox{\columnwidth}{!}{
\begin{tabular}{|c||c|c||c|c||c|c|} \hline 

 &  \multicolumn{2}{|c||}{$F_T = -2$ pN }   &\multicolumn{2}{|c||}{$F_T =
0$ pN
}    &\multicolumn{2}{|c|}{$F_T = 2$ pN}\\\hline
 & \textbf{Simulation}&  \textbf{Theory} &\textbf{Simulation}  & \textbf{Theory}&
 \textbf{Simulation} &\textbf{Theory }\\ \hline
  & \multicolumn{6}{|c|}{\textbf{kinesin-1/kinesin-1}} \\ \hline
 $\Vrun$ &$1130\pm20  $& 2120 &$798\pm2 $&781 
&$573\pm 3$ & 625\\
\hline
$\Vpool$ & $1101\pm 8$& 2120&$794\pm 1$ &781 &$576\pm 2$&
625\\\hline
$\Drun$ & $16000\pm2000$ & 132000&$4300\pm600$&$10000$ &$4000\pm 500$&
2565\\\hline
$\Dpool$ &$12900 \pm 700$& 79000 &$4300\pm200 $&10200
&$3700\pm 200$ & 2487\\
\hline
$\E[\Ncyc+1]$ & $1.52\pm.03$& 1.51&$5.82\pm .2$ &6.87&$4.20\pm .1$&
4.25\\\hline
$\E [\Rtimegen] $ & $.39\pm.01$ & .12&$4.7\pm.1$&$2.56$
&$3.1\pm .1$&
.69\\\hline
$\E [\Rlengthgen]$ & $431\pm 9$ & 255&$3700\pm100$&$2000$
&$1760 \pm 50$&
432\\\hline  
 & \multicolumn{6}{|c|}{\textbf{kinesin-2/kinesin-2}} \\\hline
 $\Vrun$ &$860\pm10 $& 1022 &$494 \pm2 $&495
&$0 \pm 10$ & 350\\
\hline
$\Vpool$ & $860 \pm6$& 1022&$494\pm1$ &495 &$16 \pm 8$&
350\\\hline
$\Drun$ & $16000 \pm2000$ & 39700&$3200\pm600$&$5290 $ &$17000\pm 2000$&
1325\\\hline
$\Dpool$ &$13900\pm800$& 27700 &$2800\pm300 $&5225
&$16000 \pm 1000$ & 1128\\
\hline
$\E [\Ncyc]$ & $3.7\pm.1$& 3.88&$19.0 \pm.8$ &18.7 &$3.26 \pm .09$&
3.28\\\hline
$\E [\Rtimegen] $ & $ .78\pm.02$ & .17&$11.4\pm.5$&9.23 &$.61\pm .02$&
.20\\\hline
$\E [\Rlengthgen]$ & $670\pm20$ & 178&$5600\pm300$&$4572$ &$10 \pm 5$&
$70$\\\hline
 & \multicolumn{6}{|c|}{\textbf{kinesin-1/kinesin-2}} \\\hline
 $\Vrun$ &$960\pm 20$& 4521 &$673\pm3 $&663
&$429\pm7$ & $472$ \\
\hline
$\Vpool$ & $1000\pm10 $& 4480&$662\pm 2$ &$653$&$401
\pm 4$&372\\
\hline
$\E [\Ncyc]$ & $2.24\pm .05$& 3.00&$9.9\pm .3$ &$17.7$ &$3.20\pm
.08$& 4.14\\\hline
$\E[\Rtimegen] $ & $.54\pm.01$ & .05&$5.1\pm.2$&$1.86 $ &$1.39\pm
.04$& .27\\\hline
$\E [\Rlengthgen]$ & $520\pm10$ & 228&$3400\pm100$&$1218$
&$560\pm 20$&
128\\\hline  
\end{tabular}
}
\caption{\textbf{Comparison of Model Simulations with Previous Theory from~\citet{klobusicky2020cooperative}}. Units of time and distance are measured
in seconds and nanometers, respectively. 
For the simulated estimates of the mean number of complete cycles $\Ncyc$, 
run time $\Rtimegen$,
and run length $\Rlengthgen$ we report the sample mean and standard
error
from 1,000 samples.  The errors in the ensemble and long-run
velocities ($\Vpool$ and $\Vrun$) and diffusivities ($\Dpool$ and $\Drun$)
are obtained
through bootstrap sampling as described in the main text.  \label{resultsde}}
\end{table}

\section{Comparison of detachment models and parameter selection}
\label{app:compdetmodels}

\begin{figure}
  \includegraphics[width=\linewidth]{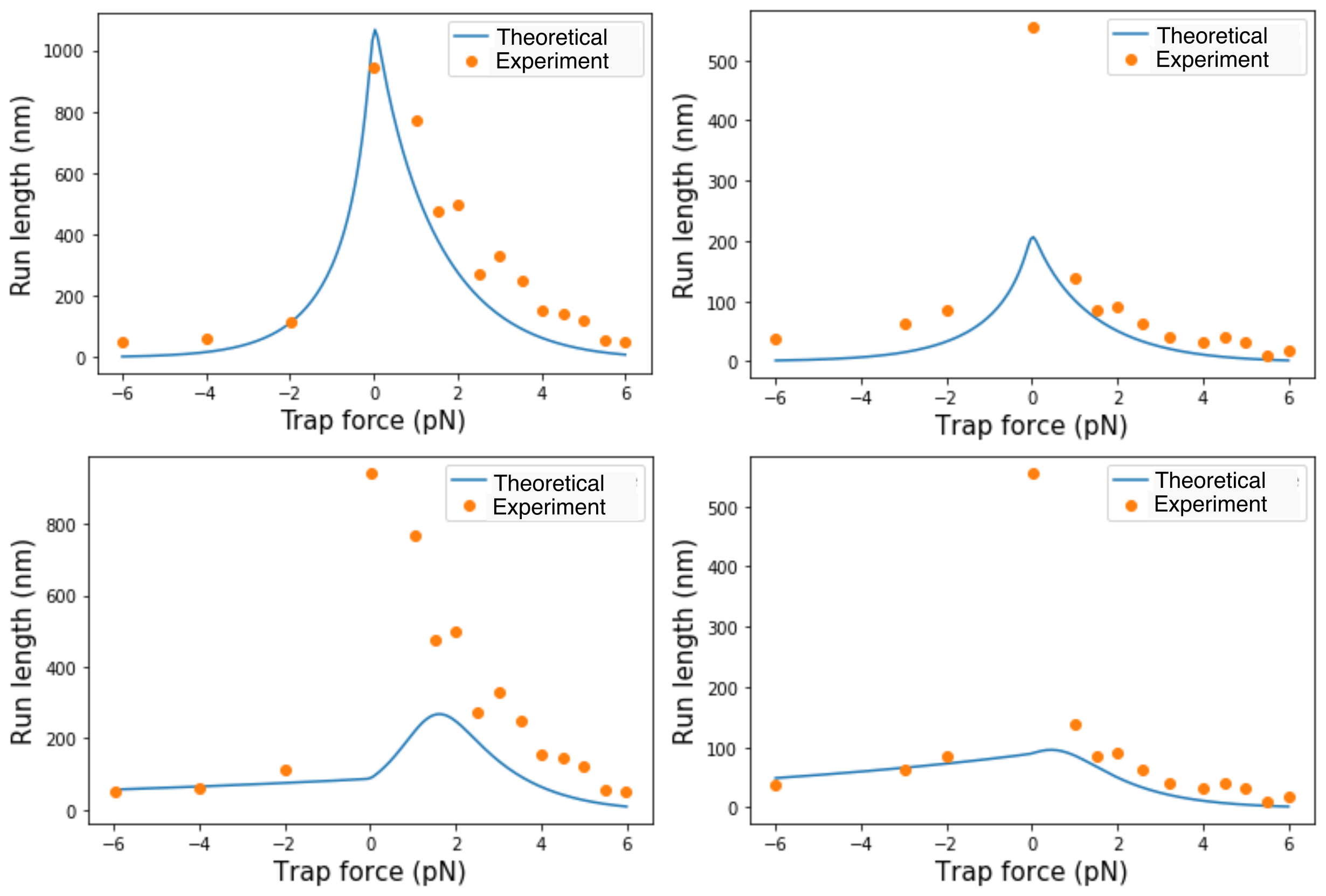}
\caption{Comparison between run length predictions based on double exponential
detachment rate models based on parameterizations from~\citet{milic2017intraflagellar,milic2014kinesin}.
 Experimental data is plotted as discrete points, with kinesin-1 in the left
column and kinesin-2 in the right column.  The continuous curves indicate
the theoretical prediction for run length using:  (\textbf{top}) continuous
detachment rate model~\eqref{detmod} with parameters $d_{0}^{(1)}
 = .7$/s, $ d_{0}^{(2)} = 2.3$/s,   $F_{d+}^{(1)} = 2.1$ pN, $F_{d+}^{(2)}
= 2.0$ pN,  $F_{d-}^{(1)} = 15.2$ pN ,and  $F_{d-}^{(2)}
= 9.8$ pN. (\textbf{bottom}) discontinuous detachment rate model~\eqref{detmoddisc}
with parameters   $ d_{0+}^{(1)} = .7$/s, $ d_{0+}^{(2)} = 2.3$/s, $ d_{0-}^{(1)}=
9.1$/s, $ d_{0-}^{(2)}= 5.6$/s, $d_{0}^{(1)}
 = .7$/s, $ d_{0}^{(2)} = 2.3$/s,   $F_{d+}^{(1)} = 2.1$ pN, $F_{d+}^{(2)}
= 2.0$ pN,  $F_{d-}^{(1)} = 15.2$ pN ,and  $F_{d-}^{(2)}
= 9.8$ pN.  \label{fig:milicexp2}  }
 \end{figure}

We  present here results from two numerical experiments which illustrate how
cargo fluctuations affect run lengths under different parameterizations of double exponential detachment rate models.  We use the same approach as in Section~\ref{sec:opt} to take a relation between detachment rate and instantaenous force together with the fixed relation~\eqref{gbari} between velocity and instantaneous force to produce a theoretical relation between the run length and applied trap force.  For our first example, we use the  continuous detachment rate model (\ref{detmod}), but now using parameters
from \cite{milic2014kinesin,milic2017intraflagellar} which were obtained
for  their discontinuous model
\begin{equation}
d^{(i)}(F) = \begin{cases}d_{0+}^{(i)} \exp(-F/F_{d-}^{(i)}) & F \le 0, \\
d_{0-} ^{(i)}\exp(F/F_{d+}^{(i)}) & F>0.  \label{detmoddisc}
\end{cases}
\end{equation}
In particular  we  used a zero-force detachment rates  of $d_{0}^{(1)}
 = .7$/s for kinesin-1 and $ d_{0}^{(2)} = 2.3$/s for kinesin-2,   hindering force scales
  $F_{d+}^{(1)} = 2.1$ pN and $F_{d+}^{(2)} = 2.0$ pN, and assisting force scales  $F_{d-}^{(1)} = 15.2$ pN and  $F_{d-}^{(2)}
= 9.8$ pN. We note that for fitting assisting
force scales $F_{d-}^{(i)}$, run lengths from zero force trap forces were
not used in    \cite{milic2014kinesin,milic2017intraflagellar}.  
A comparison
of run lengths found from the averaging theory and experimental data is shown in the top panels of Fig.
\ref{fig:milicexp2}.  The agreement between them for kinesin-1 is somewhat reasonable, although experimental
values are consistently larger for  hindering forces. The differences are
more pronounced for kinesin-2.  In particular, at zero trap force  cargo
fluctuations  allow for the motor to feel a nonzero force, resulting
in reduced run lengths. The decay of run length for kinesin-1 for small forces
is slower than for kinesin-2, so that cargo fluctuations do not affect run lengths
as drastically.
 
For our second experiment, we adopt the discontinuous model (\ref{detmoddisc}) using the parameterization from~\citet{milic2014kinesin,milic2017intraflagellar}, which differs from our first experiment only in using distinct zero-force limiting detachment rates for assisting and hindering forces:  $ d_{0+}^{(1)} = .7$/s and $ d_{0-}^{(1)}= 9.1$/s for kinesin-1 and 
 $ d_{0+}^{(2)} = 2.3$/s and $ d_{0-}^{(2)}= 5.6$/s for kinesin-2.
 Results for the theoretical predictions for run length as a function of applied trap force are shown in Fig. \ref{fig:milicexp2}.  As is evident from the figure,
the effects of cargo averaging at zero trap force are more drastic than those
seen in the continuous detachment rate model  in both kinesin-1 and 2.  This is readily
explained from the large upward jump in detachment rate when changing from a
small hindering force to small assisting force which is typically felt by motors  when considering
cargo fluctuations. 

\end{appendices}

\bibliography{motorreference}

\end{document}